\documentclass[longauth]{aa}

\usepackage[varg]{txfonts}
\usepackage{natbib}
\usepackage{graphicx}   
\usepackage[x11names]{xcolor}
\usepackage{booktabs}
\usepackage{placeins}
\usepackage{multirow}

\usepackage{hyperref}
\hypersetup{
      colorlinks = false,
      linkcolor = .,
      linkbordercolor = green,
      urlbordercolor = green
}

\bibpunct{(}{)}{;}{a}{}{,}

\newcommand{\nolinenumbers}{{}}

\newcommand{\eps}{\varepsilon}
\newcommand{\xipm}{$\xi_\pm$}
\newcommand{\cosebis}{$E_n/B_n$}
\newcommand{\cosebisE}{$E_n$}
\newcommand{\bandpowers}{$C_{\rm E,B}$}
\newcommand{\bandpowersE}{$C_{\rm E}$}
\newcommand{\dd}{{\rm d}}
\newcommand{\pp}{{\phantom{+}}}

\newcommand{\rband}{{$r$-band}}

\newcommand{\kidz}{{KiDZ}}
\newcommand{\kids}{{KiDS}}

\newcommand{\konek}{{\kids-$1000$}}
\newcommand{\kidslegacy}{{\kids-Legacy}}
\newcommand{\skills}{{SKiLLS}}

\newcommand{\omegacam}{{OmegaCAM}}

\newcommand{\planck}{{\it Planck}}
\newcommand{\euclid}{{\it Euclid}}

\newcommand{\gaia}{{Gaia}}
\newcommand{\sdss}{{SDSS}}
\newcommand{\twomass}{{2MASS}}

\newcommand{\gaap}{{\sc GAaP}}

\newcommand{\theli}{{\sc theli}}

\newcommand{\astrowise}{{\sc Astro-WISE}}

\newcommand{\cosmopipe}{{\sc CosmoPipe}}

\newcommand{\lensfit}{{{\it lens}fit}}

\newcommand{\cosmopower}{{\sc CosmoPower}}
\newcommand{\cosmosis}{{\sc Cosmosis}}
\newcommand{\treecorr}{{\sc Treecorr}}
\newcommand{\onecov}{{\sc OneCovariance}}
\newcommand{\camb}{{\sc camb}}
\newcommand{\nautilus}{{\sc nautilus}}

\newcommand{\drfour}{{\sc dr4}}
\newcommand{\drfive}{{\sc dr5}}
\newcommand{\photoz}{{photo-$z$}}

\newcommand{\nz}{{\ensuremath{N(z)}}}
\newcommand{\dz}{{\ensuremath{\delta(z)}}}
\newcommand{\zb}{{z_{\rm B}}}

\newcommand\Tstrut{\rule{0pt}{2.4ex}}

\begin{document}

\title{KiDS-Legacy: Cosmological constraints from cosmic shear with the complete Kilo-Degree Survey\thanks{\emph{
We dedicate this article to the memory of Nick Kaiser. He turned weak lensing into a powerful and practical cosmology
probe, inspiring us to conceive the Kilo-Degree Survey. The KiDS project has relied greatly on his brilliant and
wide-ranging foundational work.
}}
}
\titlerunning{\kidslegacy\ cosmic shear}

\author{ 
Angus~H.~Wright                       \inst{1}\thanks{awright@astro.rub.de} \and
Benjamin~St\"olzner                   \inst{1}    \and
Marika~Asgari                         \inst{2}  \and
Maciej~Bilicki                        \inst{3}  \and
Benjamin~Giblin                       \inst{4}  \and
Catherine~Heymans                     \inst{1,4}  \and
Hendrik~Hildebrandt                   \inst{1}  \and
Henk~Hoekstra                         \inst{5}  \and
Benjamin~Joachimi                     \inst{6}  \and
Konrad~Kuijken                        \inst{5}  \and
Shun-Sheng~Li                         \inst{1,5}  \and
Robert~Reischke                       \inst{1,7}  \and
Maximilian~von~Wietersheim-Kramsta    \inst{8,9}  \and
Mijin~Yoon                            \inst{5}  \and
Pierre~Burger                         \inst{7,10,11}  \and
Nora~Elisa~Chisari                    \inst{5,12}  \and
Jelte~de~Jong                         \inst{13}  \and
Andrej~Dvornik                        \inst{1}  \and
Christos~Georgiou                     \inst{14}  \and
Joachim~Harnois-D\'eraps              \inst{15}    \and
Priyanka~Jalan                        \inst{3}  \and
Anjitha~John~William                  \inst{3} \and
Shahab~Joudaki                        \inst{16,17}  \and
Giorgio~Francesco~Lesci               \inst{18,19}  \and
Laila~Linke                           \inst{20}  \and
Arthur~Loureiro                       \inst{21,22}  \and
Constance~Mahony                      \inst{1,23,24}  \and
Matteo~Maturi                         \inst{25}  \and
Lance~Miller                          \inst{23}  \and
Lauro~Moscardini                      \inst{18,19,26}  \and
Nicola~R.~Napolitano                  \inst{27}  \and
Lucas~Porth                           \inst{7}  \and
Mario~Radovich                        \inst{28}  \and
Peter~Schneider                       \inst{7}  \and
Tilman~Tr\"oster                      \inst{29}    \and
Edwin~Valentijn                       \inst{5} \and
Anna~Wittje                           \inst{1}  \and
Ziang~Yan                             \inst{1}  \and
Yun-Hao~Zhang                         \inst{4,5}  
}
\authorrunning{The KiDS Collaboration}
\institute{
Ruhr University Bochum, Faculty of Physics and Astronomy, Astronomical Institute (AIRUB), German Centre for Cosmological Lensing, 44780 Bochum, Germany \and 
School of Mathematics, Statistics and Physics, Newcastle University, Herschel Building, NE1 7RU, Newcastle-upon-Tyne, UK \and 
Center for Theoretical Physics, Polish Academy of Sciences, al. Lotników 32/46, 02-668 Warsaw, Poland \and 
Institute for Astronomy, University of Edinburgh, Blackford Hill, Edinburgh, EH9 3HJ, UK \and 
Leiden Observatory, Leiden University, P.O.Box 9513, 2300RA Leiden, The Netherlands \and 
Department of Physics and Astronomy, University College London, Gower Street, London WC1E 6BT, UK \and 
Argelander-Institut für Astronomie, Universität Bonn, Auf dem Hügel 71, D-53121 Bonn, Germany \and 
Institute for Computational Cosmology, Ogden Centre for Fundament Physics - West, Department of Physics, Durham University, South Road, Durham DH1 3LE, UK. \and 
Centre for Extragalactic Astronomy, Ogden Centre for Fundament Physics - West, Department of Physics, Durham University, South Road, Durham DH1 3LE, UK \and 
Waterloo Centre for Astrophysics, University of Waterloo, Waterloo, ON N2L 3G1, Canada \and 
Department of Physics and Astronomy, University of Waterloo, Waterloo, ON N2L 3G1, Canada \and 
Institute for Theoretical Physics, Utrecht University, Princetonplein 5, 3584CC Utrecht, The Netherlands. \and 
Kapteyn Astronomical Institute, University of Groningen, PO Box 800, 9700 AV Groningen, The Netherlands \and 
Institut de Física d’Altes Energies (IFAE), The Barcelona Institute of Science and Technology, Campus UAB, 08193 Bellaterra (Barcelona), Spain \and 
School of Mathematics, Statistics and Physics, Newcastle University, Herschel Building, NE1 7RU, Newcastle-upon-Tyne, UK \and 
Centro de Investigaciones Energéticas, Medioambientales y Tecnológicas (CIEMAT), Av. Complutense 40, E-28040 Madrid, Spain \and 
Institute of Cosmology \& Gravitation, Dennis Sciama Building, University of Portsmouth, Portsmouth, PO1 3FX, United Kingdom \and 
Dipartimento di Fisica e Astronomia ``Augusto Righi'' - Alma Mater Studiorum Università di Bologna, via Piero Gobetti 93/2, 40129 Bologna, Italy \and 
INAF-Osservatorio di Astrofisica e Scienza dello Spazio di Bologna, Via Piero Gobetti 93/3, 40129 Bologna, Italy \and 
Universität Innsbruck, Institut für Astro- und Teilchenphysik, Technikerstr. 25/8, 6020 Innsbruck, Austria \and 
The Oskar Klein Centre, Department of Physics, Stockholm University, AlbaNova University Centre, SE-106 91 Stockholm, Sweden \and 
Imperial Centre for Inference and Cosmology (ICIC), Blackett Laboratory, Imperial College London, Prince Consort Road, London SW7 2AZ, UK \and 
Department of Physics, University of Oxford, Denys Wilkinson Building, Keble Road, Oxford OX1 3RH, United Kingdom. \and 
Donostia International Physics Center, Manuel Lardizabal Ibilbidea, 4, 20018 Donostia, Gipuzkoa, Spain. \and 
Zentrum für Astronomie, Universitatät Heidelberg, Philosophenweg 12, D-69120 Heidelberg, Germany; Institute for Theoretical Physics, Philosophenweg 16, D-69120 Heidelberg, Germany \and 
Istituto Nazionale di Fisica Nucleare (INFN) - Sezione di Bologna, viale Berti Pichat 6/2, I-40127 Bologna, Italy \and 
Department of Physics “E. Pancini” University of Naples Federico II C.U. di Monte Sant’Angelo Via Cintia, 21 ed. 6, 80126 Naples, Italy \and 
INAF - Osservatorio Astronomico di Padova, via dell'Osservatorio 5, 35122 Padova, Italy \and 
Institute for Particle Physics and Astrophysics, ETH Zürich, Wolfgang-Pauli-Strasse 27, 8093 Zürich, Switzerland
}

\date{Received 31 March 2025 / Accepted 31 July 2025 }

\graphicspath{{./figures/}}

\abstract{ 
We present cosmic shear constraints from the completed Kilo-Degree Survey (\kids), where the cosmological parameter
$S_8\equiv\sigma_8\sqrt{\Omega_{\rm m}/0.3} = 0.815^{+0.016}_{-0.021}$ is found to be in agreement ($0.73\sigma$) with
results from the \planck\ Legacy cosmic microwave background experiment.  The final KiDS footprint spans $1347$~square
degrees of deep nine-band imaging across the optical and near-infrared (NIR), along with an extra 23-square degrees of
KiDS-like calibration observations of deep spectroscopic surveys.  Improvements in our redshift distribution estimation
methodology, combined with our enhanced calibration data and multi-band image simulations, allowed us to extend our lensed
sample out to a photometric redshift of $z_{\rm B}\leq2.0$.   Compared to previous \kids\ analyses, the increased survey
area and redshift depth results in a $\sim32\%$ improvement in constraining power in terms of
$\Sigma_8\equiv\sigma_8\left(\Omega_{\rm m}/0.3\right)^\alpha = 0.821^{+0.014}_{-0.016}$, where $\alpha = 0.58$ has been
optimised to match the revised degeneracy direction of $\sigma_8$ and $\Omega_{\rm m}$ for our current survey at higher
redshift.  We adopted a new physically motivated intrinsic alignment (IA) model that  jointly depends on the galaxy
sample’s halo mass and spectral type distributions, and which is informed by previous direct alignment measurements. We
also marginalised over our uncertainty on the impact of baryon feedback on the non-linear matter power spectrum.
Compared to previous KiDS analyses, we conclude that the increase seen in $S_8$ primarily results from our improved
redshift distribution estimation and calibration, as well as a new survey area and improved image reduction.  Our
companion paper presents a full suite of internal and external consistency tests (including joint constraints with other
datasets), finding the \kidslegacy\ dataset to be the most internally robust sample produced by \kids\ to date. 
}
 
\keywords{cosmology: observations -- gravitational lensing: weak -- galaxies: photometry -- surveys}

\maketitle
\tableofcontents

\section{Introduction} 
\label{sec:intro}

Our current understanding of the large-scale structure (LSS) of the Universe, obtained over the last three decades thanks to
increasingly sophisticated astronomical observations, can be regarded as one of the great successes of modern scientific
inquiry.
Detailed observations of the LSS probe fundamental physics,
since the structures we observe were formed by an interplay of physical effects. These primarily include\ the self-gravity of
matter, cosmic expansion, astrophysical feedback processes, and the repulsive effect of the cosmological constant or dark
energy.

When it comes to understanding the growth and evolution of LSS, the cosmological stage is set by observations of the
cosmic microwave background (CMB) that reveal the structure of dark and baryonic matter at early times. Together with
other probes, including big-bang nucleosynthesis (BBN) and baryon-acoustic oscillations (BAOs), the CMB has been used to
establish and tightly constrain the so-called standard model of cosmology, $\Lambda$CDM. It is a strikingly simple model based
on only a handful of numerical parameters \citep{planck/cosmo:2018,frank/etal:2024,ge/etal:2024} and
characterised, in particular, by two components that are at the origin of its name: a cosmological constant ($\Lambda$, which
dominates the energy landscape at late times) and a cold dark matter (CDM) component that is  dominant
over baryonic matter by a factor of roughly five. Observations of the CMB constrain the parameters of this model at high redshift ($z\sim 1100$),
when the Universe was approximately $370$ thousand years old. However, the model can be subsequently used to
predict the properties of the LSS at late times ($z\lesssim 2$; an extrapolation over more than 10 billion years), which
can then be tested with low-redshift probes of cosmic structures.

Of the low-redshift probes, the relative prevalence of CDM poses a problem for those that rely on luminous
tracers, such as galaxy clustering measurements. Modelling and constraining the relationship between luminous and dark
matter (DM), also known as galaxy bias, constitutes one of the major systematic challenges of such measurements. In contrast,
observations of the weak gravitational lensing effect of the LSS are intrinsically sensitive to all matter (dark or
otherwise), eliminating the systematic uncertainty of galaxy bias. This unique feature has turned such `cosmic shear'
experiments into a major cosmological probe that is just entering its golden age. Indeed, it is the combination of
cosmic shear with galaxy clustering -- and their cross-correlation called galaxy-galaxy lensing (GGL) -- that will
potentially yield the most precise test of the standard model.

Over the past decade, the comparison between observations of low-$z$ LSS and their CMB-based predictions has revealed a
tension in the $S_8\equiv\sigma_8\sqrt{\Omega_{\mathrm{m}}/0.3}$ parameter. Therefore, it has been the focus of extensive
study. All contemporary cosmic shear measurements to date, except one, have yielded values of $S_8$ that are lower than
CMB predictions, indicating a universe with less structure and/or matter. The significance of these measurements varies,
but there is a clear tendency towards lower values in the cosmic shear results of the past decade.
Only the Deep Lens Survey \citep[DLS;][]{jee/etal:2016} has reported a value that is higher than that of \planck, and
even then only slightly. In
contrast, lower values have been reported by the Canada-France-Hawaii Telescope Lensing Survey
\citep[CFHTLenS;][]{heymans/etal:2013}, as well as from previous results from the ongoing `stage-III' surveys by the Kilo-Degree Survey 
\citep[KiDS;][]{asgari/etal:2021,li/etal:2023b}, the Dark Energy Survey \citep[DES;][]{amon/etal:2022,secco/etal:2022},  
the Hyper-Suprime Camera survey \citep[HSC;][]{li/etal:2023,dalal/etal:2023}, and a joint analysis by  \cite{des/kids:2023}.  
Other low-$z$ LSS probes typically also share this trend but again with a varying level of significance
\citep[see][for an overview]{abdalla/etal:2022}.

Cosmic shear probes relatively low redshifts, and relatively small scales compared to other probes of LSS \citep[see
e.g. Figure~1 of][for a graphical summary]{broxterman/etal:2024}. This makes cosmic shear somewhat 
more sensitive than other low-$z$ probes to non-linear structure formation and astrophysical effects: the latter mainly involve 
feedback processes driven by supernovae and active galactic nuclei, collectively referred to below as baryonic feedback. It has therefore been suggested
that inaccuracies in the modelling of non-linear scales might be the reason for the tension in $S_8$, rather than a more
fundamental cosmological cause \citep{yoon/etal:2019,yoon/etal:2021,amon/etal:2022b,preston/etal:2023}.

A definitive answer to the reality of the $S_8$ tension is within reach. Currently, three stage-III surveys have finished their
observations and are preparing their final ana\-lyses; one of which we present here. The next generation of surveys,
called stage-IV, are just beginning, represented by the ESA \euclid\ mission \citep{euclid/overview:2024}, the Legacy
Survey of Space and Time (LSST) conducted with the Vera C. Rubin Observatory \citep{lsst:2019}, the NASA Roman Space
Telescope \citep{wfirst/roman:2015}, and the Chinese Space Station Optical Survey \citep{gong/etal:2019}. We note that
\euclid\ is expected to yield first cosmologically relevant results in 2026.  These stage-IV surveys will soon provide
datasets capable of measuring the growth of structure at low $z$ with unprecedented precision.

New facilities and new instruments invariably bring along new challenges owing to, for instance, the novelty 
of their hardware. Comparatively, the final (usually called `legacy') releases of a survey typically represent the
culmination of years or decades of analysis, understanding, and refinement regarding their survey and the associated
data. As such, the final releases of stage-III surveys will provide a crucial insight and reference for subsequent
lensing studies of LSS with stage-IV surveys.

Here, we present the cosmic shear analysis of the completed Kilo-Degree Survey (KiDS) based on the fifth and final data
release \citep[KiDS-\drfive;][]{wright/etal:2024}, dubbed \kidslegacy. With extensive wavelength coverage enabled by its
unique pairing with its near-infrared (NIR) sister survey VIKING (VISTA Kilo-Degree Infrared Galaxy Survey) and with its
superb image quality in the $r$-band, KiDS is the stage-III survey that is arguably most suited to simultaneously
leverage high-redshift galaxies and accurately calibrate their redshift distributions. Over the past decade, KiDS has
been pushing  to increasingly high redshifts, while continuously improving the redshift calibration,
shear measurements,  shear calibration through image simulations, and methodology to accurately extract
cosmological constraints from the imaging data.

This paper represents the culmination of these efforts, providing the most robust and precise analysis of \kids\ to date.
Our dataset is the largest analysed by \kids\ thus far (Sect. \ref{sec: kids}), utilising the most extensive
spectroscopic redshift compilation prepared for any cosmic shear analysis (Sect. \ref{sec: kidz}), and has been calibrated using
the most sophisticated simulations produced for this purpose (Sect. \ref{sec:redshift}).  Improvements in methodology
(Sect.~\ref{sec:theory}) have allowed us to access the highest redshifts (with the most tomographic bins) studied for cosmic
shear to date (Sect. \ref{sec:redshift}). An efficient workflow for this study was enabled by 
the new \cosmopipe\ analysis framework, which we have also made publicly available. We present null tests of our dataset in
Sect.~\ref{sec:nulltests}. The cosmological
results are presented in Sect.~\ref{sec:results}, discussed in Sect.~\ref{sec:discussion}, and summarised in
Sect.~\ref{sec:summary}. Additionally, we include supplementary information regarding our power spectrum emulation
(Appendix~\ref{sec:cosmopower}), on the definition of parameters for our fiducial intrinsic alignment model
(Appendix~\ref{sec:ia_massdep}), construction of band power spectra (Appendix~\ref{sec:bandpowers}), covariance validation
(Appendix~\ref{sec:glass}), $B_n$-mode exploration (Appendix~\ref{sec:bmodeinvestigation}), correlation function
cosmological analyses (Appendix~\ref{sec:xipmcos}), detailed posterior analyses (Appendix~\ref{sec:posteriors}), tables
of cosmological constraints (Appendix~\ref{sec:tables}), and details of changes made to the analysis and manuscript
post-unblinding (Appendix~\ref{sec:updates}). Finally, this paper has been released alongside two companion papers: the work of
\citet{stoelzner/etal:2025} presents  an extensive analysis of internal and external consistency tests of the \kidslegacy\
dataset (including joint constraints with other datasets) and the work of \citet{wright/etal:2025} describes the various redshift
distribution and bias estimation developments made for \kidslegacy.

\section{Modelling and theoretical framework}
\label{sec:theory}

The main cosmic shear observable is the ellipticity correlation between pairs of galaxies. The signal depends on the two-point
statistics of the shear and galaxy intrinsic alignment fields, which both trace the matter distribution and therefore
depend on cosmological information. 
For our predictions and sampling, we used \cosmosis, a modular cosmology pipeline
described in \citet{zuntz/etal:2015}. We included a number of new and/or modified modules in our analysis, including 
a \cosmopower\ emulator for non-linear power spectra (Appendix \ref{sec:cosmopower}) and new intrinsic alignment models (Sect.
\ref{sec:IA}).   In this section, we start with the derivation of cosmic shear power spectra from
matter power spectra and then detail our choices of intrinsic alignment models. Next, we introduce our observable
two-point statistics, briefly discussing their measurement and prediction methodology. We end this section with a brief
review of our analytical covariance modelling.

\subsection{Shear power spectrum}

Our two-point statistic predictions are calculated from shear (cross-)power spectra of galaxy samples $i$ and $j$, 
$C^{(ij)}_{\rm GG}(\ell)$, where ${\rm G}$ stands
for `gravitational', which distinguishes these power spectra from intrinsic alignment contributions to the two-point
signal.  
Shear power spectra are given by Limber-approximated projections 
\citep{kaiser:1992,loverde/afshordi:2008,kilbinger/etal:2017} taking the form
\begin{equation}
\label{eq:cs_limber}
C^{(ij)}_{\rm GG}(\ell) = \int^{\chi_{\rm hor}}_0 \!\!\! \dd \chi\;
\frac{W^{(i)}_{\rm G} (\chi)\; W^{(j)}_{\rm G} (\chi)}{f^2_K(\chi)}\; P_{\rm m, nl} \left(\frac{\ell+1/2}{f_K(\chi)},z(\chi) \right)\;,
\end{equation}
where $P_{\rm m, nl}$ is the non-linear matter power spectrum and $f_K(\chi)$ is the comoving angular diameter
distance. 
Although the integral in Eq.~\eqref{eq:cs_limber} technically runs over the entire line of sight to the horizon, $\chi_{\rm hor}$, 
in practice, the weak lensing kernel, $W^{(i)}_{\rm G} (\chi)$, 
limits the integral to the comoving distance of the furthest galaxy in our sample, expressed as 
\begin{equation}
\label{eq:kernel_lensing}
W^{(i)}_{\rm G} (\chi) = \frac{3 H_0^2 \Omega_{\rm m}}{2\, c^2}
\frac{f_K(\chi)}{a(\chi)} \int_{\chi}^{\chi_{\rm hor}} \dd
\chi'\; n^{(i)}_{\rm S}(\chi')\; \frac{f_K(\chi' - \chi)}{f_K(\chi')}\;,
\end{equation}
where $n^{(i)}_{\rm S}(\chi)$ is the distribution of comoving distances of galaxies in source sample,
$i$; in practice, we express this in terms of redshift, namely, $\nz\, \dd z=n_{\rm S}(\chi)\, \dd \chi $. Here,
$\Omega_{\rm m}$ is the total matter density parameter today, $H_0$ is the Hubble constant, $c$ is the speed of light, and
$a(\chi)$ is the scale factor normalised to unity today. 

\subsection{Matter power spectrum}

In our cosmic shear analysis, we remained mostly sensitive to Fourier scales $ 0.1\, h/{\rm Mpc} \lesssim k \lesssim 1\,
h/{\rm Mpc}$, with a weighting of scales that depends on the two-point statistic (see Sect.~\ref{sec:2pt}).
These scales are well within the (quasi-) non-linear regime and therefore require an accurate non-linear power
spectrum model for their predictions.  
Currently, there are multiple approaches that can produce non-linear matter power spectra, with varying levels of accuracy, for
this range of scales. Here, we followed the most popular approach for cosmic shear analyses, using a method inspired by the
halo model and calibrated against simulations. Specifically, we used the Boltzmann solver \camb\ \citep{lewis/etal:1999,howlett/etal:2012} to estimate the linear matter power spectrum
and the augmented halo model {\sc `HMCode2020'} \citep{mead/etal:2021} to derive its non-linear evolution.

{\sc HMCode2020} has a similar level of accuracy as popular emulators trained directly on N-body simulations 
\citep[$2.5\%$; 
see e.g.][]{knabenhans/etal:2023,arico/etal:2021b} and is twice as accurate as 
{\sc Halofit} from \citet{takahashi/etal:2012}, which has been used in a number of previous cosmic shear analyses 
\citep[see e.g.][]{amon/etal:2022,hikage/etal:2018}.  \citet{des/kids:2023} presented a comparison of 
{\sc HMCode2020} and {\sc Halofit} with {\sc EuclidEmulatorv2,}
 showing that {\sc HMCode2020} produces unbiased results when analysing mock
data created with {\sc EuclidEmulatorv2}; meanwhile {\sc Halofit} tends to overestimate power for lower values of $S_8$,
resulting in an underestimated $S_8$ value.  Similarly, an earlier version of {\sc HMCode},  {\sc HMCode2015} 
\citep{mead:2015}, which was
used in previous KiDS analyses, is $\sim 30\%$ less accurate than {\sc HMCode2020}. This can subsequently produce small
differences between cosmological constraints estimated using the two codes: \citet{troester/etal:2021} demonstrated that
the most important difference between these versions of {\sc HMCode} is in their treatment of baryon feedback, which
results in a slightly higher $S_8$ using {\sc HMCode2020}. Lastly, we note that the parameter values at which {\sc
HMCode} was fitted to simulations are somewhat more constrained than our chosen priors for $h^2\Omega_{\mathrm{m}}$
\citep[see][and Table~\ref{tab:priorrange}]{mead/etal:2021}. However, given that the fitting functions of {\sc HMCode}
are based on physically motivated parameters, we did not anticipate any significant issues just beyond this range.
Furthermore, we note that our analysis remained unaffected by this choice.

\subsection{Baryonic feedback}

The primary features of the matter power spectrum can be determined by assuming all matter is in the form of 
(pressureless) CDM, with modifications due to processes involving baryonic matter. 
At large scales ($ 0.05\; h/{\rm Mpc} \lesssim k\lesssim 0.5\; h/{\rm Mpc}$), BAOs
produce oscillations in the power spectrum. Over a wide range of intermediate scales ($ 0.1\; h/{\rm Mpc}
\lesssim k\lesssim 10\; h/{\rm Mpc}$), power is suppressed through feedback processes from active galactic
nuclei (AGN), reaching a maximum suppression at around $k\sim 5\; h/{\rm Mpc}$. At even smaller scales ($k\gtrsim 10\;
h/{\rm Mpc}$), star formation processes lead to a sharp increase in power \citep[see e.g.][]{semboloni/etal:2011}. 

To model baryonic feedback processes, {\sc HMCode2020} performs a modification to the concentration-mass relation of the
halo profiles as a function of their gas fraction; this, in turn, depends on the strength of the AGN feedback
(parametrised by $\log\left(T_{\rm AGN}\right) \in \left[7.1,8.3\right]$). This feedback model is then calibrated
against a suite of hydrodynamical simulations from BAHAMAS \citep{mccarthy/etal:2017}, which had been generated
using a wide range of effective feedback amplitudes and cosmological models. Recently, \citet[][FLAMINGO]{schaye/etal:2023} presented a new 
suite of hydrodynamical simulations, including emulators for generating power spectra \citep{schaller/etal:2024}, as a mechanism for modelling
baryonic effects. Moving beyond hydrodynamical simulations, \citet[][$\mathrm{SP}(k)$]{salcido/etal:2023}  presented alternative phenomenological
models for the same purpose, while \citet{schneider/teyssier:2015,arico/etal:2020} presented perturbative models that act directly on
the outputs of dark-matter only simulations and can be empirically calibrated
\citep{grandis/etal:2024,schneider.a/etal:2022}. 
We did not utilise these models for our cosmological analyses here, 
as these models were  published over the course of the analysis of \kidslegacy or were not (at time of writing) implemented within 
our \cosmosis\ sampling framework. However, we note that a reanalysis of our data with these models is a natural and straightforward project 
that will be undertaken in a future publication.

\subsection{Intrinsic alignments}
\label{sec:IA}

The intrinsic alignment (IA) of galaxies is correlated with the observed ellipticities of galaxy images, making them 
indistinguishable from the cosmic shear effect we wish to probe for cosmological parameter estimation. In the limit of
small shears and IA, the two-point statistics of galaxy ellipticities are given by
\begin{equation}
\label{eq:cl_cosmicshear}
C^{(ij)}_{\eps \eps}(\ell) = C^{(ij)}_{\rm GG}(\ell) +
C^{(ij)}_{\rm GI}(\ell) + C^{(ij)}_{\rm IG}(\ell) + C^{(ij)}_{\rm II}(\ell)\;,
\end{equation}
where $C^{(ij)}_{\rm GG}$ is the cosmic shear signal given by Eq.~(\ref{eq:cs_limber}). Analogous Limber equations can
be derived for IA contributions, which are typically grouped into two categories. First, intrinsic-intrinsic (II)
correlations represent the correlated alignment of galaxies that are tidally aligned by the same ambient 
LSS: 
\begin{align}
\label{eq:ia_limber_II}
C^{(ij)}_{\rm II}(\ell) &= \int^{\chi_{\rm hor}}_0 \!\!\! \dd \chi\;
\frac{n^{(i)}_{\rm S} (\chi)\; n^{(j)}_{\rm S} (\chi)}{f^2_{\rm
    K}(\chi)}\; P_{\rm II, X}^{(ij)} \left(\frac{\ell+1/2}{f_K(\chi)},z(\chi) \right)\;.
\end{align}
Second, shear-intrinsic (GI) correlations are driven by the mutual gravitational
effects of LSS on (i) galaxies within the structures, and (ii) the lensed images of background
sources that these structures produce \citep{hirata/seljak:2004}:
\begin{align}
\label{eq:ia_limber_GI}
C^{(ij)}_{\rm GI}(\ell) &= \int^{\chi_{\rm hor}}_0 \!\!\! \dd \chi\;
\frac{W^{(i)}_{\rm G} (\chi)\; n^{(j)}_{\rm S} (\chi)}{f^2_{\rm
    K}(\chi)}\; P_{\delta {\rm I}, X}^{(j)} \left(\frac{\ell+1/2}{f_K(\chi)},z(\chi) \right)\;,
\end{align}
with $C^{(ij)}_{\rm IG} = C^{(ji)}_{\rm GI}$.
In Eqs.~(\ref{eq:ia_limber_II}) and (\ref{eq:ia_limber_GI}) the power spectra $P_{\rm II, X}$ and $P_{\delta {\rm I},
X}$ are given by the chosen IA model. We note that the IA contributions feature a more compact kernel along the
line of sight than the lensing kernel in Eq.~(\ref{eq:cs_limber}), and hence the Limber approximation is poorer for
these signals. However, this effect is not significant for our analysis, as the overall IA contributions to the total
signal is small, making up only around 10 percent of the signal in our analysis. Additionally, we are only probing
scales $\ell \gtrsim 50$, where the Limber approximation is sufficient even for more compact kernels
\citep{leonard/etal:2023}.

Since IA are caused by the local matter environment, they have a different redshift scaling than lensing, which however
is insufficient to isolate the cosmological signal without substantial information loss \citep{joachimi/etal:2008}.
Therefore, cosmic shear analyses opt for physically motivated IA models that are inferred jointly with the cosmology
\citep[see][for a review of the field]{lamman/etal:2024}. 
Guided by this, we implement four IA models in
\kidslegacy: a widely used baseline with a single overall amplitude parameter (NLA, Sect.~\ref{sec:NLA}),
two minimal extensions of NLA with either an additional redshift (NLA-$z$, Sect.~\ref{sec:NLAz}) or scale dependence
(via our restricted TATT model, NLA-$k$; Sect.~\ref{sec:NLAscale}), and a version of NLA that explicitly incorporates the well-established trends of IA
with galaxy type and host halo mass (NLA-$M$, Sect.~\ref{sec:NLAmass}). The latter model is the fiducial model chosen for our
cosmological analyses.

\subsubsection{The non-linear linear alignment (NLA) model}
\label{sec:NLA}

The so-called NLA model \citep{bridle/king:2007} assumes that the intrinsic ellipticity of a galaxy depends linearly on
its surrounding tidal quadrupole \citep{catelan/etal:2001}. Moreover, it is commonly assumed that it is the tidal field
at some early time during the formation of the galaxy that sets the amplitude of alignment \citep{hirata/seljak:2004}.
This leads to the following expressions for the IA power spectra,
\begin{align}
  \label{eq:NLA}
P_{\delta {\rm I, NLA}}(k,z) &= - A_{\rm IA}\; \frac{C_1 \rho_{\rm
  cr}\, \Omega_{\rm m}}{D (z)}\; P_{\rm m, nl}(k,z)\;, \\ \nonumber
P_{\rm II, NLA}(k,z) &= A_{\rm IA}^2 \left( \frac{C_1 \rho_{\rm
  cr}\, \Omega_{\rm m}}{D (z)} \right)^2  P_{\rm m, nl}(k,z)\;,
\end{align}
where $C_1 \rho_{\rm cr} \approx 0.0134$ is a constant and $D(z)$ is the linear growth factor normalised to unity at $z=0$. The
model is empirically extended to non-linear scales by employing the full, non-linear matter power spectrum $P_{\rm m,
nl}$, encapsulating a mixture of non-linear gravitational effects as well as alignment processes on the scales of dark
matter halos \citep[cf.][]{schneider/bridle:2010}. The model has a single free parameter, namely, the global, dimensionless
amplitude $A_{\rm IA}$, which necessitates the assumption that there are no coherent trends in how galaxy ellipticities
respond to their local tidal field within the survey's galaxy population. Since $A_{\rm IA}$ is well-constrained by
tomographic cosmic shear data, we chose a wide top-hat prior for $A_{\rm IA}$ to ensure its posterior is data-driven.
All previous \kids\ cosmic shear analyses have used the NLA model as their default and, to date, there has been no evidence that
this simple prescription is insufficient in representing current-generation datasets \citep[e.g.][]{des/kids:2023}.

\subsubsection{Additional redshift dependence (NLA-$z$)}
\label{sec:NLAz}

There is currently no observational evidence of redshift evolution in the amplitude of IA
\citep{joachimi/etal:2011,fortuna/etal:2021b}, but the statistical uncertainties on measurements underpinning this 
statement are still large,
especially when extrapolated to the redshift range of cosmic shear samples. Moreover, $A_{\rm IA}$ could also acquire
an effective redshift scaling; for example, due to a combination of a luminosity dependence and Malmquist bias. Hence, we
tested whether the data prefer $A_{\rm IA}$ to evolve over the tomographic bins. Since individual IA amplitudes per
tomographic bin are poorly constrained \citep[see the detailed analysis in][]{samuroff/etal:2019}, we chose an extended
model that is linear in terms of the scale factor,
\begin{align}
P_{\delta {\rm I, zdep}}^{(i)}(k,z) &= - \left( A_{\rm IA} + B_{\rm IA} \left[ \frac{\langle a \rangle^{(i)}}{a_{\rm piv}} -1 \right] \right) \frac{C_1 \rho_{\rm
  cr}\, \Omega_{\rm m}}{D (z)}\; P_{\rm m, nl}(k,z)\;, \\ \nonumber
P_{\rm II, zdep}^{(ij)}(k,z) &= \left( A_{\rm IA} + B_{\rm IA} \left[ \frac{\langle a \rangle^{(i)}}{a_{\rm piv}} -1 \right] \right)\\ \nonumber
&\times \left( A_{\rm IA} + B_{\rm IA} \left[ \frac{\langle a \rangle^{(j)}}{a_{\rm piv}} -1 \right] \right) \left( \frac{C_1 \rho_{\rm
  cr}\, \Omega_{\rm m}}{D (z)} \right)^2  P_{\rm m, nl}(k,z)\;,
\end{align}
where $\langle a \rangle^{(i)}$ is the average scale factor in tomographic bin $i$, evaluated using the estimated \nz. 
The parameter $A_{\rm IA}$ thus is
the IA amplitude at $a_{\rm piv} \approx 0.769$, which corresponds to the commonly chosen pivot of $z=0.3$ for IA
studies \citep{joachimi/etal:2011}. We set the prior range of the second parameter, $B_{\rm IA}$, by fitting our linear amplitude dependence to the
compilation of \kids\ and \sdss\ measurements studied in \citet[see their Figure~10]{fortuna/etal:2021b}. Increasing the
resulting constraint by $50\,\%$, to account for the limited coverage of these calibration samples and for potential
additional implicit redshift dependencies, we obtain a Gaussian prior on $B_{\rm IA}$ with mean $-3.7$ and standard
deviation $4.3$. For direct comparisons between amplitudes estimated between this model and others, we define
\begin{align}\label{eq:iatot_NLAz}
  A^{{\rm NLAz},(i)}_{\rm IA,total} = \hat{A}_{\rm IA} + \hat{B}_{\rm IA} \left( \frac{\langle a \rangle^{(i)}}{a_{\rm piv}}
  -1\right)\;,
\end{align}
where $\hat{\cdot}$ denotes the best-fitted (i.e. maximum a posteriori) value of each parameter. 

\subsubsection{Restricted TATT model (NLA-$k$)}
\label{sec:NLAscale}

By design, the angular scale dependence of the NLA model is identical to that of the matter power spectrum projected
over the redshift distribution of the sample. A natural way to extend this is to consider next-to-linear alignment
effects, specifically: (i) a density weighting \citep{hirata/seljak:2004}, accounting for the fact that alignments are
only measured at local over-densities of the matter distribution where galaxies reside; and (ii) quadratic alignments as
expected for tidal torquing effects \citep{crittenden/etal:2001,catelan/etal:2001}.

These contributions are subsumed into the tidal alignment, tidal torquing (TATT) model \citep{blazek/etal:2019}, each with a free amplitude parameter. As
can be seen from Figure~1 of \citet{blazek/etal:2019}, the density-weighting and quadratic terms have very similar
dependence on physical scale. Owing to the scale-mixing of angular statistics, cosmic shear measurements poorly
distinguish small and localised scale differences, which leads to pronounced degeneracies among the four TATT parameters
\citep[e.g.][]{samuroff/etal:2019}. To avoid these, we implement the full TATT model within the analysis pipeline, but fix all but one of the free parameters that govern the non-linear terms. We treat this model as a more general modification to NLA, to capture potential deviations in scale from the model. We chose to keep the density
weighting amplitude term free, thereby adding a single additional free parameter, but note that this choice is fairly inconsequential: given 
the aforementioned similarity between the scale dependence of the terms, implementing only the density weighting or only the quadratic term would produce a similar end product. The NLA-$k$ model is then expressed as
\begin{equation}
\label{eq:scaledep}
    \begin{aligned}
P_{\delta {\rm I, scaledep}}(k,z) &= P_{\delta {\rm I, NLA}}(k,z)\\
&\!\!\!\!+ A_{\rm IA}\; b_{\rm src}\; \frac{C_1 \rho_{\rm
  cr}\, \Omega_{\rm m}}{D (z)}\; \left[ A_{0|0E}(k,z) + C_{0|0E}(k,z) \right]\;, \\ 
P_{\rm II, scaledep}(k,z) &= P_{\rm II, NLA}(k,z) \\  &\!\!\!\! + A_{\rm IA}^2 b_{\rm src}^2 \left( \frac{C_1 \rho_{\rm
  cr}\, \Omega_{\rm m}}{D (z)} \right)^2 A_{0\rm E|0\rm E}(k,z) \\ 
  &\!\!\!\!+ 2\, A_{\rm IA}^2 b_{\rm src} \left( \frac{C_1 \rho_{\rm
  cr}\, \Omega_{\rm m}}{D (z)} \right)^2 \left[ A_{0|0\rm E}(k,z) + C_{0|0\rm E}(k,z) \right]\,,
\end{aligned}    
\end{equation}
where the expressions for $A_{0|0\rm E}$, $C_{0|0\rm E}$, and $A_{0\rm E|0\rm E}$ are given in \citet{blazek/etal:2019}. We consider
these terms in the Limber approximation where only transverse modes contribute, and assume that their redshift evolution
is inherited from the linear matter power spectrum. In the original density-weighting term, the free parameter $b_{\rm
src}$ can be interpreted as an effective galaxy bias of the weak lensing source sample, albeit on fairly non-linear
scales. For our model, we chose a top-hat prior in the range $[-0.5,1.5]$ to ensure it encompasses NLA ($b_{\rm
src}=0$) and reasonable values expected for the (unknown) source galaxy bias ($b_{\rm src} \sim 1$).
For comparisons between the alignment amplitude estimated with this model 
and others, we report the alignment amplitude of the NLA component of the model from Eq.~\eqref{eq:scaledep}, namely,
\begin{align}\label{eq:iatot_scaledep}
  A^{\rm scaledep}_{\rm IA,total} = \hat{A}_{\rm IA}\;.
\end{align}

\subsubsection{Galaxy-type and mass-dependent model (NLA-$M$)}
\label{sec:NLAmass}

Two trends in IA strength are firmly established observationally: (i) a dichotomy between late- and early-type galaxies
where only the latter have yielded significant detections on cosmological scales
(\citealt{samuroff/etal:2023,johnston/etal:2019,heymans/etal:2013,georgiou/etal:2025}; see also
\citealt{mandelbaum/etal:2011,tonegawa/etal:2024,mccullough/etal:2024} for additional examples of null or marginal detections of late-type
alignments); and (ii) a scaling of early-type alignment amplitude with host halo mass that is well described by a
power-law \citep{joachimi/etal:2011,fortuna/etal:2025}. Both have solid theoretical underpinning. First, rotationally
supported late-type galaxies are expected to align through tidal torquing, which depends quadratically on the tidal
quadrupole, whereas spheroid shapes should respond linearly and thus be well described by NLA-like models. Second,
\citet{piras/etal:2018} motivated the halo mass dependence through an analytical argument and also reproduced it for
simulated dark matter haloes.

We incorporated these IA trends explicitly into the amplitude of the NLA model. We assumed that, in each tomographic bin, 
only the fraction $f_{\mathrm{r}}$ of `red' (i.e. early-type) galaxies intrinsically align, whereas the remaining galaxies have zero
alignment. The latter assumption is consistent with observations, but within a large statistical uncertainty and
involving a 
substantial extrapolation in sample properties, primarily due to the imperfect relation between galaxy colour and rotational/pressure support
(which is the property most-likely to cleanly stratify intrinsic alignment amplitudes). 
Moreover, we assumed that the red galaxy population  IA
amplitude has a power-law dependence on the average halo mass $\langle M_{\rm h} \rangle$ within a tomographic bin.
This leads to
\begin{align}\label{eq:massdep}
P_{\delta {\rm I, massdep}}^{(i)}(k,z) &= - A_{\rm IA}\; f_{\mathrm{r}}^{(i)}\; \left( \frac{\langle M_{\rm h} \rangle^{(i)}}{M_{\rm h, piv}} \right)^\beta \frac{C_1 \rho_{\rm
  cr}\, \Omega_{\rm m}}{D (z)}\; P_{\rm m, nl}(k,z)\;, \\ \nonumber
P_{\rm II, massdep}^{(ij)}(k,z) &= A_{\rm IA}^2\, f_{\mathrm{r}}^{(i)} f_{\mathrm{r}}^{(j)}\; \left( \frac{\langle M_{\rm h} \rangle^{(i)}}{M_{\rm h, piv}} \right)^\beta \left( \frac{\langle M_{\rm h} \rangle^{(j)}}{M_{\rm h, piv}} \right)^\beta \\ \nonumber
&\times \left( \frac{C_1 \rho_{\rm
  cr}\, \Omega_{\rm m}}{D (z)} \right)^2 \; P_{\rm m, nl}(k,z)\;,
\end{align}
where we adopt the pivot halo mass $M_{\rm h, piv}=10^{13.5}\,h^{-1} M_\odot$ from \citet{fortuna/etal:2025}. The same
authors also fit the power-law dependence to a compilation of early-type IA measurements (see their Figure~4), and we 
adopt their joint posterior on $A_{\rm IA}$ and $\beta$ as our prior for cosmic shear inference. Their posterior is well
approximated by a bivariate Gaussian with $A_{\rm IA} = 5.74 \pm 0.29$, $\beta = 0.44 \pm 0.03$, and a correlation coefficient of
$-0.59$. Details on how we determine $f_{\mathrm{r}}$ and halo masses for our sample of cosmic shear sources, and how we treat
their associated uncertainties, are provided in Appendix~\ref{sec:ia_massdep}.  For comparisons between the
alignment amplitude estimated with this model and others, we define 
\begin{align}\label{eq:iatot_massdep}
  A^{{\rm massdep},(i)}_{\rm IA,total} = \hat{A}_{\rm IA}\; f_r^{(i)}\; \left( \frac{\langle M_{\rm h} \rangle^{(i)}}{M_{\rm h, piv}} \right)^{\hat{\beta}}.
\end{align}

\subsection{Cosmic shear summary statistics}\label{sec:2pt} 

In cosmic shear analyses, it is most convenient to measure the signal using shear two-point correlation functions
(2PCFs) while the modelling is done through shear power spectra (see Eq. \ref{eq:cs_limber}). Alternatively, one could
take the Fourier transform of the shear field and work in harmonic space, which brings the measurements closer to the
theoretical predictions. We chose to use 2PCFs for the measurement, as they shield us from a particular form of 
$E$-/$B$-mode mixing that occurs in harmonic space in the presence of the (unavoidable) image masks. 
In addition, the shot-noise term does not impact 2PCFs (present only in their covariance), as it is only relevant at zero
lag. In harmonic space, both shot noise and masking must be modelled and corrected, prior to beginning cosmological
analyses. For more details see e.g. \citet{alonso/etal:2019}.

Our cosmological analysis here leverages two-point statistics: the complete orthogonal sets of $E/B$-integrals
(COSEBIs, Sect.~\ref{sec:cosebis}) and band power spectra (Sect.~\ref{sec:bp}), constructed 
using linear combinations of finely-binned 2PCFs (Sect.~\ref{sec:xipm}).

\subsubsection{2PCFs (\xipm)}\label{sec:xipm}

The shear two-point correlation functions \citep[$\xi_\pm$,][]{kaiser:1992} are functions of the tangential and cross
components of shear ($\gamma_{\rm t}$ and $\gamma_\times$, respectively) defined with respect to the line connecting
pairs of galaxies with angular separation $\theta$  \citep[see e.g. ][]{bartelmann/schneider:2001},
\begin{align}
\label{eq:xi-pm-def}
\xi_\pm(\theta)  =\langle\gamma_{\rm t} \gamma_{\rm t} \rangle(\theta)\pm \langle\gamma_{\rm \times} \gamma_{\rm \times} \rangle (\theta)\;.
\end{align}
The measured 2PCFs are binned in $\theta$ to reduce measurement noise and compress the data. In practice, we measure
$\xi_\pm(\theta)$ for $1000$ logarithmically spaced $\theta$-bins between $\theta_{\rm min}=2'$ and $\theta_{\rm
max}=300'$ using \treecorr\footnote{We use \treecorr\ version 4.2.3, with binslop$= 1.5$.} \citep{jarvis/etal:2004}. The
inputs to \treecorr\ are galaxy catalogues, which include galaxy ellipticities ($\eps_1$ and $\eps_2$), galaxy
positions (right ascension and declination), and galaxy weights ($w$). The weights used by \treecorr\ are equal to the
recalibrated shape weights (Sect.~\ref{sec: rowestats}) multiplied by the redshift distribution gold-weights
(Sect.~\ref{sec:redshift}). Our $\xi_\pm(\theta)$ are then estimated using
\begin{equation}
\label{eq:xipm_meaure}
\hat\xi^{(ij)}_\pm(\bar{\theta})=\frac{\sum_{ab} w_a w_b
 \left[\eps^{\rm obs}_{{\rm t},a}\,\eps^{\rm obs}_{{\rm t},b}
 \pm\eps^{\rm obs}_{\times,a}\,\eps^{\rm obs}_{\times,b}\right] \Delta^{(ij)}_{ab}(\bar{\theta}) } 
 {\sum_{ab} w_a w_b (1+m_a)(1+m_b)  \Delta^{(ij)}_{ab}(\bar{\theta}) }\ ,
\end{equation}
where $\Delta^{(ij)}_{ab}(\bar{\theta})$ is the bin selection function (which limits the sums to galaxy pairs of
separation within the angular bin labelled by $\bar{\theta}$ and the tomographic bins $i$ and $j$),  and  $\eps^{\rm
obs}_{\rm t}$ and $\eps^{\rm obs}_{\rm \times}$ are the projected tangential and radial components of observed
ellipticities between pairs of galaxies. 
The estimator includes the galaxies' multiplicative shear measurement biases $m_a$
estimated via image simulations (Sect.~\ref{sec:redshift}).

Using the flat-sky approximation, we can relate the 2PCFs to the angular (shear) power spectra by Hankel transformations:
\begin{align}
\label{eq:xipm_th}
\xi_\pm^{(ij)}(\theta) &= \int_0^\infty\frac{ \ell \mathrm{d}\ell }{2 \pi} \; 
  \left[C^{(ij)}_{\eps\eps,\mathrm{E}}(\ell) \pm C^{(ij)}_{\eps\eps,\mathrm{B}}(\ell)\right] \; 
  {\mathrm{J}}_{0/4}(\ell \theta)\;,
\end{align}
where $C^{(ij)}_{\eps\eps,\mathrm{E/B}}(\ell)$ are the $E$/$B$-mode angular power spectra for redshift bins $i$ and
$j$, and ${\mathrm{J}}_{0/4}(x)$ are the zeroth- and fourth-order Bessel functions of the first kind.
We expect cosmic shear to produce only $E$-mode signals up to first order \citep{schneider/etal:1998,hilbert/etal:2009}, while data
systematic errors can in principle create both $E$ and $B$ modes. As a result, $B$-modes have long been used as a diagnostic tool
in cosmic shear analyses \citep[see e.g.][]{pen/etal:2002,hoekstra/etal:2002,heymans/etal:2005,kilbinger/etal:2013}. 

As \xipm\ mix $E$ and $B$ modes, they are
not ideal for cosmic shear analysis. 
In addition, they 
variably weight a large range of physical scales in each angular bin, increasing sensitivity to scale-dependent effects 
such as baryonic feedback (which must be marginalised over using imperfect models). 
Figure~\ref{fig:filters} shows the range of $\ell$ probed by the \xipm\ over a range of angular separations. The
${\rm J}_0$ filter in Eq.~\eqref{eq:xipm_th} means that $\xi_+$ mixes a very wide range of scales ($1
\lesssim\ell\lesssim 10^4$) and ${\rm J}_4$ makes $\xi_-$ particularly sensitive to smaller scales ($10
\lesssim\ell\lesssim 10^5$). This breadth can create a problem when using $\xi_\pm$ for inference, as the power
spectra and intrinsic alignment models become increasingly uncertain at smaller scales. 
Consequently, we expect the
cosmological constraints to be possibly affected by a range of systematic effects that may introduce biases and, as
such, we do not include \xipm\ in our primary cosmological analyses. Nevertheless, results for \xipm\ are provided for
completeness and comparison with previous work, in Appendix~\ref{sec:xipmcos}.

\subsubsection{COSEBIs (\cosebis)}\label{sec:cosebis}

The COSEBIs \citep{schneider/etal:2010b} were originally designed to cleanly separate $E$- and $B$-modes over a finite $\theta$-range. This was achieved
through the use of filter functions $T_{\pm n}(\theta)$, which form complete bases, creating well-defined COSEBIs $E$- and
$B$-modes (\cosebis). COSEBIs can be calculated or measured via a linear transformation of $\xi_\pm$,
\begin{equation}
\begin{aligned}
\label{eq:COSEBIsReal}
 \hat E_n &= \frac{1}{2} \int_{\theta_{\rm min}}^{\theta_{\rm max}}
 \dd\theta\,\theta\: 
 [T_{+n}(\theta)\,\xi_+(\theta) +
 T_{-n}(\theta)\,\xi_-(\theta)]\;, \\ 
 \hat B_n &= \frac{1}{2} \int_{\theta_{\rm min}}^{\theta_{\rm
     max}}\dd\theta\,\theta\: 
 [T_{+n}(\theta)\,\xi_+(\theta) -
 T_{-n}(\theta)\,\xi_-(\theta)]\;.
\end{aligned}    
\end{equation}

The COSEBIs modes $n$ start from $1$ and go to infinity. However, the first few COSEBIs are sufficient to capture 
essentially
all the cosmological information in each pair of tomographic bins \citep{asgari/etal:2012}.  This means that
COSEBIs naturally compress the data into well-defined modes, once $\theta_{\rm min}$ and $\theta_{\rm max}$ are defined,
without any need for subsequent data compression. In previous \kids\ analyses \citep[e.g.][]{asgari/etal:2021}, the
first five COSEBIs modes were used for cosmological analyses. In this work, however, we add the
sixth mode, as we found that (with our increased constraining power, and larger redshift baseline) the sixth mode
contains relevant cosmological information and thus improves our constraining power on $S_8$. Conversely, the addition
of modes $7-20$ did not result in any improvement of constraining power, returning constraints identical to those
measured in the $6$ mode case. 

\begin{figure}
    \centering
    \includegraphics[width=\linewidth]{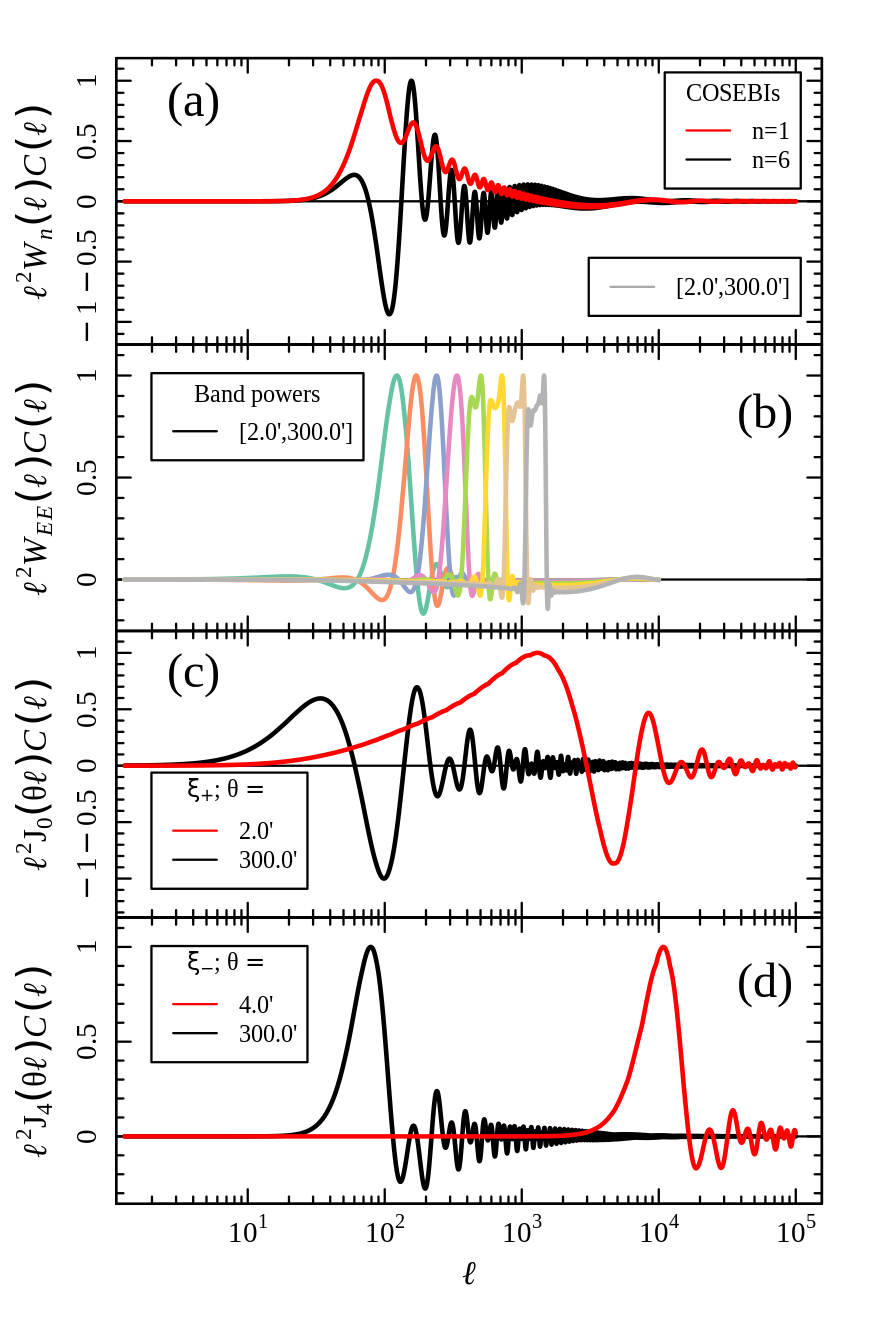}
    \caption{Integrands of the transformation between the angular power spectrum and \cosebis\
    (Eq.~\ref{eq:EnBnFourier}, panel {\em `a'}), \bandpowers\ (Eq.~\ref{eq:bp_cl}, panel {\em `b'}), 
    and \xipm\ (Eq.~\ref{eq:xipm_th}, panels {\em `c'} and {\em `d'}). All solid integrands
    are normalised by their maximum absolute value. For both COSEBIs and 2PCFs, we show the limiting ranges of the statistics
    ($n\in\{1,6\}$ for $E_n$ and $\theta\in\{2.0,300.0\}$ arcmin for \xipm). For \bandpowers\ we show all eight bins
    used in \kidslegacy, defined between $\ell=100$ and $\ell=1500$. Note that, unlike previous \kids\ analyses, our new
    treatment of $T(\theta)$ means that the \bandpowers\ initially go to zero within their desired $\ell$ ranges. 
    }
    \label{fig:filters}
\end{figure}

We use the relation between COSEBIs and shear power spectra to make our predictions,
\begin{equation}
\begin{aligned}
\label{eq:EnBnFourier}
E_n &= \int_0^{\infty}
\frac{\dd\ell\,\ell}{2\pi}C_{\mathrm{EE}}(\ell)\,W_n(\ell)\;,\\ 
B_n &= \int_0^{\infty}
\frac{\dd\ell\,\ell}{2\pi}C_{\mathrm{BB}}(\ell)\,W_n(\ell)\;,
\end{aligned}    
\end{equation}
where the weight functions $W_n(\ell)$ are
\begin{equation}
\begin{aligned}
\label{eq:Wn}
W_n(\ell) & =  \int_{\theta_{\rm{min}}}^{\theta_{\rm{max}}}\dd\theta\:
\theta\:T_{+n}(\theta) {\mathrm{J}}_{0}(\ell \theta) \\ 
& = \int_{\theta_{\rm{min}}}^{\theta_{\rm{max}}}\dd\theta\:
\theta\:T_{-n} (\theta) {\mathrm{J}}_{4}(\ell \theta)\;.
\end{aligned}    
\end{equation}
Figure~\ref{fig:filters} shows two example integrands from Eq.~\eqref{eq:EnBnFourier}. 
$W_n(\ell)$ are highly oscillatory weight functions with a limited range of support over $\ell$. This allows us to
truncate our integrals in Eq.~\eqref{eq:EnBnFourier} to a finite range in $\ell$, as the $E_n$ integrand in
Eq.~\eqref{eq:EnBnFourier} is practically zero outside $10 \lesssim\ell\lesssim 10^3$.

\subsubsection{Band powers (\bandpowers)}\label{sec:bp}

Band power spectra (hereafter `band-powers') refer to binned angular power spectra measured by integrating over
$\xi_\pm(\theta)$ \citep{schneider/etal:2002a,vanuitert/etal:2016,joachimi/etal:2021}. Ideally, we want a finite number
of bins, labelled as $l$, where each bin is only sensitive to a narrow range of $\ell$-values between $\ell_{{\rm
lo},l}$ and $\ell_{{\rm up},l}$. We also want the $E$- and $B$-mode band-powers to be pure, namely, no leakage from $B$ modes into $E$ modes and vice versa. For this ideal case, we can express the following:\ 
\begin{equation}
\label{eq:bp_ideal}
{\cal C}_{{\rm E/B},l}^{(ij)} := \frac{1}{{\cal N}_l} \int_0^\infty \dd \ell\, \ell\; S_l(\ell)\; C_{\eps \eps, {\rm E/B}}^{(ij)}(\ell)\;,
\end{equation}
for redshift bins $i$ and $j$, where $S_l(\ell)$ is the response function for bin $l$. Here we have normalised the integral with
\begin{equation}
\label{eq:bp_norm_ideal}
{\cal N}_l = \int_0^\infty \dd \ell\, \ell^{-1}\; S_l(\ell)\;.
\end{equation}
In this work, we consider the ideal case of a top-hat response function between $\ell_{{\rm lo},l}$ and $\ell_{{\rm up},l}$ which simplifies the normalisation to
\begin{equation}
\label{eq:normalisation}
{\mathcal N}_l = \log(\ell_{{\rm up},l})-\log(\ell_{{\rm lo},l})\;,
\end{equation}
and produces a band-power that traces $\ell^2 C(\ell)$ at the logarithmic centre of the bin. We define eight band-powers
logarithmically binned between $\ell_{\rm min}=100$ and $\ell_{\rm max}=1500$.  In practice, measuring band-powers with
perfect top-hat functions is infeasible, as $\xi_{\pm}(\theta)$ is only measured for a finite $\theta$-range. This has
the adverse effect that our band-powers will inevitably mix $E$ and $B$ modes\footnote{Other binning options are also
explored in the literature, which can alleviate some of the issues with the top-hat response function at the expense of
broadening the range of $\ell$-values that contribute to each bin \citep[see e.g.][]{becker/etal:2016}.}.  The mixing
increases for narrower $\ell$-bins and when the available range of $\theta$-scales is smaller \citep{asgari/etal:2015}.
With this definition for
band-powers, we forgo perfect $E$-/$B$-mode separation in favour of more control over the $\ell$-values that contribute
to our analyses. We present the detailed derivation of our band-power windows ${\cal C}_{{\rm E},l}$ in
Appendix~\ref{sec:bandpowers}, and show the integrands of ${\cal C}_{{\rm E},l} $ in Fig.~\ref{fig:filters}. 
Note, however, that our integrands behave differently to those in \cite{asgari/etal:2021}, due to our different
treatment of the edges of $T(\theta)$ (Appendix~\ref{sec:bandpowers}) and due to our different scale-cuts.

\subsection{Analytic covariance modelling}\label{sec: onecov}

We  presented the covariance modelling for \kidslegacy\ and the {\sc{OneCovariance}} code\footnote{\href{https://github.com/rreischke/OneCovariance}{https://github.com/rreischke/OneCovariance}} in \citet{reischke/etal:2024}. In general, the methodology for the cosmic shear part still follows closely the covariance
modelling of KiDS-1000 \citep{joachimi/etal:2021}. Major changes to the previous analysis include: $(i)$ the possibility of
including non-uniform source distributions in the mixed term of the covariance from the triplet counts of the survey;
$(ii)$ including non-binary masks for the area calculation and survey response;  and$(iii)$ offering a more generalised framework
for a consistent treatment of all summary statistics used in KiDS-Legacy.

The cosmic shear covariance, $\boldsymbol{\tens{C}}$, generally consists of four contributions:
\begin{equation}
    \label{eq:covariance_general}
    \boldsymbol{\tens{C}} = \boldsymbol{\tens{C}}_{\mathrm{G}} + \boldsymbol{\tens{C}}_{\mathrm{NG}} +  \boldsymbol{\tens{C}}_{\mathrm{SSC}} + \boldsymbol{\tens{C}}_{\mathrm{mult}}\;, 
\end{equation}
with the Gaussian contribution G, the non-Gaussian term NG due to non-linear in-survey modes,  the super-sample covariance SSC from modes
larger than the survey, and lastly the contribution due to the uncertainty in the
multiplicative shear bias calibration, $\boldsymbol{\tens{C}}_{\mathrm{mult}}$. \citet{asgari/etal:2021,secco/etal:2022} included the multiplicative shear bias in the theory prediction instead and marginalised over in the sampling process. If the residual uncertainties on $m_a$ are small, these two approaches are equivalent.
The Gaussian contribution can further be
split into three parts:
\begin{equation}
    \boldsymbol{\tens{C}}_{\mathrm{G}} = \boldsymbol{\tens{C}}_{\mathrm{sva}} + \boldsymbol{\tens{C}}_{\mathrm{sn}} + \boldsymbol{\tens{C}}_{\mathrm{mix}}\;,
\end{equation}
describing the sample variance term `sva' due to the finite number of available modes in the survey volume, the pure shot
noise term `sn' due to the finite number of sources and their composition, and the mix term `mix'.

We calculated the covariances for angular power spectra and transform those to the three summary statistics
considered. The only exception was the pure noise term, which was directly computed from the weighted number of pairs in the
catalogue. SSC and NG terms were modelled with a halo model approach and the survey variance was directly calculated from
the survey mask. Our analytical prescription was tested against GLASS \citep{tessore/etal:2023} mocks to test the impact of variable depth and the
overall agreement. These mocks are discussed in more detail in
Appendix~\ref{sec:glass}.
In \citet{reischke/etal:2024}, we showed that the agreement between our analytic covariance and that constructed from GLASS 
is within $10$ percent at large angular separations and below $5$ percent for scales below $0.5$ degrees. Furthermore,
we found that our idealised mix term (which does not account for non-uniformity of the source distribution) did not
affect our inference results significantly, and as a result was neglected in our analyses.

\section{Data and analysis} 
\label{sec: data}

In this section, we summarise the KiDS-Legacy lensing data and its calibration, the associated image simulations, and our
analysis blinding strategy. 
Our analysis was performed using a new version of the \cosmopipe\ infrastructure presented in
\citet{wright/etal:2020b}. Our fiducial pipeline includes all steps required for a catalogues-to-cosmology analysis,
leveraging three primary inputs: a (blinded) shear catalogue (Sect. \ref{sec: blinding}), a spectroscopic compilation (Sect.
\ref{sec: kidz}), and a simulated galaxy lightcone (Sect. \ref{sec:redshift}). With these data products, the pipeline
performs nine primary functions: pre-processing of the wide-field catalogue, 
simulated catalogue construction, redshift calibration,
\nz\ construction, shear calibration, data vector construction, covariance construction, cosmological inference, and
post-processing. Details of each processing step are provided alongside the public code base\footnote{\url{github.com/AngusWright/CosmoPipe}}. 

\begin{figure*}
    \centering
    \includegraphics[width=\textwidth]{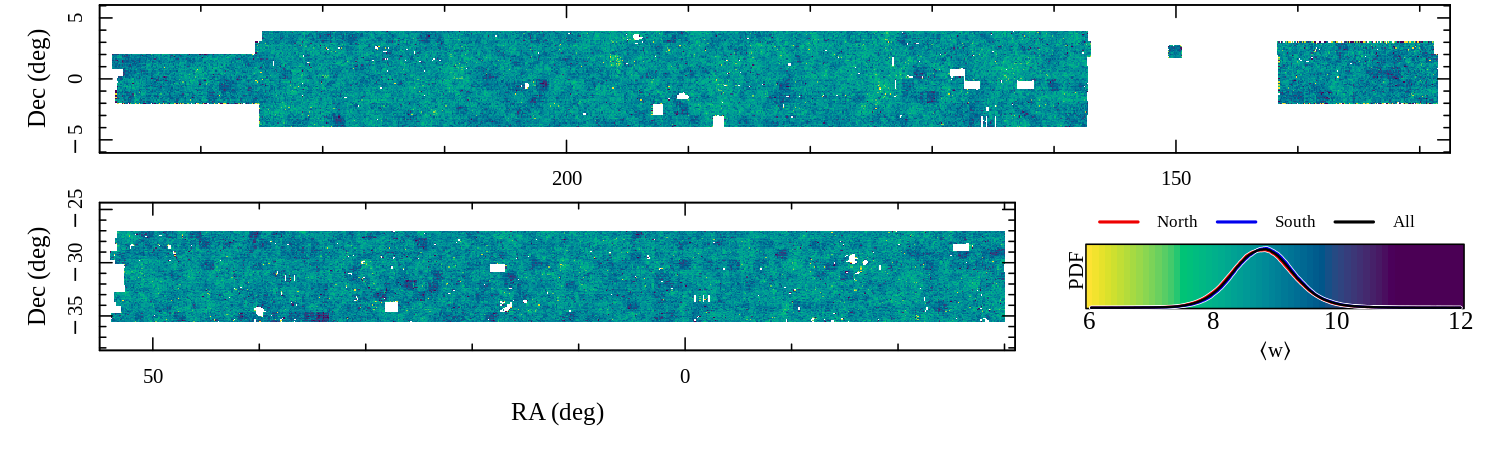}
    \caption{Distribution of the mean weight of the \kidslegacy\ source galaxies in on-sky bins ($0.1\times 0.1$ degrees). 
    The weights here include both
    shape measurement weights (Sect.~\ref{sec: rowestats}) and redshift distribution gold weighting
    (Sect.~\ref{sec:redshift}). }
    \label{fig:kidslegacy}
\end{figure*}

Figure~\ref{fig:kidslegacy} shows the footprint of the \kidslegacy\ sample, split between the two survey patches of
\kids\ (one in each of the northern and southern Galactic caps). The figure is coloured by the mean effective weight
(i.e. including both shape measurement and redshift distribution estimation weights) of
sources on sky, which is an approximate tracer of survey properties such as the size of the point-spread function and
variable depth (see Appendix~\ref{sec:glass} and \citealt{yan/etal:2024}). The weight is defined as described in Section 7 of
\citet{wright/etal:2024}, and has the range $w\in[0,\sigma_{\rm pop}^{-2}]$, where $\sigma_{\rm pop}=0.255$ is an  
approximate intrinsic ellipticity dispersion of all sources in \kids\ (without e.g. tomographic or brightness
selections) defined for Legacy in \citet{li/etal:2023a}. 

\begin{table*}
  \centering
  \caption{\label{tab:bins}Properties of the six \kidslegacy\ tomographic bins. }
  \begin{tabular}{ccccccccc}
    \hline
    \hline
    Tomo. & \photoz & \multicolumn{1}{c}{$N_{\rm gold}$} &
    \multicolumn{1}{c}{$n_{\rm eff,gold}$} & \multicolumn{1}{r}{$\sigma_{\eps,{\rm gold}}$}
    & \multicolumn{1}{c}{$\langle z \rangle_{\rm est}$} & \multicolumn{1}{c}{$\delta z$} & \multicolumn{1}{c}{$m_{\rm gold}$} \Tstrut\\
    Bin & Selection &                                    &
    \multicolumn{1}{c}{$[\mathrm{arcmin}^{-2}]$} &                                                   
    &                                & \multicolumn{1}{c}{$[ \times 10^{-2}]$} &   \multicolumn{1}{c}{$[ \times 10^{-2}]$}  \Tstrut\\
\hline\Tstrut
    1    & $0.10<\zb\le0.42$ &  \,7\,416\,371 & 1.77 & 0.28 & 0.335 &$   -2.6\pm0.2\pm1.0$ & $  -2.3\pm0.6 $ \\
    2    & $0.42<\zb\le0.58$ &  \,7\,359\,911 & 1.65 & 0.27 & 0.477 &$ \pp1.4\pm0.1\pm1.0$ & $  -1.6\pm0.6 $ \\
    3    & $0.58<\zb\le0.71$ &  \,6\,799\,681 & 1.50 & 0.29 & 0.587 &$   -0.2\pm0.2\pm1.0$ & $  -1.1\pm0.7 $ \\
    4    & $0.71<\zb\le0.90$ &  \,6\,880\,432 & 1.46 & 0.26 & 0.789 &$ \pp0.8\pm0.1\pm1.0$ & $\pp2.0\pm0.7 $ \\
    5    & $0.90<\zb\le1.14$ &  \,6\,477\,538 & 1.35 & 0.28 & 0.940 &$   -1.1\pm0.2\pm1.0$ & $\pp3.0\pm0.8 $ \\
    6    & $1.14<\zb\le2.00$ &  \,5\,960\,461 & 1.07 & 0.30 & 1.224 &$   -5.4\pm0.4\pm1.0$ & $\pp4.5\pm0.9 $ \\
    1--6 & $0.10<\zb\le2.00$ &   40\,894\,394 & 8.79 & 0.28 & 0.680 &         --           &       --        \\
    \hline
  \end{tabular}
\tablefoot{
    All values are computed for the final gold-selected sample of 
    sources, and with the appropriate shape-measurement and gold weights applied. 
    Where relevant values are computed using an effective area of $967.4$ deg$^2$.
    Values of $\sigma_{\epsilon}$ and $n_{\rm eff}$ are computed using
    Eqs.~C.9 and C.12 of \citet{joachimi/etal:2021}, respectively. The
    $\sigma_{\epsilon}$ values are the ellipticity
    dispersion averaged over both components. $m_{\rm gold}$ is the
    multiplicative shear measurement bias, averaged over both ellipticity components, 
    estimated from our \skills\ simulations.  
    $\sigma_{\epsilon}$ values include the appropriate correction for multiplicative shear
    biases (i.e. the $m_{\rm gold}$ column). 
}
\end{table*}

\subsection{KiDS and VIKING}\label{sec: kids}

The KiDS-Legacy lensing sample was drawn from the fifth and final data release of the KiDS survey\footnote{\url{https://www.eso.org/sci/observing/phase3/news.html##KIDS_DR5}}, KiDS-\drfive, described in
\cite{wright/etal:2024}. It includes 1347 square-degree tiles observed in the $u,g,r,$ and $i$ filters with the VST
\citep{capaccioli/etal:2005} and OmegaCAM \citep{kuijken:2011} between 2011 and 2019, as well as the 2009--2018 VIKING
survey VISTA/VIRCAM observations \citep{edge/etal:2013} of the same area in $Z,Y,J,H,$ and $K_{\rm s}$. Additionally, the
dataset contains spectroscopic calibration fields (the `KiDZ' fields, 27 square-degree tiles in total, including four
that are part of the main survey).

KiDS-\drfive\ significantly extends the previous (fourth) data release \citep{kuijken/etal:2019}, which was the basis
for the KiDS-1000 analyses \citep{asgari/etal:2021,heymans/etal2021,troester/etal:2021}. It represents a substantial
increase in survey area (by 34\%) and in $i$-band depth (doubling the exposure time with a second pass). The volume of
photometric redshift calibration data has also grown, including photometry for 5 times more spectroscopic sources than
in the original KiDS-1000 analysis \citep{hildebrandt/etal:2021}. The deeper $i$-band depth and more extensive
calibration data has allowed us to push the photometric redshift limit for galaxies in the lensing sample from 1.2 to 2.
The combined effect of the increase in area and depth is an increase in the survey volume of a factor of 3.5.

Several aspects of the data analysis were changed between \drfour\ and \drfive, as detailed in \cite{wright/etal:2024}.  The
astrometry reference catalogue for the \theli\ reduction moved from the Sloan Digital Sky Survey
\citep[SDSS][]{sdss_dr16:2020} and the 2-micron All Sky Survey \citep[2MASS][]{skrutskie/etal:2006} to \gaia\ DR2
\citep{gaia/dr2}, and the astrometric solutions and photometric zero-point determinations in the \astrowise\ $u$-band
reductions were made more robust. Also, the manual and automated masking of artefacts in the images, mostly due to ghost
reflections, artificial satellite tracks, and detector instabilities, was revisited and improved for the final release.
Most importantly, the weak lensing shape measurement method uses an updated version of \lensfit\ (v321), with
improvements in the sampling of the ellipticity likelihood surface and of the treatment of residual PSF leakage in the
shapes and associated weights \citep{li/etal:2023a}.  Details and tests of these new implementations are given in
\cite{wright/etal:2024}, and in the sections below. A summary of the properties of the \kidslegacy\ sample is given in
Table~\ref{tab:bins}, including details of tomographic bin limits, final effective number densities of galaxies (after
redshift distribution calibration), and redshift and shape measurement bias parameters.  

Lastly, we note that a pernicious astrometric issue was identified during the investigation of our B-mode null tests (Sect.~\ref{sec:
bmodes}), which resulted in additional masking of approximately 4\% of the survey sources to be used in the KiDS-Legacy analysis. 
More details are given in Sect.~\ref{sec: bmodes} and Appendix~\ref{sec:bmodeinvestigation}. The total area after masking is
$967.4$ deg$^2$, with $n_{\rm eff}=8.79$ per arcminute within the photometric redshift limits of \kidslegacy. 

\subsection{KiDZ}\label{sec: kidz}

The KiDZ data consist of VST and VISTA images that target well-studied spectroscopic survey fields, including the Cosmic
Evolutions Survey field \citep[COSMOS;][]{scoville/etal:2007}, the DEEP2 Galaxy Redshift Survey
\citep[DEEP2;][]{newman/etal:2013} 02h and 23h fields, the {\em Chandra} Deep Field South (CDFS), the Visible
Multi-object Spectrograph (VIMOS) Very Large Telescope (VLT) Deep Survey \citep[VVDS;][]{lefevre/etal:2005} 14h field,
and the VIMOS Public Extragalactic Redshift Survey \citep[VIPERS;][]{guzzo/etal:2014} W1 and W4 fields (a full list is
given in \citealt{wright/etal:2024}). They were purposely taken under very similar observing conditions, and with the
same observing setup, as the main KiDS/VIKING survey and serve as the calibration sample for photometric redshifts of
the faint KiDS sources.  A total of $126\,085$ sources with spectroscopic redshifts have nine-band KiDZ photometry.
Additional photometric redshift calibration via cross-correlations uses the overlap between KiDS and shallower,
wide-angle spectroscopic redshift surveys, principally: SDSS, the Galaxy And Mass Assembly \citep{driver/etal:2022}
survey, 2-degree Field Lensing Survey \citep[][]{blake/etal:2016}, WiggleZ survey
\citep{blake/etal:2008}, and the Dark Energy Spectroscopic Instrument \citep[][]{desi/etal:2023}. 

\subsection{Blinding} 
\label{sec: blinding}

We used the same blinding methodology as the one implemented in \citet{asgari/etal:2021}, first presented in
\citet{kuijken/etal:2015}. This process involves the double-blind generation of three realisations of our shape
measurement catalogues. One of these is the true catalogue, while in the other two there are induced systematic differences in measured shapes that result in (up to) $\pm 2\sigma$
change in the inferred $S_8$.  The true catalogue is equally likely to be the one that yields the highest, middle, or lowest $S_8$. These blinded data vectors were analysed
throughout the project until the analyses were finalised, cosmological chains were run,  the results written up in
this publication, and the manuscripts had undergone internal review within the \kids\ collaboration. 
As such, this manuscript (for example) was written containing results for all three blinds, with
only the discussion and conclusion sections written after unblinding.

The authors note (for the reference of future surveys) that this blinding scheme effectively  hampered the diagnosis of subtle
systematic effects in the dataset (see Appendix~\ref{sec:bmodeinvestigation}), as we were limited in our ability to directly compare measured 
shapes between previously constructed catalogues and our \kidslegacy\ catalogues. Such comparisons are extremely useful for the validation of 
spatially localised systematic patterns in the shape catalogue, which are obfuscated when translated into two-point statistics. 
This blinding strategy, however, makes such direct comparisons between shape catalogues problematic, as it can also lead to unblinding. 

\subsection{Redshift distributions and calibration} 
\label{sec:redshift}

\begin{figure*}
  \centering
  \includegraphics[width=1.25\columnwidth]{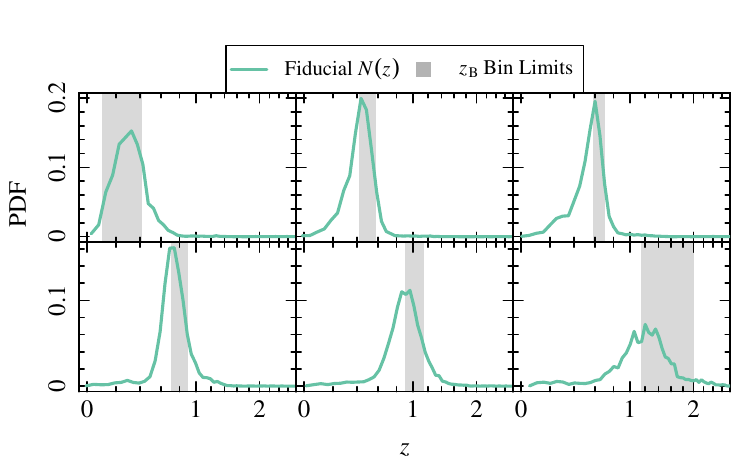}
  \includegraphics[width=0.75\columnwidth]{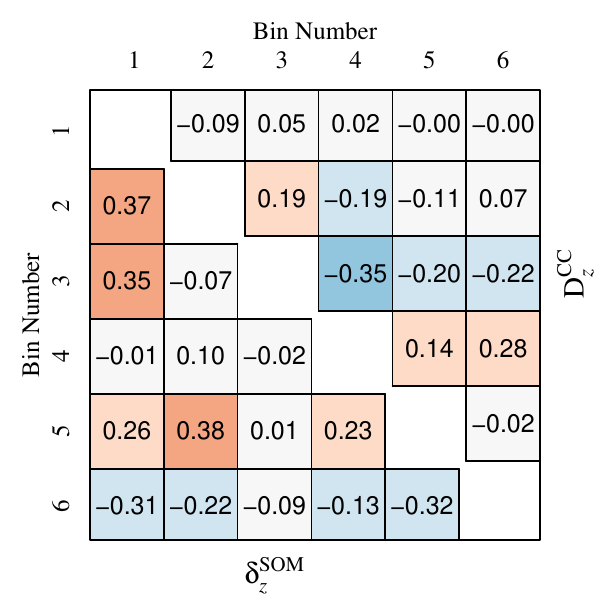}
  \caption{{\em Left:} \nz\ used in the fiducial cosmic shear analyses of \kidslegacy. Each panel is one tomographic bin whose
  definition is determined by the grey band in \photoz. 
The \nz\ have been shifted according to the mean bias estimated using our \skills\ simulations, making
  each \nz\ a correct representation of the effective \nz\ that is used in the cosmological analyses. Fiducial biases
  are provided in Table~\ref{tab:bins}.
{\em Right:}  $\delta z$ correlation
  matrix for our fiducial \nz ($\delta^{\rm SOM}_z$, {\em lower triangle}) and the correlation matrix inferred from our cross correlation \nz\
  on the data ($D^{\rm CC}_z$, {\em upper triangle}).}\label{fig: Nz}
\end{figure*}

Redshift distributions (hereafter `\nz') used in \kidslegacy\ were estimated and calibrated leveraging both colour-based
direct calibration using self-organising maps \citep[SOMs][]{kohonen:1982} and spatial cross-correlations (CC) between tracer spectroscopic
samples and the \kidslegacy\ sources. For a detailed description of the \nz\ construction and validation process used in
this work, we refer  to \citet{wright/etal:2025}. In brief, the \nz\ here differ in their
construction (compared to previous work from \kids) in three main regards: a considerably larger redshift calibration
sample from \kidz\ (see Sect.~\ref{sec: kidz}); a new weighting scheme that replaces the binary `gold-class' with a continuous `gold-weight';
and through the optional inclusion of weights for spectroscopic objects, such as our so-called `prior-volume weighting'.
Gold-weights primarily account for noise in the colour-based association between wide-field sources and their
calibrating spectra, and are computed by repeating the gold classification $N$ times (where we arbitrarily chose $N=10$). 
Similarly, our prior-volume
weighting applies an a priori weighting to the distribution of calibrating spectra as a function of redshift, to reduce
bias caused by colour-redshift degeneracies and the complex redshift selection function of our spectroscopic compilation.  

The redshift distribution calibration for \kidslegacy\ made use of two suites of simulations. For the calibration of the SOM redshift distributions (and
calibration of source shape measurements, Sect.~\ref{sec: rowestats}), we used the multi-colour \skills\ simulation of \citet{li/etal:2023a}. For additional 
SOM redshift distribution calibration and cross-correlation redshift distribution calibration, we used the MICE2
simulated lightcone
\citep{fosalba/etal:2015,crocce/etal:2015,fosalba/etal:2015b,carretero/etal:2015,hoffmann/etal:2015} adapted to KiDS noise levels and sample selections \citep{vandenbusch/etal:2020}.

    Figure \ref{fig: Nz} shows the fiducial redshift distributions used in this cosmological analysis. These \nz\ 
include the appropriate $\delta z$ shifts in each of their mean, as inferred using our simulations and described in
\citet{wright/etal:2025}. As such, the \nz\ presented are shown as-used in the generation of the theory/models during
parameter inference. In \citet{wright/etal:2025} we quantified the impact of a series of different analysis choices for
\nz\ estimation and calibration, demonstrating that they have no significant impact on the final \nz\ that we used for
cosmological inference.
The figure also presents the off-diagonal elements 
of the correlation matrices for $\delta z$, used to construct our correlated $\delta z$ 
priors. The lower triangle shows the correlation matrix estimated from our SOM redshift distribution calibration, while
the upper triangle shows the correlation matrix for the CC estimated redshift distributions.

\subsection{Shape measurement and calibration}\label{sec: rowestats}

The measurement and calibration of source shapes in \kidslegacy\ were performed with the latest \kids\ methods, which have
already been established in the literature. We refer to \citet{li/etal:2023a} for an extensive description 
of the state-of-the-art for shape measurement and calibration in \kids. Topics discussed therein include: 
the \lensfit\ algorithm \citep[v321; see also ][]{miller/etal:2007,miller/etal:2013,fenechconti/etal:2017}, used for
measuring shapes of galaxies (including modelling of the PSF); shape calibration using \skills\ image simulations; and 
PSF leakage corrections \citep[with minor
updates to the PSF leakage estimation as described in][]{wright/etal:2024}. The \skills\ image simulations are multi-band
image simulations, constructed specifically to match the observational properties of \kidslegacy\ in all available
bands, and featuring realistic clustering and correlations between galaxy properties and environment, allowing
for realistic treatment of blending effects. The simulations also feature redshift-dependent shear
\citep{maccrann/etal:2022}, although this is determined to be of negligible importance to the computation of
multiplicative bias \citep{li/etal:2023a} and redshift distribution bias \citep{wright/etal:2025} at the sensitivity of
\kidslegacy.  

The multiplicative bias estimation in \kidslegacy\ was performed using the same suite of multi-colour image simulations from
\skills used for redshift distribution calibration. Redshift distribution calibration utilises many
realisations of the sample to estimate uncertainties on tomographic bias parameters \citep[see][]{wright/etal:2025}.
Afterwards, these samples were also used to estimate the multiplicative shear measurement bias, $m$, per tomographic bin. 
This means that, for the first time, we have been able to directly estimate the 
covariance between our redshift distribution bias calibration and our multiplicative shear bias estimation. 
The resulting covariance matrix (and the corresponding correlation matrix) are functionally block-diagonal: there is no cross covariance between $m$ and $\delta z$ estimates.

We checked for the degree of PSF leakage measured in the lensing sample through a direct empirical calculation of the
correlation between PSF and galaxy ellipticities. Here, we document the measured PSF contamination in our \kidslegacy\
sample, and the measured residual one- and two-dimensional constant ellipticity terms (`$c$-terms') 
in the dataset. We adopted the first-order systematics model from \citet{heymans/etal:2006b}:
\begin{equation}\label{eq:leakage}
  \eps_k^{\rm obs} = (1+m_k)(\eps^{\rm int}_k + \gamma_k) + \alpha_k\eps_k^{\rm PSF} + c_k,\; k\in\{1,2\}\;,
\end{equation}
where $k$ denotes the individual ellipticity components, $\gamma$ is the (true) shear, and the 
superscripts `obs', `int', and `PSF' refer to a
source's observed, intrinsic, and modelled PSF ellipticity, respectively. In the limit of large numbers of sources
averaged over a large sky area, we can assume $\langle(1+m_k)(\eps^{\rm int}_k + \gamma_k)\rangle=0$, and therefore model the
residual PSF contamination using a simple linear regression between observed source ellipticities \citep[i.e. after
recalibration to remove PSF leakage][]{li/etal:2023a,wright/etal:2024} and modelled PSF ellipticities. 

\begin{figure}
  \centering
  \includegraphics[width=\columnwidth]{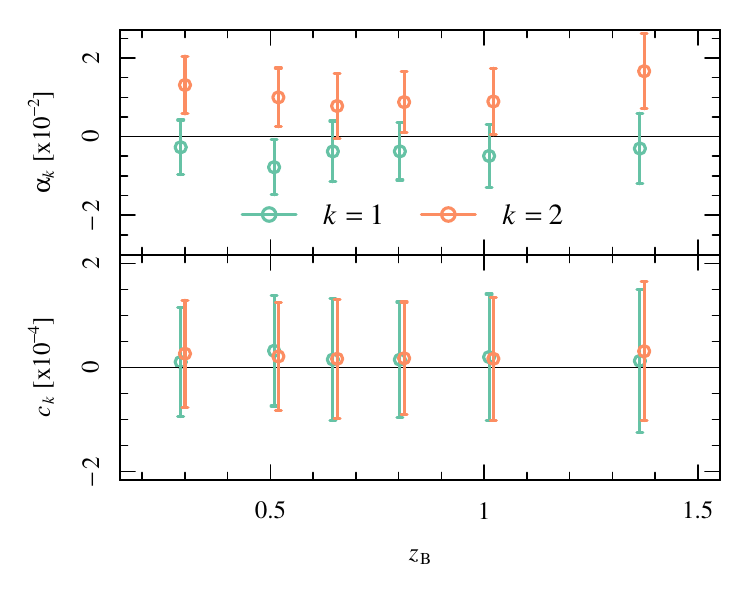}
  \caption{PSF leakage measured in the \kidslegacy\ lensing sample after recalibration of shapes, computed per
  tomographic bin and ellipticity component $k$, and following the first-order systematics model in Eq.~\eqref{eq:leakage}.
  }\label{fig: c_alpha}
\end{figure}

Figure \ref{fig: c_alpha} shows the $\alpha_k$ and $c_k$ PSF leakage terms in each ellipticity component of the
\kidslegacy\ sample, measured in each tomographic bin. Note that the shape recalibration
process used in \kidslegacy\ necessitates that the measured PSF leakage factors be approximately zero, as the recalibration
specifically addresses any residual in this parameter, in bins of source signal-to-noise ratio (SN) and resolution
\citep{li/etal:2023a}. The residual $\alpha_k$ are all negligible, with the $\alpha_2$ having 
a slightly larger residual amplitude overall ($\alpha_2\approx 0.01$, roughly three orders of magnitude 
smaller than that measured before any recalibration effort).  
Furthermore, we note that the constant term shown in the figure ($c_k$) is not the same as the additive bias which is 
removed from our galaxy sample. 
In practice, the additive component of our ellipticities was computed directly (using a weighted mean) from the sample of 
galaxies per tomographic bin and 
hemisphere: $\hat{c}_k = \langle \eps_k\rangle_w$. The $\hat{c}$-terms that were subtracted from our recalibrated shape 
catalogues prior to computation of correlation functions are provided in Table~\ref{tab:cterm}. 
In contrast, the figure presents the constant component of our linear regression (Eq~\eqref{eq:leakage}) to all 
data per tomographic bin after removal of the additive bias. As such it is expected that the $c_k$-terms are
small, but not necessarily zero.  

\begin{table}
  \caption{Subtracted additive shear per ellipticity component, tomographic bin, and hemisphere.} \label{tab:cterm}
  \centering
  \begin{tabular}{cc|cc}
    \hline
    Hemi & Tomo& $\hat{c}_1$ & $\hat{c}_2$ \\
         & Bin & $(\times 10^{-4})$ & $(\times 10^{-4})$ \\
    \hline
    \multirow{6}{10mm}{\centering North}
    & 1 & $ 3.372\pm1.528$ & $ 7.941\pm1.442$  \\
    & 2 & $ 8.852\pm1.642$ & $ 5.594\pm1.531$  \\
    & 3 & $ 4.523\pm1.747$ & $ 4.533\pm1.777$  \\
    & 4 & $ 4.722\pm1.713$ & $ 5.368\pm1.665$  \\
    & 5 & $ 6.658\pm1.887$ & $ 5.532\pm1.890$  \\
    & 6 & $ 4.224\pm2.252$ & $ 10.26\pm2.400$  \\
    \midrule
    \multirow{6}{10mm}{\centering South}
    & 1 & $-3.398\pm1.626$ & $-8.002\pm1.572$ \\
    & 2 & $-9.536\pm1.519$ & $-6.026\pm1.590$ \\
    & 3 & $-4.755\pm1.835$ & $-4.766\pm1.731$ \\
    & 4 & $-4.532\pm1.653$ & $-5.152\pm1.594$ \\
    & 5 & $-6.117\pm1.910$ & $-5.082\pm1.834$ \\
    & 6 & $-3.717\pm2.151$ & $-9.027\pm2.282$\\
    \hline
  \end{tabular}
\end{table}

\begin{figure}
  \centering
  \includegraphics[width=0.9\columnwidth]{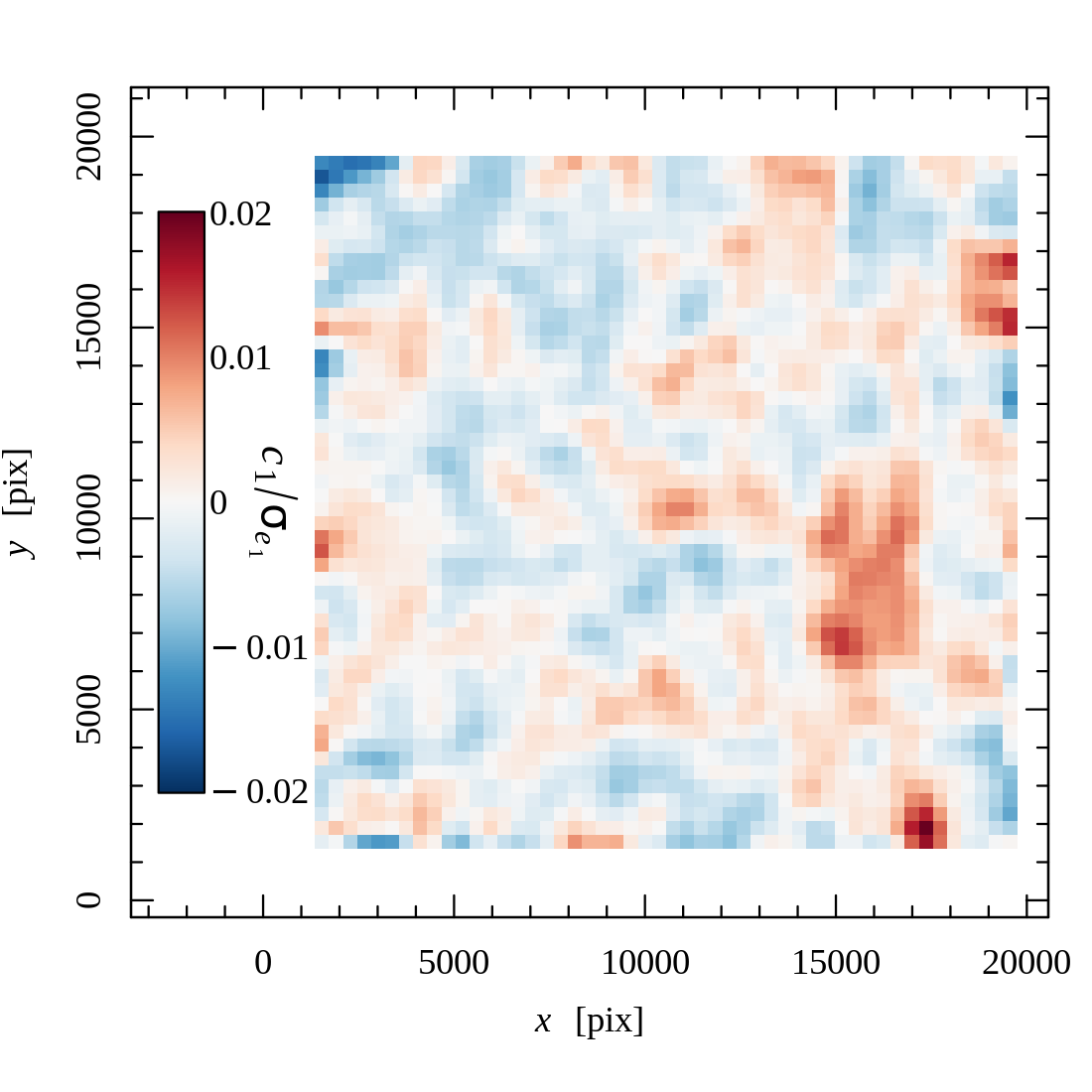}
  \includegraphics[width=0.9\columnwidth]{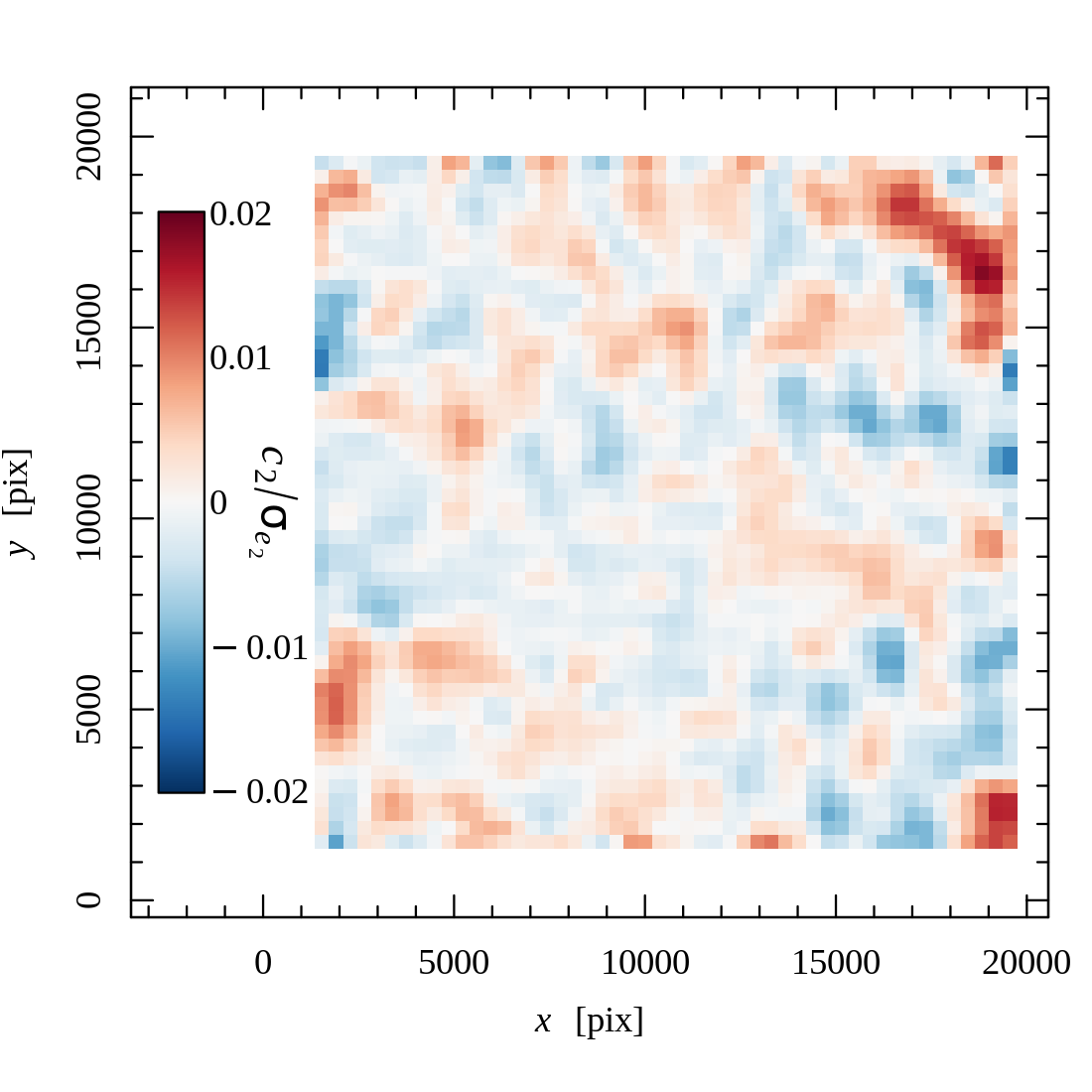}
  \caption{Average of each ellipticity component in bins of the focal plane, for sources in all tomographic bins, relative to 
  the (weighted) dispersion of ellipticities in the same bin. The weighted ellipticity dispersion per component over the
  bins in the focal plane are $0.287\pm0.001$ and $0.288\pm0.001$, indicating that the observed pattern is not driven by
  variable ellipticity dispersion across the focal plane. The figures still indicate the presence of the two-dimensional
  pattern reported in Figure 2 of \protect\cite{hildebrandt/etal:2020}.}\label{fig:2dcterm}
\end{figure}

As a further quality test, we computed the non-tomographic 2D $c$-terms in camera coordinates per ellipticity component, which is shown in Fig. \ref{fig:2dcterm}.
The 2D $c$-terms were calculated as a weighted mean per bin (using both shape- and gold-weights), with the resulting map
smoothed with a Gaussian with one-bin standard deviation (approximately $1\farcm2$), and the computed $c$-term 
scaled by the (also smoothed) measured ellipticity dispersion in each bin. The residual $c$-term can be seen to have
amplitudes $|c_i/\sigma_{\eps_i}|\leq 1\%$, indicating that these 2D patterns are negligible in the context of our cosmic
shear measurements. The primary structure that is visible  in $c_1$ was identified in \cite{hildebrandt/etal:2020} as
being due to an electronic effect that introduces an anomalous signal in the read-out direction of one detector on \omegacam.  
However, as shown in \cite{asgari/etal:2021}, the amplitude of this signal ($\lesssim 1\%$ of the ellipticity dispersion) is 
negligible in the context of our cosmic shear measurements. 

\section{Null tests} \label{sec:nulltests}

In this section, we outline the various null tests that were completed prior to the unblinding of the cosmological
analysis of \kidslegacy; the reader interested in cosmological parameter estimates only may wish to skip to 
Section~\ref{sec:results}. 
We performed three primary tests: an analysis of the \citet{paulin-henriksson/etal:2008} systematics model,  which quantifies the impact of PSF modelling residuals on correlation functions; an analysis of \citet{bacon/etal:2003} statistics, which 
test specifically for contamination of the galaxy shape measurements by the PSF; and an analysis of $B$ modes, 
which tests for a systematic signal that cannot be generated by cosmic shear. 

\subsection{\citet{paulin-henriksson/etal:2008} model}

To validate the fidelity of our shape measurement catalogue, we first compute the \cite{paulin-henriksson/etal:2008}
systematics model for \kidslegacy, which aims to quantify the impact of PSF modelling residuals on cosmic shear
correlation functions. This systematics model, in the context of the shear two-point correlation function, reduces to the
following: 
\begin{equation} 
    \begin{aligned}
 \label{eq: phmodel}
\langle \eps_{\rm obs} \eps_{\rm obs} \rangle \simeq & \left( 1 + 2 \left[{\frac{\overline{\delta T_{\rm
PSF}}}{T_{\rm gal}}}\right] \right) \left\langle \eps_{\rm obs}^{\rm perfect} \eps_{\rm obs}^{\rm perfect} \right\rangle \\ 
+ \,\phantom{1} & \, \left[ \,\overline{\frac{1}{T_{\rm gal}}}\,\right] ^2 \langle (\eps_{\rm PSF} \, \delta T_{\rm
PSF}) \,  (\eps_{\rm PSF} \, \delta T_{\rm PSF}) \rangle\\ 
+ \,2 & \, \left[ \,\overline{\frac{1}{T_{\rm gal}}}\,\right] ^2 \langle (\eps_{\rm PSF} \, \delta
T_{\rm PSF}) \,  (\delta \eps_{\rm PSF} \, T_{\rm PSF}) \rangle\\ 
+ \,\phantom{1} & \, \left[ \,\overline{\frac{1}{T_{\rm gal}}}\,\right] ^2 \langle (\delta \eps_{\rm
PSF} \, T_{\rm PSF}) \,  (\delta \eps_{\rm PSF} \,T_{\rm PSF}) \rangle \; ,
  \end{aligned}
\end{equation}
where we only keep terms that correlate to first order. It should be noted that we use a shorthand notation in Eqs.
(\ref{eq: phmodel}), (\ref{eq:totsys}), and  (\ref{eqn: xisys}) as in Section 3.3 of \citet{giblin/etal:2021}, so that
$\langle ab\rangle$ denotes $\xi_\pm$. Here, $\eps_{\rm obs}$ is the observed ellipticity of a source, $\eps^{\rm
perfect}_{\rm obs}$ is the noiseless and unbiased measurement of source ellipticity (which we approximate as being equal
to the modelled shear: $\eps^{\rm perfect}_{\rm obs}=\gamma_{\rm theory}$), and $T$ is the $R^2$ object size estimate. 

\begin{figure}
  \centering
\includegraphics[clip=true,width=\columnwidth]{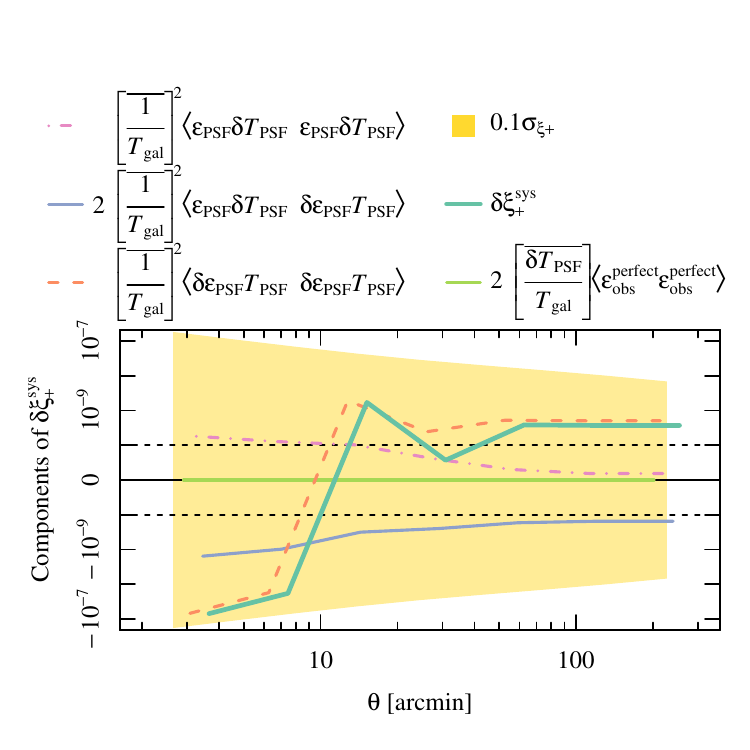}
  \caption{Contributions to the additive systematic $\delta\xi^{\rm sys}_{+}(\theta)$ from the
  \cite{paulin-henriksson/etal:2008} systematics model. The various lines show the four components, with their summation
  in thick blue-green. The yellow band depicts $10\%$ of the cosmic shear uncertainty on $\xi_+$. The figure is linear between 
  the black dashed lines, and logarithmic outside. This demonstration is for tomographic
  bin six, but all bins show quantitatively similar results.}\label{fig: rowestats}
\end{figure}

Figure \ref{fig: rowestats} shows the various contributions to the additive systematic, $\delta\xi^{\rm
sys}_{+}(\theta)$, from the \cite{paulin-henriksson/etal:2008} 
systematics model. The four terms from Eq. (\ref{eq: phmodel}), shown in various colours, cause $\langle \eps_{\rm obs}
\eps_{\rm obs}\rangle$ to deviate from $\langle \eps_{\rm perfect} \eps_{\rm perfect}\rangle$. The total systematic $\delta
\xi^{\rm sys}_+ $: 
\begin{align}\label{eq:totsys}
    \delta\xi^{\rm sys}_+(\theta) \coloneqq \langle \eps_{\rm obs} \eps_{\rm obs} \rangle - \bigg\langle \eps_{\rm obs}^{\rm perfect} \eps_{\rm obs}^{\rm perfect} \bigg\rangle\;,
\end{align}
is given by the summation of these four terms\footnote{The terms in Eq.~\eqref{eq: phmodel} are related to the often-used and so-called `$\rho$'-statistics, as recommended by \cite{rowe:2010} and \citet{jarvis/etal:2016}.  In contrast to the total systematic term $\delta \xi_+^{sys}$ (Eq.~\ref{eq:totsys}), the systematic requirements on the `$\rho$-statistics, relative to the statistical noise on the measured cosmic shear signal, are unspecified.  We therefore chose not to explore the `$\rho$'-statistics directly.} and can be compared to the yellow band which encloses $10\%$ of the
uncertainty on the cosmic shear correlation function $\xi_+$, from the \onecov\ (Sect. \ref{sec: onecov}).
The figure presents the analysis of our sixth tomographic bin (quantitatively similar results are obtained for all other
bins), and demonstrates that errors in our PSF modelling are sufficiently small (consistently less than $10\%$ of our 
correlation function uncertainty) that there is no significant contamination of our two-point statistics from this source of bias. 
As such, we conclude that PSF modelling errors are negligible for the analysis of cosmic shear in \kidslegacy. 

\subsection{\citet{bacon/etal:2003} statistics}\label{sec:baconstats} 

\begin{figure*}
  \centering
  \includegraphics[width=\textwidth,clip=T]{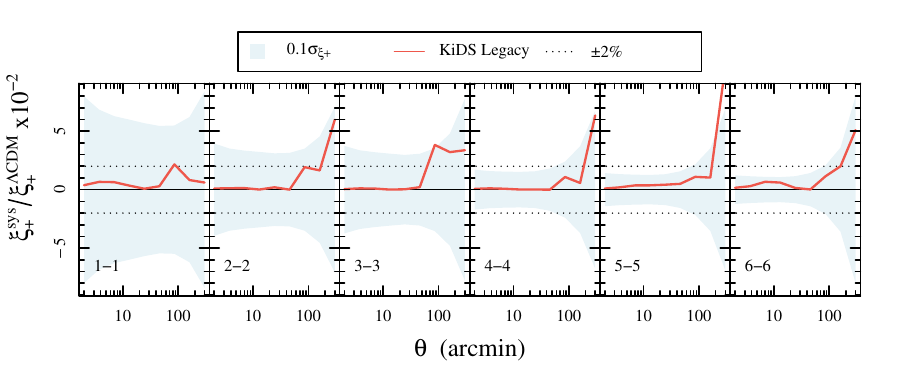}
  \caption{Ratio of PSF contamination in the cosmic shear correlation function, $\xi^{\rm sys}_{+}$, to the predicted
  best-fit cosmic shear correlation function, $\xi^{\rm \Lambda CDM}_{+}$. The statistic is shown relative to $10\%$ of the 
  $1\sigma$ uncertainty on our fiducial $\xi_+$ correlation functions (blue). We show only tomographic bin auto-correlations 
  here, as the cross-correlation signals are extremely similar to the auto-correlations (i.e. combinations
  $i-j;\;\forall\, j>i$ are quantitatively similar to the $i-i$ signal).}\label{fig: xisys}
\end{figure*}

We also used the method from \citet{bacon/etal:2003} to infer the degree of
PSF leakage into the two-point correlation function measurement using 
\begin{equation}\label{eqn: xisys}
  \xi^{\rm sys}_{\pm} = \frac{\langle\eps^{\rm obs}\eps^{\rm PSF}\rangle^2}{\langle\eps^{\rm PSF}\eps^{\rm PSF}\rangle}\;. 
\end{equation}
In Fig. \ref{fig: xisys}, we show the value of $\xi^{\rm sys}_+$ in each tomographic bin autocorrelation, as a fraction of
the theory prediction of the cosmic shear signal from our best-fit \xipm\ cosmology ($\xi^{\rm \Lambda
CDM}_{+}$). Results from the cross-correlation bins are quantitatively similar to the autocorrelation. 
As a reference, we also show $10\%$ of the $1\sigma$ uncertainty on the cosmic shear signal as reported by
the \onecov\ (Sect.~\ref{sec: onecov}). The measured fractional PSF contamination is less than the benchmark $10\%$ of
the cosmic shear correlation function uncertainty for all but five data points (out of $189$ across all tomographic bin
combinations). As such, we conclude that the PSF contamination of cosmic shear correlation functions is negligible in
\kidslegacy, as expected from our PSF leakage analysis (Sect.~\ref{sec: rowestats}).

\subsection{$B$ modes}\label{sec: bmodes} 

\begin{figure*}
  \centering
  \includegraphics[width=\textwidth]{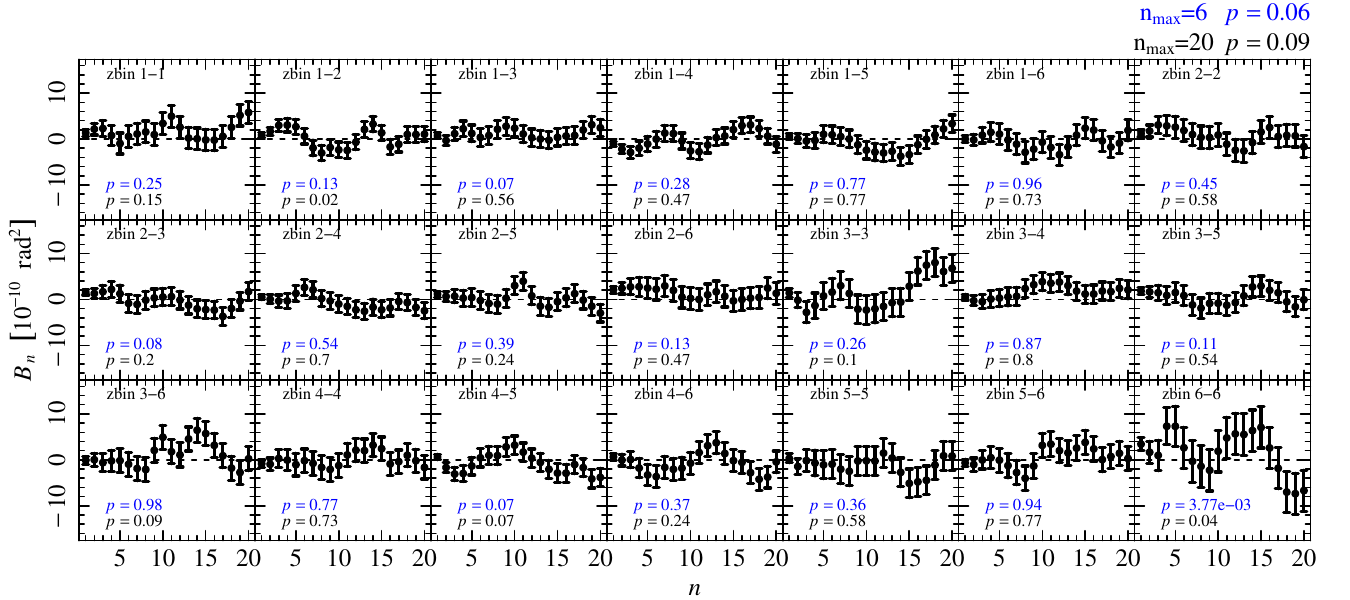}
  \caption{COSEBIs $B$ modes measured in \kidslegacy. Each panel is annotated with the $p$-value of the $B$-mode signals for twenty
  modes (black) and our fiducial 6 modes (blue). The total $p$-value for the full data vector, also for six and twenty
  modes, is annotated in the upper right corner of the figure. The $B$-mode signals are highly correlated across modes 
  within a tomographic bin, so readers are cautioned against so-called `$\chi$-by-eye'.}\label{fig: bmodes}
\end{figure*}

Our third null test examined the significance of $B$-mode signals in the data vector used for cosmological analyses. This
test was based on the understanding that, at the sensitivity available to stage-III cosmological imaging surveys,
$B$-mode signals of cosmological and astrophysical origin ought to be negligible. As such, any significant detection of
$B$ modes implies a contamination of the data vector by systematic effects, which may similarly (but undiagnosably) also
impact the $E$-modes used for cosmological inference. 

We used a tomographic analysis, equivalent to that of the fiducial cosmological analysis, to derive the
significance of $B$ modes (estimated via the COSEBIs $B_n$) in our data vector. In the early stages of our blinded cosmological
analysis, it became clear that our initial \kidslegacy\ sample failed this null test, at high significance. This led to
a re-examination of the cosmological sample, which in turn led to a redefinition of the survey footprint to
conservatively exclude $46.6$ square-degrees with increased astrometric scatter between exposures in our lensing imaging 
(see Appendix~\ref{sec:bmodeinvestigation}). 

The $B$-mode signal is presented in Fig. \ref{fig:
bmodes}. The $p$-values for the measured $B$ modes across all tomographic bin combinations are $0.04$ in the case of 
$6$ $B_n$ modes, and $0.09$ in the case of $20$ modes, and thus both pass our required threshold of $p>0.01$ 
(adopted from previous work within both the \kids\ and DES teams). Looking at the individual tomographic bins, only the sixth bin 
autocorrelation, with 6 $B_n$ modes, has a $p$-value below $0.01$. However, with 21 tests (one per bin combination) it is expected that 
one will occasionally find $p<0.01$ even with correlated data drawn from the null hypothesis; indeed, this is essentially the calculation 
that is performed in the computation of the `all bins' $p$-value, which passes.

\FloatBarrier

\section{Results}
\label{sec:results}

In this section, we present the cosmological results from \kidslegacy\ split between our fiducial results (Sect.~\ref{sec:fidcosmo}),  
and variations of analysis and modelling choices (Sect.~\ref{sec:varcosmo}). Furthermore, we refer the reader to our companion paper 
\citep{stoelzner/etal:2025} for detailed analyses of the internal consistency of \kidslegacy, and for analysis of the external consistency 
between \kidslegacy\ and datasets from DES, DESI, Pantheon+ \citep{brout/etal:2022}, and \planck.

\subsection{Fiducial constraints}\label{sec:fidcosmo} 

The fiducial analysis of \kidslegacy\ used two statistics (\cosebisE, \bandpowersE) measured over $2\farcm0
\leq\theta\leq300\arcmin$, an NLA-$M$ intrinsic alignment model, correlated $\delta z$ prior marginalisation, emulated
\cosmopower\ $P(k)$ trained on \camb\ and {\sc HMCode2020}, covariances constructed with the \onecov\ and augmented with
the uncertainty in the shear calibration, and sampling performed by \nautilus\ \citep{lange:2023} within \cosmosis, using a 
Gaussian likelihood. We  chose the \nautilus\ sampler due to its superior efficiency, 
robustness of uncertainties, and accurate evidence
estimation \citep[see e.g.][]{lemos/etal:2023,des/kids:2023}.
Priors for the various cosmological and nuisance parameters are given in Table~\ref{tab:priorrange}. 
Correlated priors used for our redshift distribution bias parameters are given in Sect.~\ref{sec:redshift}, 
and for the IA model are given in Appendix~\ref{sec:ia_massdep}.  

\begin{table}
  \caption{Fiducial prior ranges used in the analysis of \kidslegacy.  }\label{tab:priorrange}
  \centering
  \begin{tabular}{c|c}
    Parameter & Prior range \\
    \hline
    $S_8$ & $\left[0.5,1.0\right]$ \\
    $\Omega_{\mathrm{c}}h^2$ & $\left[0.051,0.255\right] $ \\
    $\Omega_{\mathrm{b}}h^2$ & $\left[0.019,0.026\right] $ \\
    $\Omega_{{K}}$ & $0.0$ \\
    $n_{\mathrm{s}}$ & $\left[0.84,1.10\right]$ \\
    $h$ & $\left[0.64,0.82\right]$ \\
    $\log_{10}(T_{\rm AGN}/K)$ & $\left[7.3,8.3\right]$ \\
    $\sum m_\nu$ & $0.06$eV \\
    $w_0$ & $-1.0$ \\
    $w_{\mathrm{a}}$ & $0.0$ \\
    \hline
    $\delta z^{(i)}$ & $\mathcal{N}(\mu_z,\mathbf{C}_{\delta z})$ \\
    $A_{\rm IA},\;b_{\rm IA}$ & $\mathcal{N}(\mu_{\rm IA,Ab},\mathbf{C}_{\rm IA,Ab})$ \\
    $\log_{10}(\langle M \rangle^{(i)})$ & $\mathcal{N}(\mu_{\rm IA,M},\mathbf{C}_{\rm IA,M})$ \\
    \hline
  \end{tabular}
  \tablefoot{
  Parameters in the first block are inputs to the $P(k)$
  emulator, and all use flat priors denoted by their lower and upper limits
  $\left[\mathrm{lower},\mathrm{upper}\right]$. 
  We sample over five cosmological parameters: $S_8$, $\Omega_{\mathrm{c}}$ and
  $\Omega_{\mathrm{b}}$ 
  (the density parameters for cold dark matter and baryonic matter, respectively); $h$ (the 
  dimensionless Hubble parameter); and $n_{\mathrm{s}}$ (the spectral
  index of the primordial power spectrum). 
  Additional inputs to the power spectrum emulator include four fixed cosmological parameters: 
  $\Omega_{{K}}$ (the energy density parameter associated with spatial curvature); $w_0$ and $w_{\mathrm{a}}$ (describing the equation 
  of state of dark energy and its redshift evolution); and $\sum m_\nu$ (the total mass of all families of neutrinos). 
  We also sample over the baryon feedback parameter from {\sc HMCode2020} $\log_{10}(T_{\rm AGN}/K)$. 
  Parameters in the second block are implemented within additional \cosmosis\ modules, and 
  are used in the intrinsic alignment model ($A_{\rm IA},\,b_{\rm IA}$, and $\log_{10}(\langle M \rangle^{(i)})$) or our 
  redshift distribution bias mitigation ($\delta z^{(i)}$). 
  These sets of parameters are specified with multivariate
  Gaussian priors denoted by $\mathcal{N}(\mu_x,\mathbf{C}_x)$, where $\mu$ is
  the vector of prior means and $\mathbf{C}$ is the related covariance matrix.
  Values of $\mu_{{\rm IA},x}$ are given in Appendix~\ref{sec:ia_massdep}. Values of $\mu_z,{\rm diag}(\mathbf{C}_{\delta z})$ 
  are given in Table~\ref{tab:bins}, and the correlation matrix of ${\mathbf{C}_{\delta z}}$ is shown in Fig.~\ref{fig: Nz}. 
  }
\end{table}

\begin{figure*}
  \centering
  \includegraphics[clip=true,trim={0 0 0 0},width=0.47\textwidth]{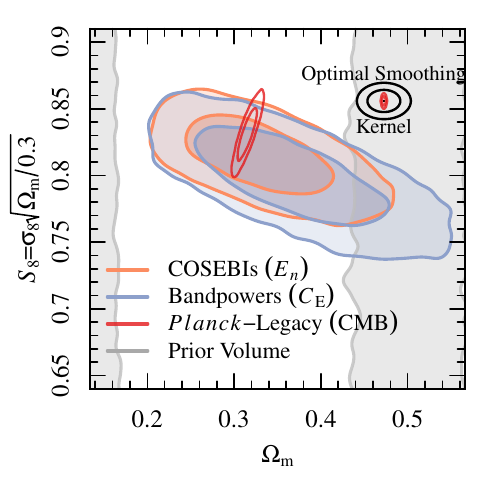}
  \includegraphics[clip=true,trim={0 0 0 0},width=0.47\textwidth]{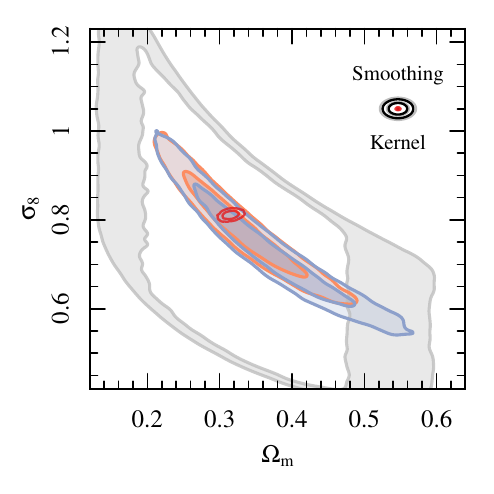}
  \caption{Fiducial $\Omega_{\rm m} ,S_8$ ({\em left}), and $\Omega_{\rm m} ,\sigma_8$ ({\em right}) constraints for our
  fiducial two-point statistics (\cosebisE\ and \bandpowersE, orange and purple respectively), compared to the CMB results from {\em Planck}-Legacy
  (red). The grey contours outline the marginal distribution of our a priori volume, with (inner) $1\sigma$ and (outer) $2\sigma$ boundary levels. }\label{fig:fiducialContours}
\end{figure*}

\begin{figure*}
  \centering
  \includegraphics[clip=true,trim={0 0 0 0},width=0.95\textwidth]{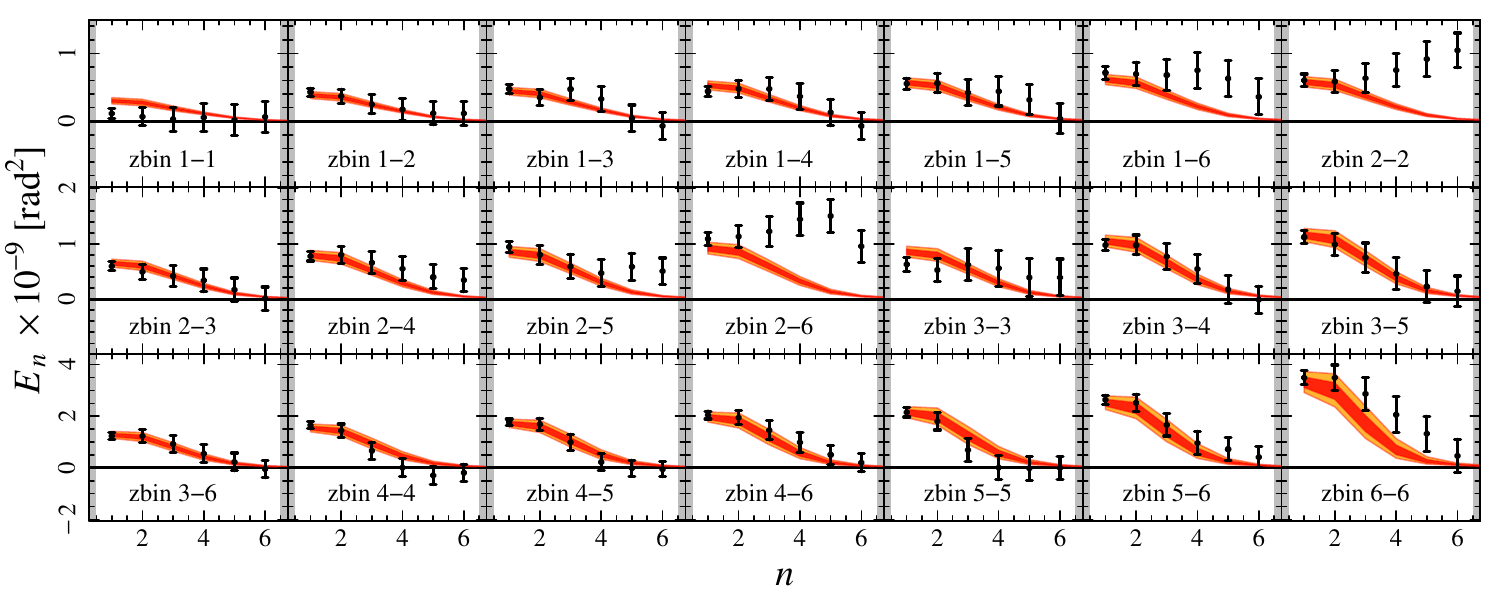}
  \includegraphics[clip=true,trim={0 0 0 0},width=0.95\textwidth]{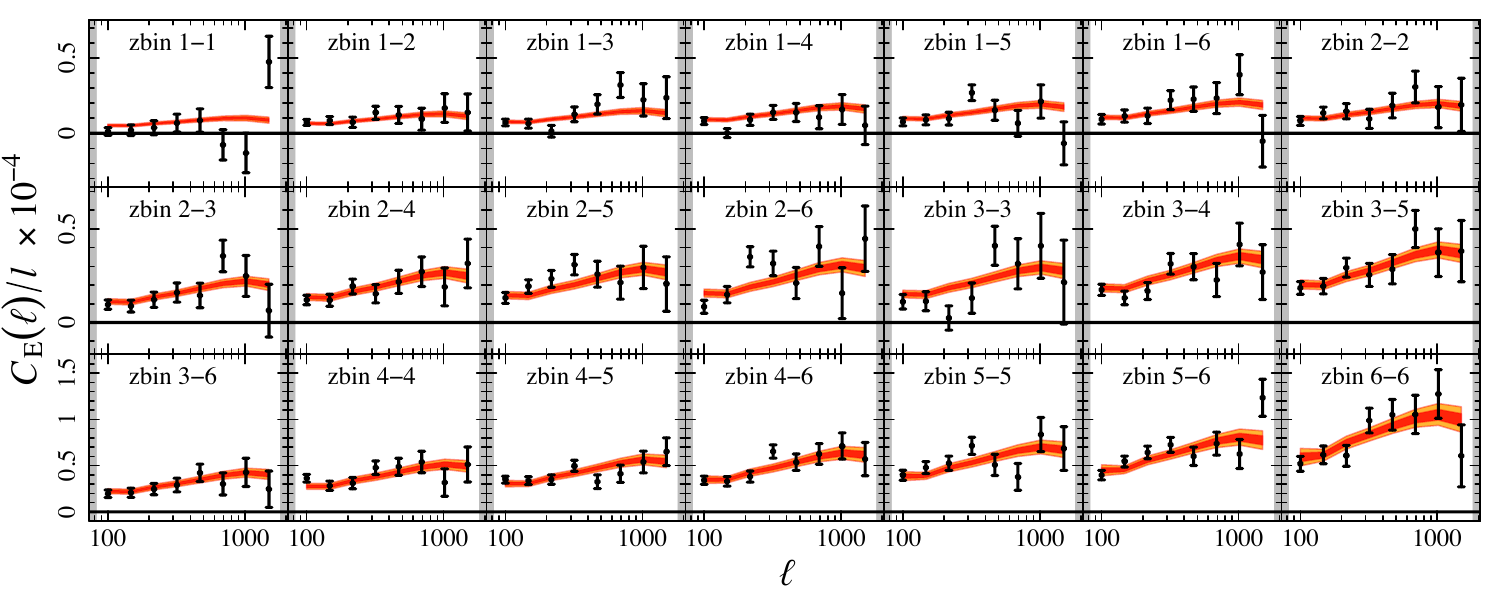}
  \caption{Data vector and TPDs for the fiducial analysis of \cosebisE\ ({\em upper panel}) 
  and \bandpowersE\ ({\em lower panel}).  
  TPDs are shown as polygons spanning the 
  $68^{\rm th}$ (red), and $95^{\rm th}$ (orange) percentiles of the posterior models. 
  The $E_n$ signals are highly correlated across modes
  within a tomographic bin, so readers are cautioned against so-called `$\chi$-by-eye'. The bin 2 autocorrelation 
  signal, for example, is consistent with the best fit model (PTE$=0.09$), despite the apparent divergence of the data from the model. 
  The overall PTEs for the full tomographic dataset in each statistic are provided in Table~\ref{tab:S8constraints}.} \label{fig:tpd_combined}
\end{figure*}

Figure \ref{fig:fiducialContours} presents the fiducial marginal constraints in $\sigma_8$, $S_8$, and $\Omega_{\mathrm{m}}$.
Marginal constraints over both $S_8$ and $\Sigma_8=\sigma_8(\Omega_{\rm m}/0.3)^{\alpha}$ (with the exponent fitted to
the $\Omega_{\rm m}$ vs. $\sigma_8$ contour) are presented in Table~\ref{tab:S8constraints}, for three estimation 
types: marginal mean and standard deviation, marginal mode and 
1D $68\%$ highest posterior density interval (HPDI), and maximum a posteriori (MAP) and $68\%$ projected joint-distribution 
highest posterior density interval (PJ-HPD). \citet[Sect.~2.6]{des/kids:2023} present a discussion of the benefits and detriments of each 
metric. We also report the 
$\chi^2$ of our cosmological model evaluated at the MAP, and the model
probability to exceed (PTE). The MAP was evaluated using an adaptive Nelder-Mead
(downhill simplex) algorithm, as described in \cite{joachimi/etal:2021}. 

Data vectors for the fiducial \cosebisE\ and \bandpowersE\  are given in Fig. \ref{fig:tpd_combined}, alongside 
translated posterior distributions \citep[TPDs; see][for the first application of these consistency metrics to cosmic
shear datasets]{koehlinger/etal:2019} drawn from the fiducial posterior computed for each statistic.

Our fiducial cosmological results, from COSEBIs, estimate a value of $S_8=0.813\pm0.018$, evaluated using the marginal
mean and standard deviation. This is the tightest constraint made with our COSEBIs measurements, although all
constraints are relatively consistent in their constraining power: the marginal mode and HPDI estimate is
$S_8=0.815^{+0.016}_{-0.021}$, and the maximum a posteriori and PJ-HPD estimate is $S_8=0.811^{+0.022}_{-0.015}$. The
consistency between the measurements is a reflection of the Gaussianity of the marginal posterior, and of the
$N$-dimensional posterior (in the case of the PJ-HPD). This is despite the necessary smoothing of the posterior
distribution that is required of the HPDI computation, which is performed using a Gaussian kernel with standard
deviation defined using the method of \cite{sheather/jones:1991}. Testing of the HPDI method indicates that, for our
posterior distributions in $S_8$, the chosen smoothing kernel is unlikely to introduce additional spurious uncertainty
to our cosmological parameter constraints.  As such, we expect that the constraints returned from our various
summarisation methods to be robust. In the following, all values and uncertainties correspond to the marginal mode and
HPDI estimate, unless otherwise specified. 

The above discussion about the stability of our summarisation statistics is contrasted by the observed variance between
our $S_8$ constraints from different statistics. The variance in $S_8$ between our statistics is seemingly large, with
our \bandpowersE\ and \cosebisE\ differing by $0.55\sigma$. However, examination of the 2D posterior distribution in
$\Omega_{\rm{m}}$ and $S_8$ (Fig.~\ref{fig:fiducialContours}) indicates that the
optimal degeneracy direction for \kidslegacy\ is somewhat different than the
assumed $\alpha=0.5$ parametrised by $S_8$. This is expected, given the
different scales probed by each of our statistics, and has the effect of
exacerbating apparent differences between our statistics when marginalising on
the $S_8$ parameter (as translation along the degeneracy can give the false
impression of internal inconsistency between final parameter constraints). 

When fitting for the optimal $\alpha$ under the more general $\Sigma_8 = \sigma_8 (\Omega_{\mathrm{m}} / 0.3)^\alpha$,
we found the best fitting $\alpha$ is $0.58$ for $E_n$ and $0.60$ for $C_{\rm E}$. These
significant deviations from $\alpha=0.5$ can cause offsets between derived values of $S_8$, as is seen
between our \bandpowersE\ and \cosebisE\ constraints. Additionally, the change in degeneracy direction has a significant
impact on the apparent constraining power of each of our statistics. For \bandpowersE\ in particular, we find a
significant $\sim40\%$ reduction in the relative marginal uncertainty between our $S_8$ and $\Sigma_8$ parameters. As such,
we opt to preferentially present results in the $\Sigma_8$ parametrisation in this manuscript, especially when
quantifying the impact of analysis choices on recovered constraints. We note, however, that the different $\alpha$
values between the different statistics mean that $\Sigma_8$ values recovered using each statistic are not directly comparable to one another.

\begin{table*}
  \caption{Constraints on $S_8$ ($\alpha=0.50$) and $\Sigma_8$ ($\alpha$ free).}\label{tab:S8constraints}
  \centering
  \begin{tabular}{cc|cccccccccc}
    \toprule
            & Setup    &  Statistic & $\alpha$ & $\chi^2$  &   dof   & PTE   & $N_{\rm samp}^{\rm PJ-HPD}$ & Marginal                  & Max. Apost.               & Marginal                  \\
            &          &            &          &           &         &       &                             &
            Mode+HPDI         &               +PJ-HPD     &          Mean + CI        \\
    \midrule
               $S_8$ &             Fiducial &      $E_n$ & 0.50 & 127.8 & 120.5 & 0.307 &      18      & $0.815^{+0.016}_{-0.021}$ & $0.811^{+0.022}_{-0.015}$ & $0.813^{+0.018}_{-0.018}$ \\ [+0.1cm]
               $S_8$ &             Fiducial & $C_{\rm E}$ & 0.50 & 151.0 & 162.5 & 0.731 &       6      & $0.799^{+0.022}_{-0.025}$ & $0.804^{+0.016}_{-0.031}$ & $0.797^{+0.023}_{-0.024}$ \\ [+0.1cm]
    \midrule
          $\Sigma_8$ &             Fiducial &      $E_n$ & 0.58 & 127.8 & 120.5 & 0.307 &      23      & $0.821^{+0.014}_{-0.016}$ & $0.817^{+0.029}_{-0.006}$ & $0.820^{+0.015}_{-0.015}$ \\ [+0.1cm]
          $\Sigma_8$ &             Fiducial & $C_{\rm E}$ & 0.60 & 151.0 & 162.5 & 0.731 &      19      & $0.813^{+0.018}_{-0.016}$ & $0.822^{+0.017}_{-0.020}$ & $0.813^{+0.017}_{-0.017}$ \\ [+0.1cm]
    \bottomrule
  \end{tabular}
   \tablefoot{
   $\alpha$ in this table refers to the exponent in the equation $\Sigma_8=\sigma_8\left(\Omega_{\rm m}/0.3\right)^\alpha$. When $\alpha=0.5$ (upper section) reported constraints are for $S_8$. Otherwise, constraints are for the optimal degeneracy direction for each 
   statistic, thereby showing the maximal constraining power. We report the $\chi^2$ of the best-fit (i.e. maximum a posteriori) 
   model for each statistic, and the number of effective degrees of freedom (dof, computed as $N_{\rm data}-N^{\rm eff}_{\rm param}$ where $N^{\rm eff}_{\rm param}$ is estimated as described in Appendix~B of \citealt{stoelzner/etal:2025}). 
   We compute the best-fit probability-to-exceed (PTE), which indicates the fraction of possible data vectors, 
   drawn from a generative model described by our best fit, that would be at least as extreme as our observed data vector.  
   We provide three constraints over our primary cosmological parameters per
   setup: the marginal mode and marginal highest posterior density interval
   (HPDI), the marginal mean and standard deviation, and the maximum a posteriori
   point and projected joint highest posterior density interval (PJ-HPD). The latter statistic can be noisy in the limit of few 
   samples $N_{\rm samp}^{\rm PJ-HPD}$ residing within the PJ-HPD, and so we also provide the number of samples in the PJ-HPD region as a guide. }
\end{table*}

\subsection{Analysis variations}\label{sec:varcosmo} 
In addition to the fiducial constraints, we provide a range of additional cosmological constraints computed with various
permutations of our modelling and optimisation choices. Specifically, we present: the impact of the emulation of the 
matter power spectrum (Sect.~\ref{sec:camb}), modelling of the signal with only the Gaussian component of the 
covariance (Sect.~\ref{sec:gausscov}), iterative
computation of covariances (which require a pre-defined cosmology; Sect. \ref{sec:iterative}), the impact of different
choices of intrinsic alignment models (Sect. \ref{sec:iamodelling}), the role of observational systematics (Sect.
\ref{sec:systematics}), the impact of different redshift calibrations (Sect. \ref{sec:alternate_nz}), 
the influence of different scale cuts (Sect.  \ref{sec:scalecuts}), and the impact of individual tomographic bins on our
estimated cosmological parameters (Sect. \ref{sec:dropbins}). 

\begin{figure*}
  \centering
  \includegraphics[clip=true,trim={0 0 0 0},width=0.77\textwidth]{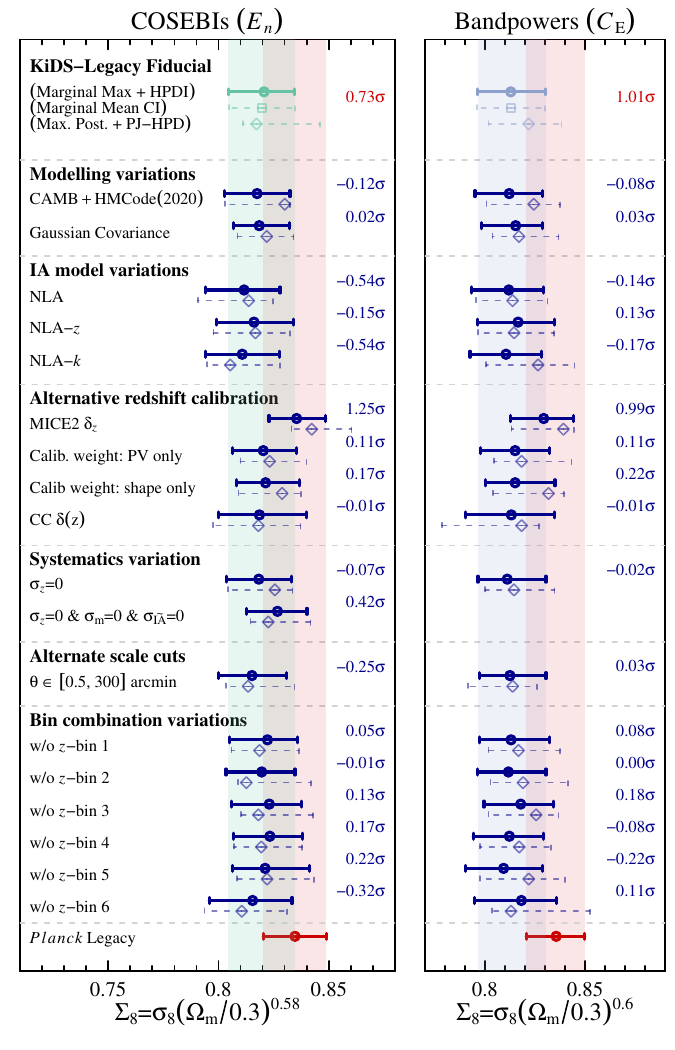}
  \caption{Whisker diagram of $\Sigma_8$ cosmological results presented in this work. The best-fitting $\alpha$ used for 
  each statistic is annotated on the x-axis. Each panel shows the compilation 
  of results for one of our two-point statistics, as annotated. In each panel, we show the fiducial constraints in the 
  relevant statistic with coloured points (colour-coded by the statistic), and highlight the fiducial marginal HPDI 
  with a band of the equivalent colour. We also highlight the marginal HPDI in $\Sigma_8$ from {\em Planck}-Legacy (shaded red).
  All variations to the analysis are annotated, with the difference relative to the fiducial result quantified using the
  metric $(\Sigma_8^{\rm fid}-\Sigma_8^{\rm var}) / \sigma_{\Sigma_8^{\rm var}}$ (listed in blue). 
  The Hellinger tension between the fiducial analyses and \planck\ is
  annotated in red.
  }\label{fig:whisker_sig8}
\end{figure*}

Figure \ref{fig:whisker_sig8} presents a whisker diagram of all the various $\Sigma_8$ (that is, for the optimal
$\Omega_{\mathrm{m}},\sigma_8$-degeneracy) results presented in this work.  
The figure includes our fiducial results in each statistic, with constraints computed using the three methods previously 
described in Table~\ref{tab:S8constraints}. All discussions below refer to our fiducial metric, the marginal mode and HPDI summary, unless 
stated otherwise.  
Additionally, we have annotated the figure with a measure of the difference between 
our various results, computed using the metric $(\Sigma_8^{\rm fid}-\Sigma_8^{\rm var}) / \sigma_{\Sigma_8^{\rm var}}$. Differences are computed between the $\Sigma_8$ values of each setup and the fiducial.
We also annotate the fiducial constraints with their Hellinger distance \citep[see e.g. Appendix G.1 of ][]{heymans/etal2021}
to the constraints from the public, fiducial {\em Planck}-Legacy analysis \citep{planck/cosmo:2018}. 
Finally, we present a similar figure for $S_8$ in Appendix~\ref{sec:posteriors}, and tables of all constraints are presented 
in Appendix~\ref{sec:tables}.

\subsubsection{Modelling $P(k)$ with \camb}\label{sec:camb} 
We tested the impact of the \cosmopower\ emulator by re-running our fiducial analysis directly with \camb\
and comparing the cosmological constraints between these runs in Fig.~\ref{fig:whisker_sig8}. 
We found that, for all statistics, the \camb\ posteriors were consistent with those estimated with the emulated power 
spectrum, showing maximal changes in the marginal value of $\Sigma_8$ of $0.03\sigma$. We consider this level of change to be 
consistent with chain-to-chain variations, and therefore negligible.

\subsubsection{Gaussian covariance}\label{sec:gausscov} 
We tested the impact of our covariance modelling on the recovered value of $\Sigma_8$ in Fig.~\ref{fig:whisker_sig8}, by 
performing the fiducial measurement with only the Gaussian component of the covariance activated in the \onecov\ code. 
We found a reduction in the uncertainty on $\Sigma_8$ of between $12\%$ (\bandpowersE) and $19\%$ (\cosebisE), and no
change in the value of $\Sigma_8$ with respect to the fiducial ($\Delta \Sigma_8 = 0.02\sigma$). 
This is an indication that, while shape noise is still sufficient to shroud information on non-linear scales, 
\kidslegacy\ has a non-negligible contribution to its constraining power from these non-Gaussian 
scales.

\subsubsection{Iterative covariance}\label{sec:iterative} 
The covariance used in our cosmological inference modelling assumes a fixed cosmology. If this fixed cosmology differs
from the best-fit cosmology preferred by the dataset, this can add a source of inconsistency in our
fiducial results. Therefore, as with previous \kids\ analyses \citep[see e.g.][]{vanuitert/etal:2017}, we verified the
fidelity of our cosmological inference by recomputing the fiducial sample covariances at the maximum a posteriori
cosmology for each chain, and rerun our cosmological inference. This iterative evaluation of the covariance and
cosmological parameters do not need to converge after a single iteration, and so we ran this process for many ($\lesssim 10$)
iterations. We found that the cosmological constraints (and constraining power) converged after two iterations, even for
our most complex model choices.  This is in good agreement with previous investigations into the impact of varying
cosmological parameters on covariances in the context of stage-III surveys \citep{reischke/etal:2017,kodwani/etal:2019}.

\subsubsection{Intrinsic alignment modelling} \label{sec:iamodelling} 

\begin{figure}
  \centering
  \includegraphics[width=\columnwidth]{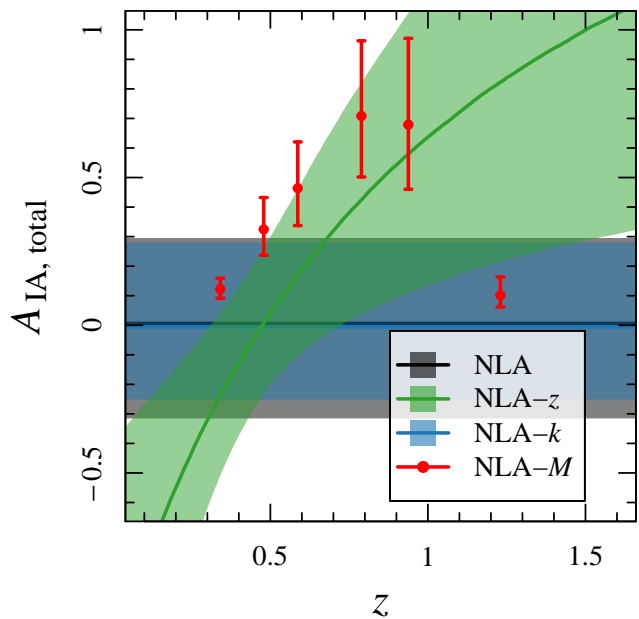}
  \caption{Constraints on the total intrinsic alignment amplitude from the various intrinsic alignment models, per
  tomographic bin (where appropriate), computed using the \cosebisE\ posteriors. Bands and error bars represent the
  $1\sigma$ posterior constraint on each model. Our NLA-$M$ model (red) is shown per tomographic bin, with the point
  positioned at the mean photo-$z$ of the bin. 
  }\label{fig:ia_amps}
\end{figure}

We tested the cosmological parameter constraints recovered when modelling intrinsic alignments with our four different
models. The resulting cosmological parameters are presented in Fig.~\ref{fig:whisker_sig8}. 

First, we see that the impact of changing our IA models on the marginal constraints of $S_8$ and $\Sigma_8$ are
mild. For our \bandpowersE\ statistic, we found maximal differences of $0.17\sigma$ across all model
variations. Also, \cosebisE\ values are more sensitive to our chosen IA model, showing
maximal differences of $0.54\sigma$ (for the NLA and NLA-$k$ models and $\Sigma_8$). Second, we see that changing the IA
model has a minor impact on the constraining power of our survey, with systematic changes in constraining power being
most evident for the NLA-$z$ model: we see a consistent reduction in constraining power (compared to the fiducial
NLA-$M$ model) of $12-13\%$ for $\Sigma_8$ across all statistics. For our NLA and NLA-$k$ models, the reduction in
constraining power compared to the fiducial is slightly smaller, at $\sim5\%$.  This reduced constraining power (despite
the fact that e.g. the NLA model has fewer parameters) is attributed to the informative prior applied in our NLA-$M$
model (see Sect.~\ref{sec:NLAmass} and Appendix~\ref{sec:ia_massdep}). Finally, we also note mild changes to our
recovered goodness-of-fit under the various IA models. In all cases, the $\chi^2$ of the modified IA model is consistent
with the fiducial (changes in PTE at the order of $1-5\%$).

The stability of our cosmological constraints under variation of the IA modelling provides confidence that any possible
misspecification of the IA model is unlikely to introduce bias in our cosmological parameter constraints. Furthermore,
we can compare the implied amplitude of IA under each of our models by examining the $A_{\rm IA,total}$ values (see Eqs.
\ref{eq:iatot_NLAz}, \ref{eq:iatot_scaledep}, \ref{eq:iatot_massdep}).  Figure~\ref{fig:ia_amps} shows the value of
$A_{\rm IA,total}$ inferred from our $E_n$ analysis, as a function of redshift. All statistics show quantitatively
similar behaviour. In Appendix~\ref{sec:posteriors}, we present the posterior distributions in all parameters
(Figs.~\ref{fig:nzpost} and \ref{fig:cosmopost}) for each IA model. 

For our NLA and NLA-$k$ models, there is no explicit dependence on tomographic bin number (or, more accurately, redshift),
and thus the IA amplitude is necessarily constant across the redshift axis for these models. Both of these models are consistent with IA
amplitudes of zero. A change in IA amplitude with redshift is nominally expected, though, as the combination of
Malmquist bias and the mass-scaling of IA amplitudes combine to increase the effective measured IA with increasing
redshift, thereby motivating our NLA-$z$ and NLA-$M$ models. 

For our fiducial NLA-$M$ model, the IA do change with tomographic bin, due to their varying fraction of passive/red
galaxies and their typical mass (blue galaxies have zero IA in this model). In our analysis, the alignment amplitudes
initially climb through bins one to five, but rapidly drop in bin six where there are very few galaxies with an early-type
SED (see Appendix~\ref{sec:ia_massdep} and Table~\ref{tab:ia_massdep_pars}).\footnote{This may be a (suboptimal) feature of our 
SED based classification, which is unable to reliably identify the precursor sources of modern pressure-supported systems using 
uncertain SED models, especially at high-$z$.}
As such, the NLA-$M$ model represents a
functional form that is not accessible to any of our other models. It is therefore interesting to note the similarity
between the inferred values of $A_{\rm IA,total}$ in our NLA-$z$ and NLA-$M$ models. Both show the expected increase in
alignment amplitudes with increasing redshift, although the NLA-$z$ model is at slightly lower amplitudes overall. We
attribute this difference to the sixth tomographic bin: as the NLA-$z$ model requires that all bins follow the same
(monotonic) trend, any reduction in $A_{\rm IA,total}$ between the fifth and sixth tomographic bins will require a
compensatory reduction in alignment amplitude across all bins.

\subsubsection{Impact of observational and systematic nuisance parameters} \label{sec:systematics} 

To demonstrate the impact of the various nuisance parameters and the relative importance of marginalisation over
various priors, we ran a posterior analysis with our intrinsic alignment model uncertainty ($\sigma_{\rm IA}$) and
redshift and shear calibration uncertainty ($\sigma_z,\sigma_m$) nuisance parameters set to zero. The associated parameters to these
uncertainties were fixed to the centre of their fiducial prior. We repeated the posterior calculation using only the
remaining (cosmological and physical) parameters. Figure~\ref{fig:whisker_sig8} shows these results. We found that the
posterior constraints on $\Sigma_8$ in the absence of marginalisation over these observational and systematic nuisance
parameters are tighter by $10\%$ ($20\%$) for our \cosebisE\ (\bandpowersE) measurements, and are shifted with respect
to the fiducial by $0.28\sigma$ ($0.14\sigma$). This suggests that \kidslegacy\ is statistics-limited.

\subsubsection{Alternative redshift distribution calibration} \label{sec:alternate_nz}

We analysed the changes in parameter constraints found under different
redshift calibration scenarios. In particular, we demonstrate the constraints found when we switch off shape-measurement
weighting of the calibration sample, and when we switch off prior volume weighting of the calibration sample. 
In both cases, we found that the
cosmological constraints are consistent with our fiducial constraints, showing a mild increase of $\Delta \Sigma_8
\leq 0.17\sigma$. This is, to some degree, expected: if our redshift distribution estimation changed appreciably when
modifying our calibration method, this would also be reflected in our estimated bias parameters $\delta z$. As such,
this test is particularly relevant for demonstrating possible sensitivity to source galaxies that reside outside the
redshift baseline of our \skills\ simulations. The lack of significant change in our cosmological parameters under
changing the redshift calibration methodology suggests that our analysis is not strongly influenced by sources outside
the redshift limits of \skills.

Conversely, we also show in Fig.~\ref{fig:whisker_sig8} the result of calibrating 
our redshift distributions using simulations from MICE2. These are the simulations that were used for \konek\ \nz\ calibration, 
but differ here in that we applied our \kidslegacy\ calibration methodology to these simulations.
\citet{wright/etal:2025}  
discuss the accuracy of the redshift distribution bias parameters estimated from MICE2, in particular possible biases due to the MICE2
redshift limits of $0.1 \lesssim z \lesssim 1.4$. As our bin six \nz\ (in particular) extends beyond this upper limit, it is expected 
that our redshift distribution bias parameters there will be inaccurate. Therefore, we include them here predominantly
as a demonstration of the effect of switching from MICE2 to \skills.
The maximum difference in the mean redshifts between our fiducial (effective) \nz\ and that from MICE2 is
$\Delta \mu_{z,{\rm eff}}=0.016$, in bin 3. However, the difference found is coherent between bins $3-6$, which is therefore 
expected to cause a shift in $S_8$. As seen in Fig.~\ref{fig:whisker_sig8}, our analysis using \nz\ calibrated with the
MICE2 simulations increases the central value of our $\Sigma_8$ constraints, by up to $1.25\sigma$ (for \cosebisE). 
We consider these results inferior to the fiducial redshift calibration, though,  given the considerably less realistic
simulations (compared to \skills), and the smaller redshift baseline that truncates the sixth bin to half of its
supposed width.  

In addition to our SOM calibrated redshift distributions, we also show results using our redshift distributions calibrated to cross-correlation 
measurements \citep[][]{wright/etal:2025}. Our cosmological constraints from this analysis are essentially unchanged in their central value for both 
\cosebisE\ and \bandpowersE, showing a $0.01\sigma$ shift in the central value of $\Sigma_8$ (computed with the fiducial $\alpha$ for each statistic). 
However, the results have lower constraining power, given the larger uncertainty in (particularly) the sixth tomographic
bin: uncertainties in the optimal projection of $\Sigma_8$ are $23\%$ ($29\%$) larger for our CC \nz/\dz\
marginalisation than in the fiducial case for \cosebisE\ (\bandpowersE). Nonetheless, the strong agreement between the
estimates made with our CC and SOM \nz\ provides confidence that redshift distribution biases in
\kidslegacy\ are under control.

\subsubsection{Scale cuts} \label{sec:scalecuts} 

\begin{figure}
    \centering
    \includegraphics[width=0.95\columnwidth]{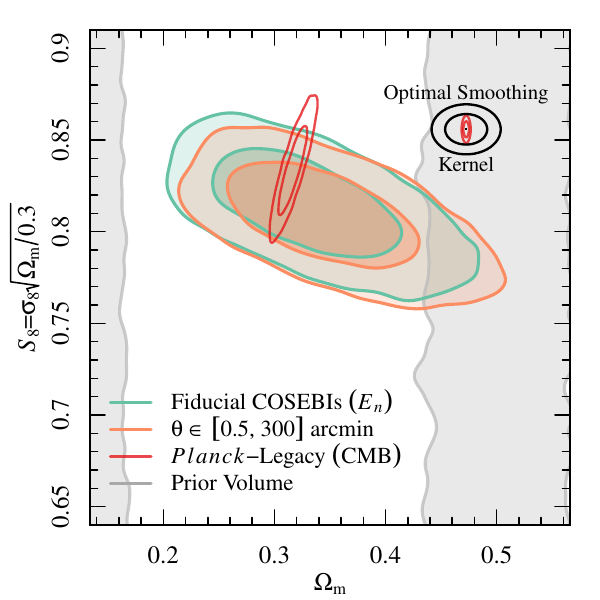}
    \includegraphics[width=0.95\columnwidth]{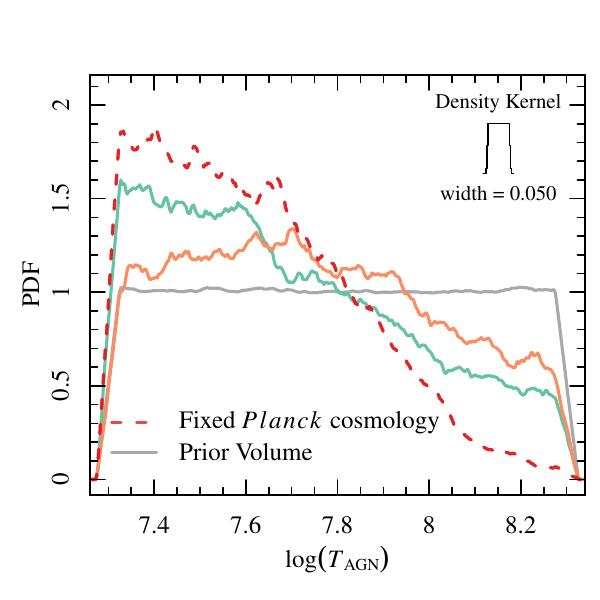}
    \caption{Comparison of the marginal constraints on $\Omega_{\mathrm{m}}$ and $S_8$ ({\em top}) and  
    $\log_{10}(T_{\rm AGN}/K)$ ({\em bottom}) when using a wider range of scales ($0\farcm5\leq \theta \leq 300\arcmin$, orange), 
    compared to the fiducial (blue-green). We compare the constraints to the full fiducial \planck-Legacy posterior
    (solid red, left) and the marginal constraint on $\log_{10}(T_{\rm AGN}/K)$ when analysing our fiducial data vector with cosmology
    fixed to the \planck-Legacy best-fit and marginalisation over nuisance parameters only (dashed red, right).}
    \label{fig:scalecut}
\end{figure}

In \konek, \cite{asgari/etal:2021} presented cosmological constraints using a range of angular scales from $0\farcm5
\leq \theta \leq 300\arcmin$.
 Subsequent
investigation by \cite{des/kids:2023}, however, found that an optimal treatment of baryonic systematic effect mitigation
(for \cosebis) was achieved using $2\farcm0 \leq \theta \leq 300\arcmin$. For \konek\ \citep{asgari/etal:2021}, this
resulted in a shift in the marginal parameter constraints of $S_8$ at the level of $0.8\sigma$. Reanalysis of \konek\
by \citet{li/etal:2023b}, with improved shape measurement and calibration from \citet{li/etal:2023a}, showed a reduced
sensitivity to the choice of scale cut, seeing a shift of $0.3\sigma$.  We also tested the influence of changing our scale
cuts, shown in the 
marginal projection of $\Omega_{\mathrm{m}},S_8$ in Fig.~\ref{fig:scalecut}, finding that including smaller
scales results in a slight ($0.1\sigma$) decrease in the marginal value of $\Sigma_8$ and $S_8$.  We also saw a
slight change in the preferred value of our baryonic feedback parameter (see Fig.~\ref{fig:scalecut}) where there is a reduction in posterior
probability mass at negligible feedback amplitudes (i.e. $\log_{10}(T_{\rm AGN}/K)\approx 7.3$, consistent with dark-matter only)
when adding in smaller scale information. However, this must be interpreted with caution, as both
posteriors are formally prior dominated in this parameter. This extra data is expected to be sensitive to baryon feedback, but it is difficult to distinguish from a simple change in the noise
realisation. In either case, though, it is clear that the additional scales have at most a minor impact on our
cosmological constraints. 

We note that the expanded scale cut has a non-negligible effect on the significance
of our $B$ mode null test, causing it to once again fail our $p>0.01$ requirement.  A full reanalysis of the additional
astrometric selections required to reach non-significance in our null test with this expanded scale cut has not been
explored here, and so the accuracy of the constraints with these expanded scales is not guaranteed. Nonetheless, the
stability of the cosmological constraints in the presence of both baryonic feedback contamination and possible
systematics contamination is encouraging. 

\subsubsection{Removing tomographic redshift bins} \label{sec:dropbins} 

In previous analyses of \kids, the removal of individual tomographic bins (including all
cross-correlations) introduced changes in marginal cosmological
constraints of up to $1.8\sigma$ in $\Sigma_8$.\citep[see e.g.][]{asgari/etal:2021}. 
We performed the same test here, and show the results in
Fig.~\ref{fig:whisker_sig8}.  We found that our cosmological results are more robust to the
removal of individual tomographic bins than previous work from \kids: 
we observe a maximal difference in $\Sigma_8$ when removing individual bins of
$0.24\sigma$ for $E_n$, $0.17\sigma$ for \bandpowersE. We attribute this decreased sensitivity at least partly to the 
reduction in relative importance of each tomographic bin (as signal is shared more evenly between the bins using our 
approximately equal-number tomography, and sixth bin).  
Furthermore, in our companion paper \citep{stoelzner/etal:2025}, we performed a series
of internal consistency tests aimed at specifically estimating the constraints from different combinations of
tomographic bins, finding that they were all consistent with each other, unlike KiDS-1000.  As such, this represents a considerable
improvement in internal consistency for \kids\ that we attribute to the enhanced redshift calibration samples and methodology, 
as well as with the general improvements in data quality seen in Legacy compared to previous \kids\ data releases.

\section{Discussion }
\label{sec:discussion}

The results from \kidslegacy\ are consistent with the results of \planck\
Legacy, with a marginal $\Sigma_8$ consistency of $0.73\sigma$ for \cosebisE\
and $1.01\sigma$ for \bandpowersE. The constraints from \kidslegacy\ are also 
of comparable constraining power to those from \planck\ when projected into $\Sigma_8$, 
with \planck\ having a marginal uncertainty of $\pm0.015$ compared to Legacy's $0.016$. 

Compared to previous cosmological constraints from \kids, Legacy represents a decrease in the marginal 
tension with \planck\ (computed using the Hellinger distance) of $2.2\sigma$, compared to previous work from 
\citet{li/etal:2023b}, who utilised methods and modelling choices that most resemble those used in
Legacy. Such a difference in marginal constraints from individual KiDS analyses was deemed acceptable prior to
unblinding (indeed, such a judgement call was required prior to unblinding), however was nonetheless surprising to the
authors. This prompted an investigation into the underlying drivers of the change in cosmological constraints between
the two samples. 

A complete description of the investigation into the consistency between KiDS-1000 and \kidslegacy\ is provided in
Appendix~\ref{sec:updates}. 
In summary, we found that improvements to our redshift distribution estimation and calibration methodology are responsible for 
approximately two thirds ($1.32\sigma$) of the reduction in marginal tension with \planck, as reported by KiDS-1000 and Legacy. 
An additional reduction is attributable to a reduction in statistical noise from the sources outside the KiDS-1000
footprint. 
These results are shown graphically in Fig.~\ref{fig:changes_whisker}. 

\begin{figure*}
    \centering
    \includegraphics[width=0.355\textwidth]{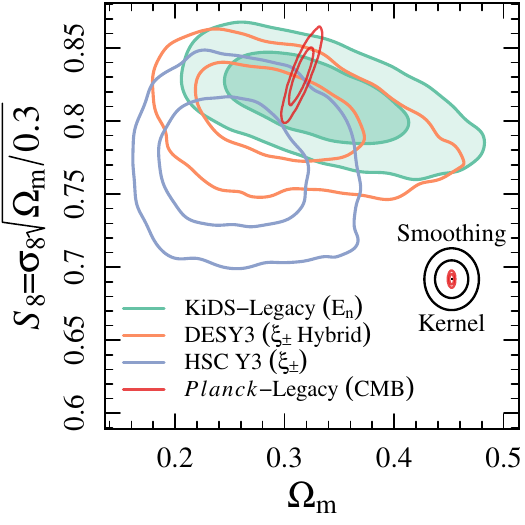}
    \includegraphics[width=0.55\textwidth]{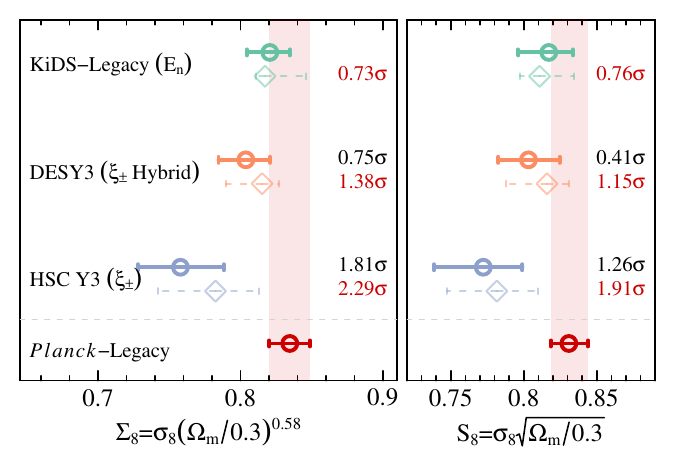}
    \caption{Comparison between \kidslegacy\ \cosebisE\ cosmological constraints and recent stage-III cosmic shear
    survey results from the literature. We show the 2D posteriors for each analysis $\Omega_{\mathrm{m}}, S_8$ ({\em left}) 
    as well as the marginal projections in $\Sigma_8$ ($\alpha=0.58$, {\em middle}), and $S_8$ ({\em right}). The 1D
    marginal `whisker' plots are annotated as previously: circular points represent marginal mode and HPD intervals,
    while diamonds represent maximum a posteriori points and their associated PJ-HPD. Values are all calculated directly
    from the chains released by each analysis/survey, and in all cases except Legacy we take the MAP directly from the
    chain. We see that all surveys presented here are consistent with both KiDS-Legacy (black annotations) and \planck\
    (red annotations) using our fiducial consistency metric of $|\Delta X| \leq 2.6\sigma$ (where $X$ is either $\Sigma_8$ or $S_8$), 
    as estimated using the Hellinger tension between marginal mode summaries. 
    }\label{fig:surveycompar}
\end{figure*}

We also explored the consistency between our new Legacy cosmological parameters with recent results from the literature. 
Figure~\ref{fig:surveycompar} presents the \cosebisE\ cosmological constraints from Legacy in the context of 
two recent stage-III cosmic shear analyses: DES-Y3 hybrid-pipeline results from \cite{des/kids:2023}, and 2PCF
measurements from HSC-Y3 \citep{li/etal:2023}. 
DES-Y3 is a largely independent analysis (there is a small on-sky overlap), which is also most consistent with Legacy:
differences between marginal cosmological parameter constraints are $0.75\sigma$ in $\Sigma_8$ and $0.41\sigma$ in
$S_8$. Comparisons to the joint analysis of DESY3 and KiDS-1000, or to HSC, are both more covariate (due to area
overlap) and show larger differences ($1.5\sigma$ and $1.8\sigma$ in $\Sigma_8$ respectively), but are otherwise
consistent. Overall, we found good consistency between all samples presented here, and with \planck. To this end, there 
is no evidence of significant tension in $S_8$ (or in the more constraining $\Sigma_8$) from \kidslegacy.
Furthermore, the consistency that we found between \kidslegacy\ and other low-redshift cosmological probes motivates the
joint analysis of cosmological parameters with these surveys, which can be found in our companion paper
\citep{stoelzner/etal:2025}. Lastly, as mentioned before, we do not include $\xi_\pm$ into our fiducial analysis.
However, the constraints and resulting consistency metrics are very similar and is discussed in Appendix
\ref{sec:xipmcos}.

\begin{figure}
    \centering
    \includegraphics[width=\linewidth]{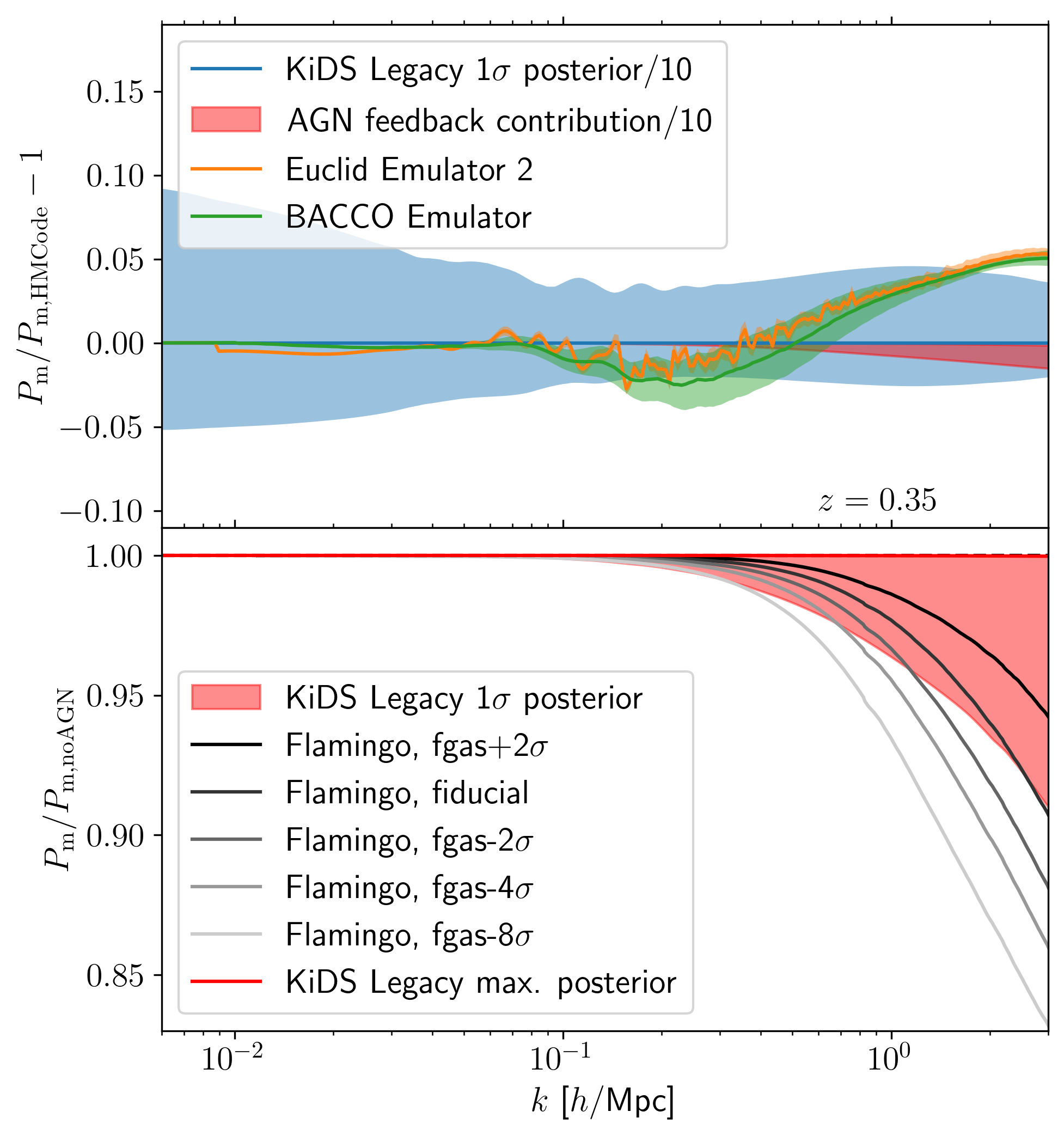}
    \caption{{\em Upper panel:} Estimate of the systematic uncertainty in our modelling of the matter power spectrum
    using \camb\ and {\sc HMCode2020}, compared to two $P(k)$ emulators \textsc{BACCO} (green) and
    \textsc{EuclidEmulator2} (orange). The figure shows the ratio between the power spectrum from \camb\ and {\sc
    HMCode2020} to those returned by each emulator, evaluated at $10\,000$ representative samples from the fiducial
    \cosebis\ posterior with a tenth of a standard-deviation. {\em Lower panel:} Constraints on
    baryonic feedback suppression from \kidslegacy. The marginal HPDI of feedback amplitudes from our fiducial \cosebis\
    is shown in red, relative to a range of hydrodynamical models of baryonic feedback suppression from {\sc Flamingo}
    \citep{schaye/etal:2023}.}
    \label{fig:pk_tagn}
\end{figure}

We performed an analysis of the effective constraining power provided by \kidslegacy\ over the matter power spectrum
$P(k)$ and the amplitude of baryon feedback. Figure~\ref{fig:pk_tagn} shows our constraint over $P(k)$ from the fiducial
\cosebisE, estimated by sampling the full posterior of \kidslegacy\ and generating posterior predictive distributions
for $P(k)$ using \camb\ and {\sc HMCode2020}. We show the $P(k)$ constraint normalised to the mean of all posterior
samples, with a $0.1\sigma$ region shaded around this central reference. This allows us to directly compare our $P(k)$
constraint to the distribution of $P(k)$ returned by two power spectrum emulators, \textsc{BACCO} \citep{arico/etal:2021b} and
\textsc{EuclidEmulator2} \citep{knabenhans21}, also evaluated at the same posterior samples. We note, though, that these emulators have a 
reduced range of validity compared to our fiducial priors, particularly in $n_s$. Nonetheless, we see that the
differences between the emulated power spectra and those returned by \camb\ are within the $0.1\sigma$ credible interval
of our posterior constraint on $P(k)$, demonstrating that differences between the various $P(k)$ models (evaluated at
the same cosmological and nuisance parameters) are consistently less than $10\%$ of our constraining power, and are thus
unlikely to be a significant source of bias in our analysis. With increased constraining power (from e.g. stage-IV
surveys), however, the uncertainty in our modelling of $P(k)$ will become a significant fraction of the final
uncertainty. 

We also explored the importance of baryon feedback modelling on our cosmological constraints, by examining the
posterior on our feedback amplitude parameter $\log_{10}(T_{\rm AGN}/K)$ in the context of our $P(k)$ constraints. We
found that the
contribution to the power spectrum from our inferred feedback amplitude is negligible for scales $k\lesssim 0.2\,h/{\rm Mpc}$. Above 
this scale our posterior constraint on the feedback model introduces a $1-10\%$ suppression in power (compared to the
dark-matter only case), and is consistent with the fiducial baryon feedback model from FLAMINGO at all scales. Above
approximately $k=1\,h/{\rm Mpc}$, our marginal posterior on $P(k)$ excludes the more extreme FLAMINGO feedback models. Our MAP
estimate of $\log_{10}(T_{\rm AGN}/K)$ is at the lower boundary limit of the prior, indicating that our best-fitting model prefers 
negligible baryon feedback amplitudes. Our marginal constraint over $\log_{10}(T_{\rm AGN}/K)$ is
unconstrained (see Fig.~\ref{fig:scalecut}), but is nonetheless skewed towards the MAP value. 

Finally, given the agreement between our cosmological constraints and those from \planck, we investigated the inferred
amplitude of baryon feedback with cosmological parameters fixed to those reported by \planck. This has the effect of 
placing a strong informative prior that the cosmological model is as described by \planck, and used cosmic shear to
marginalise effectively over only observational systematic effects (IA, $\delta z$, and baryon feedback). Our
posterior constraints over all systematic parameters are essentially unchanged with respect to the fiducial, with the
only small change of note being that our baryon feedback posterior skewed slightly more strongly towards the lower prior
boundary of $\log_{10}(T_{\rm AGN}/K)=7.3$, while the MAP remains at the boundary (see Fig.~\ref{fig:scalecut}). In the fiducial case our marginal 
posterior on $\log_{10}(T_{\rm AGN}/K)$ was unconstrained \citep[using the metric of][see their Appendix A]{asgari/etal:2021}, however with 
the informative \planck\ prior we were able to place an upper limit constraint of $\log_{10}(T_{\rm AGN}/K)\leq 7.71\;(8.01)$ at 
one-sided $68\%$ $(95\%)$ confidence. 
As such, we argue that there is little evidence in KiDS-Legacy for significant amplitudes of baryon feedback, even when
placing strong informative priors on cosmological parameters. 

\section{Summary}\label{sec:summary}
We present a cosmological analysis of the completed Kilo-Degree Survey (KiDS). 
Utilising data from KiDS Data Release 5 \citep{wright/etal:2024}, we constructed our cosmic shear source sample from
1347 square degrees of optical and near-infrared imaging, containing deeper imaging, new astrometric and photometric
calibrations, and cleaner masking of spurious sources and artefacts. Our improved data was then coupled with a greatly 
expanded spectroscopic calibration dataset named `KiDZ' (consisting of $2.4-5.0$ times the spectra as used in the analysis of
KiDS-1000), along with an updated 
redshift distribution calibration methodology \citep[`gold-weight'; see][]{wright/etal:2025}. This allowed us to
calibrate sources at higher redshifts than those in any previous KiDS analysis. We defined six tomographic bins,
spanning photometric redshifts between $0.1<z_{\rm B}\leq 2.0$, and calibrated the source redshift distributions with a
precision up to six times higher than previously possible. We calibrated the source shapes using multi-band image
simulations \citep{li/etal:2023a}, finding that the \kidslegacy\ sample has the lowest systematic contamination (by e.g.
PSF leakage) of any KiDS dataset to date. As per standard practices in the field, our cosmological analysis was
performed fully blind. 

Our fiducial cosmological parameter constraints were estimated with COSEBIs (\cosebisE), utilising
conservative $\theta\in[2.0,300.0]$ arcmin scales and a new physically-motivated intrinsic alignment (IA) model. 
Our fiducial \cosebisE\ measurement constrains $S_8\equiv\sigma_8\sqrt{\Omega_{\rm m}/0.3} = 0.815^{+0.016}_{-0.021}$, 
which is in full agreement (Hellinger distance of $0.76\sigma$) with the marginal prediction $S_8=0.834^{+0.016}_{-0.016}$ from the \planck\
Legacy cosmic microwave background experiment.
Our fiducial measurement represents a $\sim15\%$ improvement in constraining power over the previous results from \kids. 
However, we also found a significant change in optimal degeneracy between $\sigma_8$ and $\Omega_{\rm m}$, with 
$\Sigma_8\equiv\sigma_8\left(\Omega_{\rm m}/0.3\right)^\alpha$ and $\alpha=0.58$ being preferred by our \cosebisE\
measurement. We found $\Sigma_8 = 0.821^{+0.014}_{-0.016}$, which is a $\sim32\%$ improvement in optimal constraining
power compared to previous \kids\ analyses. In this projection, the corresponding Hellinger distance to \planck\ is
$0.73\sigma$. We further computed the cosmological parameters using band power spectra (\bandpowersE) and found full agreement
with our fiducial COSEBIs. We performed a number of analysis
variations, with our cosmological parameter constraints shown to be consistent across all modelling choices, statistics,
tomographic bins, and summary statistics. Our companion paper \citep{stoelzner/etal:2025} presents a full suite of
internal and external consistency tests, finding the \kidslegacy\ dataset to be the most internally robust sample
produced by \kids\ to date.

We tested for systematic bias with respect to previous results from KiDS, finding that most of
the differences in parameter constraints can be attributed to improved redshift distribution estimation and calibration,
as well as new survey area and improved image reduction. 
We leveraged our consistency with \planck\ to estimate the preferred baryon feedback amplitude of
\kidslegacy, allowing us to place an upper limit on $\log_{10} T_{\rm AGN}\leq 7.71\;(8.01)$ at $68\%$ ($95\%$)
confidence. As such, the results from \kidslegacy\ paint a harmonious picture of cosmology at the end of stage-III, with full agreement between the cosmological parameters estimated by both low- and high-redshift probes of cosmic structure.
 
\begin{acknowledgements}
The authors would like to sincerely thank Matthias Bartelmann, who has acted as the external blinding coordinator for every blinded analysis performed by KiDS in the last $10$+ years.
AHW and HHi are supported by the Deutsches Zentrum für Luft- und Raumfahrt (DLR), made possible by the Bundesministerium
  für Wirtschaft und Klimaschutz, under project 50QE2305, and acknowledge funding from the German Science Foundation
  DFG, via the Collaborative Research Center SFB1491 ``Cosmic Interacting Matters - From Source to Signal''.
BS, CH, \& ZY acknowledge support from the Max Planck Society and the Alexander von Humboldt Foundation in the framework of the Max Planck-Humboldt Research Award endowed by the Federal Ministry of Education and Research.
MA is supported by the UK Science and Technology Facilities Council (STFC) under grant number ST/Y002652/1 and the Royal Society under grant numbers RGSR2222268 and ICAR1231094.
MB \& PJ are supported by the Polish National Science Center through grant no. 2020/38/E/ST9/00395. MB is also supported by grant no. 2020/39/B/ST9/03494.
BG acknowledges support from the UKRI Stephen Hawking Fellowship (grant reference EP/Y017137/1).
CH also acknowledges support from the UK Science and Technology Facilities Council (STFC) under grant ST/V000594/1.
HHi, CM, RR, \& AD are supported by an ERC Consolidator Grant (No. 770935).
HHi is also supported by a DFG Heisenberg grant (Hi 1495/5-1).
HHo \& MY acknowledge support from the European Research Council (ERC) under the European Union’s Horizon 2020 research and innovation program with Grant agreement No. 101053992.
BJ acknowledges support by the ERC-selected UKRI Frontier Research Grant EP/Y03015X/1 and by STFC Consolidated Grant ST/V000780/1.
KK acknowledges support from the Royal Society and Imperial College.
SSL is receiving funding from the programme ``Netzwerke 2021'', an initiative of the Ministry of Culture and Science of the State of North-Rhine Westphalia.
MvWK acknowledges the support by the UKSA and STFC (grant no. ST/X001075/1).
PB acknowledges financial support from the Canadian Space Agency (Grant No. 23EXPROSS1) and the Waterloo Centre for Astrophysics.
CG is funded by the MICINN project PID2022-141079NB-C32.
JHD acknowledges support from an STFC Ernest Rutherford Fellowship (project reference ST/S004858/1)
SJ acknowledges the Ram\'on y Cajal Fellowship (RYC2022-036431-I) from the Spanish Ministry of Science and the Dennis Sciama Fellowship at the University of Portsmouth.
LL is supported by the Austrian Science Fund (FWF) [ESP 357-N].
AL acknowledges support from the research project grant `Understanding the Dynamic Universe' funded by the Knut and Alice Wallenberg Foundation under Dnr KAW 2018.0067.
CM acknowledges support from the Beecroft Trust, and the Spanish Ministry of Science under the grant number PID2021-128338NB-I00.
LM acknowledges support from STFC grant ST/N000919/1.
LMo acknowledges the financial contribution from the grant PRIN-MUR 2022 20227RNLY3 “The concordance cosmological model: stress-tests with galaxy clusters” supported by Next Generation EU and from the grant ASI n. 2024-10-HH.0 `Attività scientifiche per la missione Euclid – fase E'.
NRN acknowledges financial support from the National Science Foundation of China, Research Fund for Excellent International Scholars (grant n. 12150710511), and from the research grant from China Manned Space Project n.  CMS-CSST-2021-A01.
LP acknowledges support from the DLR grant 50QE2002.
MR acknowledges financial support from the INAF grant 2022.
TT acknowledges funding from the Swiss National Science Foundation under the Ambizione project PZ00P2\_193352.
YZ acknowledges the studentship from the UK Science and Technology Facilities Council (STFC).
Based on observations made with ESO Telescopes at the La Silla 
Paranal Observatory under programme IDs 179.A-2004, 177.A-3016, 177.A-3017, 
177.A-3018, 298.A-5015. This work was performed in part at Aspen Center 
for Physics, which is supported by National Science Foundation grant PHY-1607611.
\\
\textit{Author Contributions:} 
All authors contributed to the development and writing of this paper. The
authorship list is given in three groups: the lead authors (AHW, BS), 
followed by two alphabetical groups. The first alphabetical
group includes those who are key contributors to both the scientific analysis
and the data products of this manuscript and release. The second group covers those who 
have either made a significant contribution to the preparation of data products or to the 
scientific analyses of \kidslegacy. 
\end{acknowledgements}

\bibliographystyle{aa}
\bibliography{library}

\clearpage
\begin{appendix} 

\section{Emulating matter power spectra with \cosmopower}
\label{sec:cosmopower}

The most computationally expensive section of our modelling pipeline is the prediction of the matter power spectra with
\camb. Fortunately, we can circumvent this time-consuming computation through the use of artificial neural network
emulators, which act to provide a direct non-linear mapping between input cosmological parameters and the corresponding
output power spectrum. Such an emulator can be created with relative ease using the \cosmopower\ framework of
\citet{spuriomancini/etal:2022}. The original \cosmopower\ emulator was trained to reproduce the linear matter power
spectrum, $P_{\rm m,lin}(k)$, and the perturbations caused by non-linear evolution, $P_{\rm m,nl}(k)/P_{\rm m,lin}(k)$,
given a set of relevant input parameters. We updated this process to
 emulate residuals $\log_{10} [P(k)/P_{\rm ref}(k)]$ between logarithmic power spectra  (for both linear and non-linear spectra) and a single
reference power spectrum instead. The reference spectrum was chosen to be
near the centre of our sampling/prior volume, which is defined by the ranges of the various parameters in Table
\ref{tab:priorrange}. These ranges subsequently also define the range of validity of our emulator.

Construction of the emulator requires a large sample of training and testing power-spectra, which were generated by
\camb, over the seven-dimensional\footnote{Four of our parameters were kept fixed to fiducial values in our power
spectra: the curvature contribution to the total energy density ($\Omega_{\mathrm{K}}$) was fixed to zero, the sum of the neutrino
masses ($\sum m_\nu$) was fixed to its minimum standard value, and the equation-of-state parameters of dark energy
($w_0$, $w_{\mathrm{a}}$), were fixed at the values for a cosmological constant. See Table~\ref{tab:priorrange}.} 
volume that we wish to sample. For this purpose, we
generated $450\,000$ training and $50\,000$ testing spectra, produced by a Latin hypercube sampling over our volume.

The consistency between our emulated spectra and the target spectra returned by \camb\ and {\sc HMcode2020} is validated
in the publicly available emulator repository on GitHub\footnote{\url{https://github.com/KiDS-WL/CosmoPowerCosmosis}}, through computation of 
the distribution of absolute fractional residuals between the target
and estimated spectra for all $50\,000$ testing samples.  We found that our updated method of emulation produces a more
uniform error distribution than the original emulation method from \citet{spuriomancini/etal:2022}, across the wide
range of scales that we emulate ($k\in[10^{-5},20]\; h/{\rm Mpc}$), and generally reduces the maximum error over all
scales.  Our emulator exhibits a maximal error below $0.1\%$ for $68\%$ of our test
samples, in both the linear and non-linear power spectra. 

\FloatBarrier
 
\section{Galaxy sample properties for intrinsic alignment modelling}
\label{sec:ia_massdep}

For our fiducial galaxy-type and mass-dependent intrinsic alignment (IA) model, we need to measure the average host halo
  mass $M_{\rm h}$ and the red galaxy fraction $f_r$ in each tomographic bin. These, and relevant intermediate
  properties, are summarised in Table~\ref{tab:ia_massdep_pars}. We note that, where appropriate, we computed averages as
  the means over logarithmic quantities as these propagate straightforwardly to the averages of derived quantities in
  power-law relations. Moreover, all averages were weighted by the shear measurement weight and redshift estimation gold
  weight.

\begin{table}
  \caption{Early-type galaxy sample properties for the mass-dependent intrinsic alignment model per tomographic bin.}
  \label{tab:ia_massdep_pars}
  \centering
    \begin{tabular}{cllll}
\hline\hline 
Bin & $\langle z \rangle$ & $\log_{10} \left( \frac{\langle L_r\rangle }{ L_0 } \right)$ & $\log_{10} M_{\rm eff} [h^{-1} M_\odot]$ & $f_r$ \\ [0.1cm]
\hline 
1 & 0.34 & $-1.32$ & 11.69 & 0.15 \\ [0.1cm]
2 & 0.48 & $-0.72$ & 12.46 & 0.20 \\ [0.1cm]
3 & 0.59 & $-0.47$ & 12.76 & 0.17 \\ [0.1cm]
4 & 0.79 & $-0.28$ & 12.93 & 0.24 \\ [0.1cm]
5 & 0.94 & $-0.14$ & 13.08 & 0.19 \\ [0.1cm]
6 & 1.23 & $\phantom{+}0.01$ & 13.21 & 0.03 \\ [0.1cm]
\hline
\end{tabular}
  \tablefoot{Columns, from left to right: tomographic bin number, mean redshift, logarithmic normalised galaxy luminosity in the $r$-band, logarithm of the effective halo mass, and the red galaxy fraction.}
\end{table}

To determine $f_r$, we used the output of the photometric redshift estimation from the template-fitting code \textsc{bpz}
\citep{benitez:2000}. \textsc{bpz} employs a set of six model templates, ordered in terms of star formation activity,
and interpolates linearly between adjacent template to determine a best-fitting spectral energy distribution. The
\textsc{bpz} output $T_{\rm B}$ encodes the best-fit templates and their admixture in steps of 0.1. We chose $T_{\rm
B}<1.9$ to define early-type galaxies, encompassing all galaxies with some contribution of an elliptical galaxy spectrum
(template no.~1). We validated that, at least on a matched sample between the KiDS and GAMA \citep{driver/etal:2011}
surveys (i.e. for relatively bright and low-redshift galaxies), this $T_{\rm B}$ cut is capable of isolating the red
sequence.

We found that in the first five tomographic bins $f_r$ fluctuates around 0.2, broadly in line with expectations for a
flux-limited sample. The sixth bin contains a negligible fraction of early-type galaxies, a feature that we also see in
our simulated sixth bin from  \skills. The statistical error on $f_r$ is very small. We repeated the measurement varying
$T_{\rm B}$ by $\pm 0.1$ and obtained changes in $f_r$ of less than 0.02. Therefore, we decided to consider $f_r$ as fixed
in the IA model.

To compute halo masses, we made use of the $r$-band luminosities derived as part of the KiDS Data Release~5
\citep{wright/etal:2024}. \citet{vanuitert/etal:2015} measured effective halo masses for SDSS (LOWZ and CMASS)
early-type galaxies via weak lensing, as a function of luminosity and redshift. We used their reported halo mass results
and galaxy properties to construct the following linear relation
\begin{align}   
\label{eq:vuitert_halomass}
    \log_{10} M_{\rm h}^{(i)} [h^{-1} M_\odot] &= \gamma \left( \frac{1}{1+\langle z \rangle^{(i)}} - \frac{1}{1+z_{\rm piv}} \right)\\ \nonumber
    & \;\; + \delta\,  \log_{10} \left( \frac{\langle L_r\rangle^{(i)} }{ L_0 } \right) + \Delta M\;, 
\end{align}
where we chose $z_{\rm piv}=0.3417,$ such that the parameters $\gamma$ and $\Delta M$ are uncorrelated to a good
approximation. An additional parameter $\delta$ describes the dependence on the average luminosity $\langle L_r\rangle$,
normalised by the reference luminosity $L_0 = 10^{11} (h/0.7)^{-2} L_\odot$. \citet{vanuitert/etal:2015} showed that the
luminosity dependence evolves only mildly with redshift, so that we determine $\delta$ as the inverse variance-weighted
average scaling over the four samples they considered. Thus, we obtained $\gamma = 1.25 \pm 0.33$, $\delta = 1.42 \pm
0.16$, and $\Delta M = 13.56 \pm 0.02$. In line with expectations, both the average luminosity and halo mass increase
monotonically by more than an order of magnitude from the first to the sixth tomographic bin (see
Table~\ref{tab:ia_massdep_pars}).

Statistical uncertainties on the halo mass are propagated via Monte Carlo sampling from the fit and measurement
uncertainties. The standard deviation of $\log_{10} M_{\rm h}$ is approximately $0.2$ in the first tomographic bin, driven
by the extrapolation of Eq.~(\ref{eq:vuitert_halomass}) to fainter samples, and of order $0.1$ in the remaining bins,
where a reduced error due to luminosities more compatible with the range fitted by \citet{vanuitert/etal:2015} is
gradually traded off with larger uncertainty because of extrapolation to higher redshifts. Halo mass estimates are
strongly correlated (correlation coefficient greater than $0.9$) between tomographic bins due to the extrapolation from the
luminosity and redshift ranges considered by \citet{vanuitert/etal:2015}. Therefore, we propagated the uncertainty in
halo masses into our IA model by assuming a multivariate Gaussian prior with standard deviation of $0.2$ in $\log_{10}
M_{\rm h}$ for all bins and the correlation matrix as obtained from the Monte Carlo error propagation.

Throughout the construction of our astrophysics-informed galaxy IA model, we have not distinguished between central and
satellite galaxies, although their alignment mechanisms are believed to be different \citep{schneider/bridle:2010}. The NLA
model assumes a fixed relation between linear and non-linear scales where centrals and satellites tend to dominate IA
signals, respectively, although our scale-dependent IA model tests whether the data prefers a different ratio on average
over the whole survey. However, the satellite fraction in our early-type galaxy sample is unknown, yet almost certainly
evolves with halo mass and redshift \citep[see e.g.][]{shuntov/etal:2022}. For a given mass and redshift, early-type satellite
fractions are typically twice as large as those of late types \citep{mandelbaum/etal:2006}, so that the impact of potential IA
signals of blue satellites is less of a concern.

Moreover, the satellite fractions in most galaxy samples used in direct IA constraints are also unknown.
\citet{fortuna/etal:2025} estimated a satellite fraction of $30\,\%$ and possibly higher \citep[see
also][]{vanuitert/etal:2016} in a subset of the GAMA-KiDS sample and subsequently excluded it from their fit.
\citet{johnston/etal:2019} studied alignments amongst and between centrals and satellites. They tentatively found a
steeper slope of the red satellite IA signal on small transverse scales. The mass scaling of the IA amplitude in red
satellites is qualitatively similar to that of centrals \citep{fortuna/etal:2021}.

In summary, neither cosmic shear surveys nor direct measurements to calibrate IA currently provide sufficient
information to treat the IA signals of satellites and their associated trends separately. We are forced to treat
satellite effects as part of an effective small-scale IA model, with any discrepancies in the combined dependencies on
scale and galaxy sample properties expected to be comfortably absorbed in the still wide priors. Once surveys reach the
statistical power to significantly constrain IA models beyond an overall amplitude, calibration measurements of IA will
need to be (re-)analysed separately for central and satellite populations. Halo models \citep{fortuna/etal:2021} will be
a natural choice to accurately represent different IA mechanisms and scalings on large and small scales.

\FloatBarrier
 
\section{Band-powers derivation}\label{sec:bandpowers}
To arrive at the correct response functions for our band-powers, we first write ${\mathcal C}_{{\rm E/B},l} $ as a
function of $\xi_\pm(\theta)$, in the same manner as for the COSEBIs in Eq.~\eqref{eq:COSEBIsReal},
\begin{equation}
\label{eq:bp_xipm}
{\mathcal C}_{{\rm E/B},l} = \frac{\pi}{{\mathcal N}_l}\; \int_0^\infty \dd \theta\, \theta\; T(\theta) \left[ g_+^l(\theta)\;\xi_+(\theta)\; \pm g_-^l(\theta)\;\xi_-(\theta)\right]\;, 
\end{equation}
where $g_\pm^l(\theta)$ are the band-power filter functions for bin $l$ and $T(\theta)$ is a function that limits the range of $\theta$-scales that contribute to the integral. If $\xi_\pm$ is known for all scales then $T(\theta)=1$ everywhere. We now write $g_\pm^l(\theta)$ in terms of the idealised response function, 
\begin{equation}
g_\pm^l(\theta) =  \int_0^\infty \dd \ell\, \ell\, S_l(\ell)\; {\rm J}_{0/4}(\ell \theta)\;.
\end{equation}
If $S_l(\ell)$ is a top-hat then,
\begin{align}
\label{eq:bp_kernel_cosmicshear}
g_+^l(\theta) &= \frac{1}{\theta^2} \left[\theta \ell_{{\rm up},l}\;
  {\rm J}_1(\theta \ell_{{\rm up},l}) - \theta \ell_{{\rm lo},l}\;   {\rm J}_1(\theta \ell_{{\rm lo},l}) \right] \;, \\ \nonumber
g_-^l(\theta) &= \frac{1}{\theta^2} \left[{\cal G}_-(\theta \ell_{{\rm
      up},l})  - {\cal G}_-(\theta \ell_{{\rm lo},l}) \right]\;,
\end{align}
with,
\begin{equation}
    {\cal G}_-(x) = \left(x - \frac{8}{x}\right) {\rm J}_1(x) - 8 {\rm J}_2(x)\;.
\end{equation}
Inserting for $\xi_\pm$ from Eq.~\eqref{eq:xipm_th} into Eq.~\eqref{eq:bp_xipm} we arrive at,
\begin{align}
\label{eq:bp_cl}
    {\cal C}_{{\rm E},l}^{(ij)} &=  \frac{1}{2 {\cal N}_l} \int_0^\infty \dd
\ell\; \ell \left[ W^l_{\rm EE}(\ell)\; C^{(ij)}_{\eps \eps, \rm
    E}(\ell) + W^l_{\rm EB}(\ell)\; C^{(ij)}_{\eps \eps, \rm B}(\ell) \right]\; , \\ \nonumber
{\cal C}_{{\rm B},l}^{(ij)} &=  \frac{1}{2 {\cal N}_l} \int_0^\infty \dd
\ell\; \ell \left[ W^l_{\rm BE}(\ell)\; C^{(ij)}_{\eps \eps, \rm
    E}(\ell) + W^l_{\rm BB}(\ell)\; C^{(ij)}_{\eps \eps, \rm B}(\ell) \right]\;,
\end{align}
where the weight functions are, 
\begin{equation}
\begin{aligned}
W^l_{\rm EE}(\ell) &= \!\! \int_0^\infty \!\! \dd \theta\, \theta\; T(\theta) \left[ {\rm J}_0(\ell \theta)\; g_+^l(\theta) + {\rm J}_4(\ell \theta)\; g_-^l(\theta) \right]\;, \\
W^l_{\rm EB}(\ell) &=  \!\! \int_0^\infty \!\! \dd \theta\, \theta\; T(\theta) \left[{\rm J}_0(\ell \theta)\; g_+^l(\theta) - {\rm J}_4(\ell \theta)\; g_-^l(\theta) \right]\;,
\end{aligned}    
\end{equation}
with $W^l_{\rm EE}(\ell) = W^l_{\rm BB}(\ell)$ and $W^l_{\rm EB}(\ell) = W^l_{\rm BE}(\ell)$.
Since $\xi_\pm$ are measured between $\theta_{\rm min}$ and $\theta_{\rm max}$, $T(\theta)$ has to be defined such that
no information beyond this range is included. To avoid ringing, we apodise $T(\theta)$ with a Hahn window  and smooth
its transitions to zero:
\begin{align}
\label{eq:apodisation}
T(\theta) = \left\{ 
\begin{aligned}  0\,; &\; x < x_{\rm lo} - \frac{\Delta_x}{2} \\ 
\cos^2 \left[\frac{\pi}{2} \frac{x- (x_{\rm lo}+\Delta_x/2)}{\Delta_x} \right]\,;  &\; x_{\rm lo} -\frac{\Delta_x}{2} \leq x <  x_{\rm lo} +\frac{\Delta_x}{2} \\ 1 \,; &\;  x_{\rm lo} +\frac{\Delta_x}{2} \leq x <  x_{\rm up} -\frac{\Delta_x}{2} \\  
\cos^2 \left[\frac{\pi}{2} \frac{x-(x_{\rm up} - \Delta_x/2)}{\Delta_x} \right]\,; &\;  x_{\rm up} -\frac{\Delta_x}{2}  \leq x <  x_{\rm up} +\frac{\Delta_x}{2}\\ 0 \,; &\; x \geq  x_{\rm up} +\frac{\Delta_x}{2}\;. \end{aligned} 
\right. 
\end{align}
Here $x=\log(\theta)$, $\Delta_x$ is the log-width of the apodisation, $x_{\rm lo}-\Delta_x/2 = \log(\theta_{\rm min})$
and $x_{\rm up}+\Delta_x/2 = \log(\theta_{\rm max})$. We require $\Delta_x>\log(\theta_{\rm up}/\theta_{\rm lo})$, and
in practice have used $\Delta_x=0.5$ for all \kidslegacy\ \bandpowers\ analyses. We note, also, that this apodisation
method is different to how we applied the apodisation for KiDS-1000 analysis, where we instead had $x_{\rm lo} =
\log(\theta_{\rm min})$ and $x_{\rm up} = \log(\theta_{\rm max})$. This change was made as it was deemed more
intuitive/appropriate that the band power windows be contained primarily within the requested $\ell$ ranges after
apodisation, rather than before. 

\FloatBarrier
 
\section{KiDS-Legacy variable depth mocks and covariance validation}\label{sec:glass}
\begin{figure}
    \centering
    \includegraphics[width = .45\textwidth]{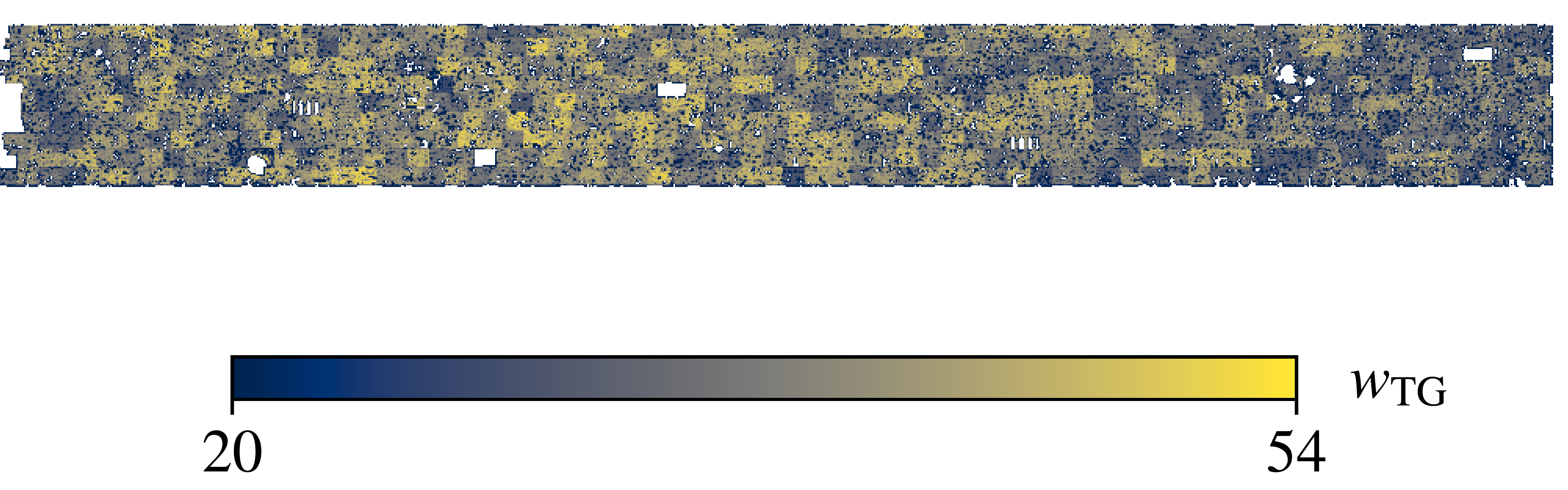}
    \caption{Cartesian spatial map (corresponding $N_{\mathrm{side}}=1024$, showing a local flat projection of the
    \texttt{healpix} map) of the TG weights, $w_{\mathrm{TG}}$, throughout a single field of our KiDS-Legacy-like mock
    catalogue.  Larger TG weights correspond to a higher local source density. }
    \label{fig: tg_map}
\end{figure}
\begin{figure}
    \centering
    \includegraphics[width = .45\textwidth]{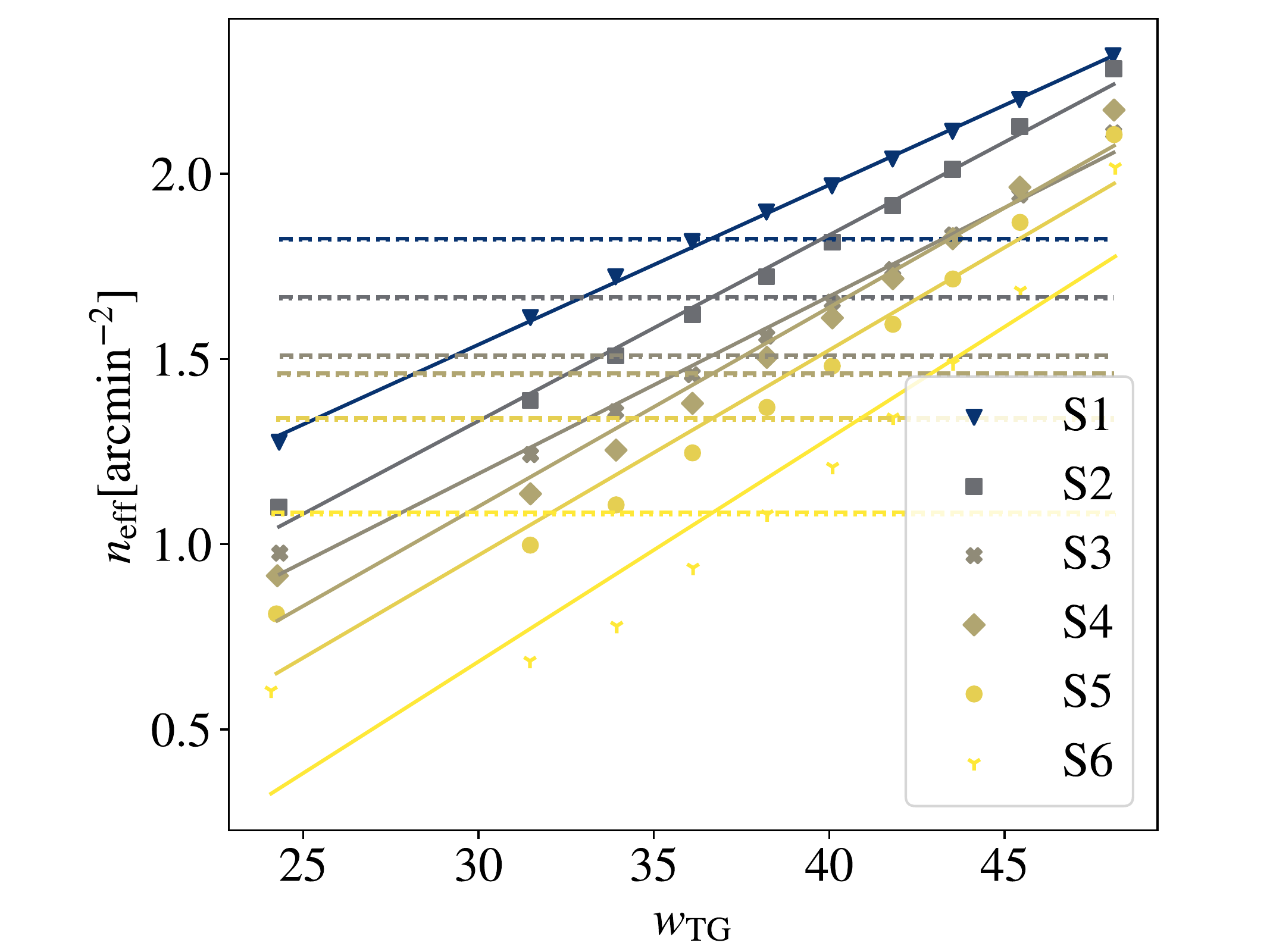}
    \caption{Dependence of the galaxy density, $n_{\mathrm{eff}}$, on TG weights, $w_{\mathrm{TG}}$ in the
    KiDS-Legacy-like mock catalogue. The data points represent the mean $n_{\mathrm{eff}}$ of ten equi-populated bins in
    $w_{\mathrm{TG}}$. The solid line shows the linear fit to the aforementioned data points of their respective
    tomographic bin, labelled ${\rm S}1$-${\rm S}6$. The dashed horizontal lines show the mean values of
    $n_{\mathrm{eff}}$ calculated from the galaxy samples with variable depth per tomographic bin, while the dashed
    horizontal lines show the values of $n_{\mathrm{eff}}$ for the respective galaxy samples without any spatial
    variations in the observational depth.}
    \label{fig: neff_tg}
\end{figure}

Our analysis used an analytical covariance matrix, as summarised in Sect.~\ref{sec: onecov}. Here, we describe the mocks
which were used to validate the covariance matrix \citep[see][]{reischke/etal:2024} in more detail.

The log-normal simulations are based on
GLASS\footnote{\href{https://glass.readthedocs.io/stable/}{https://glass.readthedocs.io/stable/}}  \citep[Generator for
Large-Scale Structure,][]{tessore/etal:2023} with a detailed description given in \citet{wietersheim/etal:2024} as
well as Section~$8.1$ in \citet{reischke/etal:2024}. The resulting density field realisations, in concentric shells, were
subsequently integrated along the line-of-sight weighted by the weak lensing kernel, Eq.~(\ref{eq:kernel_lensing}), to
create a total of 4224 realisations of the shear field.

For each realisation, galaxies were sampled, factoring in systematic effects such as the survey footprint and variable
depth (as caused by seeing, sky transparency fluctuations, and differences in the number of overlapping pointings) using
SALMO\footnote{\url{https://github.com/Linc-tw/salmo}} \citep[Speedy Acquisition for Lensing and Matter
Observables,][]{joachimi/etal:2021}. The inclusion of a sixth tomographic bin in KiDS-Legacy incorporates more faint
galaxies in the sample, enhancing the overall magnitude of variable depth and boosting the lensing signal relative to
the noise. Although previous analyses suggested that depth variability effects on the signal in KiDS-1000 remain below
the noise threshold \citep{heydenreich/etal:2020}, this might not be the case for KiDS-Legacy.

To model variable depth, we rely on organised randoms \citep[ORs, introduced in][]{johnston/etal:2021,yan/etal:2024}. ORs
use SOMs to generate random samples that reflect the spatial variability caused by a high-dimensional systematic space.
The ORs provide Tiaogeng (TG) weights, $w_{\mathrm{TG}}$, \citep[see][for a detailed explanation]{yan/etal:2024},
quantifying the spatial variability of the selection function of the galaxy sample as a function of position on the sky.
The ORs robustly trace the variations in depth for \kidslegacy. \citet{yan/etal:2024} demonstrated that the TG weights
are uncorrelated with $r$-band magnitude, photometric redshift, and underlying cosmological density variations. In
contrast, they are highly correlated with the magnitude limit of the survey in the $r$-band, which is used for both
source extraction and shape measurement.

Figure~\ref{fig: tg_map} shows the TG weights for a single field in our \kidslegacy-like mock, as a function of position
on the simulated sky. Additionally, Fig.~\ref{fig: neff_tg} shows the corresponding local source density of sources as
a function of TG weights, colour-scaled for the different source bins. The horizontal dashed lines depict the
effective homogeneous number density. Solid lines are linear fits to the measured effective number density in ten
equi-populated bins of the TG weights. This linear interpolation allows for an effective evaluation of
$n_{\mathrm{eff}}$ at each pixel. 

The local density variations seen here also affect the redshift distributions. This issue is addressed by reapplying
the photometric redshift calibration using SOMs \citep[see][for details]{wright/etal:2025} to each tomographic bin and each of
the ten TG weight bins. The resulting redshift distribution then enters the shear maps via the weak lensing kernel. 

On each simulation, we measured the shear two-point correlation functions as indicated in Sect.~\ref{sec:2pt}, from
which the sample covariance could be estimated directly. These mocks were then compared to the {\sc{OneCovariance}} code.
The results are presented in \citet{reischke/etal:2024} with 10 percent agreement for the variances, very
similar to what was found in previous analysis \citep{joachimi/etal:2021}. In particular, we found that the limiting
factor in the covariance modelling is not the effect of variable depth but rather the survey geometry on large scales.
All these effects, however, do not significantly affect the cosmological results presented in this paper.

\section{$B$-mode analysis}\label{sec:bmodeinvestigation}

During the initial phases of the blinded \kidslegacy\ analysis, our $B$-mode null-test analyses (see Sect.~\ref{sec:
bmodes}) failed our required significance threshold ($p\geq0.01$). This led to a dedicated effort to investigate the
possible causes of the $B$-mode signal in the dataset, starting with the highest level data products and systematically
working our way back through the data production and analysis pipelines, to identify the root cause of the $B$-mode
signal. 

As \konek\ did not show significant $B$ modes (specifically, as analysed by \citealt{li/etal:2023b} using the same
shape-measurement code and analysis choices, such as scale cuts), the initial test in our investigation was to determine
whether the observed $B$-mode significance in \kidslegacy\ was owing to a decrease in covariance (specifically shot
noise) caused by the increased area of \kidslegacy\ over \konek. We therefore started by analysing the \kidslegacy\
$B$-mode significance in the same footprint as \konek, finding that the \kidslegacy\ data in this footprint still
exceeded the $B$-mode significance requirement. 

Subsequent investigation of the $B$-mode origin focussed on PSF contamination (via Paulin-Henriksson statistics; see
Sect.~\ref{sec:nulltests}), 
one- and two-dimensional $c$-terms, covariance construction, and shape
measurement biases (as a function of, for example, galaxy morphology and position on-sky). Each of these tests was
performed locally (per square degree tile), in the detector plane (i.e. coadding all sources in the frame of the focal
plane), with spatial jackknife subsets of the full survey, and globally (e.g. at the data vector level, and looking for
population-level trends). Each of these investigations failed to illuminate the origin of the $B$-mode signal. Finally,
we investigated the lowest-level possible cause of the anomalous $B$-mode signal: systematic failures in the astrometric
solution used in the \theli\ calibration (with \gaia) of \rband\ images. The global astrometric accuracy of \kidslegacy\
is presented in \citet{wright/etal:2024}, where we demonstrated the accuracy of the \theli\ (constructed with a \gaia\
reference) and \astrowise\ (constructed with a SDSS and 2MASS reference) catalogues both globally (see their Figures 10,
11, C1, and C2) and as a function of right ascension and declination (RA and Dec; see their Figures 8 and 9), compared
to \sdss, \twomass, and \gaia. In our new investigation, we explored the astrometric accuracy in finer detail (i.e.
higher on-sky resolution) and using the individual astrometric solutions of \theli\ exposures and \astrowise\ co-adds. 

We computed two test statistics for investigation of the $B$-mode signal. The first test
statistic is sensitive to the astrometric consistency between \theli\ co-added images, which are used for source
detection, and \astrowise\ co-added images, which are used for photometric measurements (and therefore impacts
photometric redshift estimation, which is used for tomographic binning). This metric ($\delta
\mathrm{r}_{\mathrm{int}}$) measured the systematic difference between the centroid of all targets extracted from these
images. This was achieved by matching all sources (within
$2\arcsec$) extracted from \astrowise\ and \theli\ co-added \rband\ images. 
With these matches, we computed the radial separation $r$ between each source's centroid in the
two images and then calculated the median of these separations in $1\times1$ arcminute on-sky bins. The 2D map
of median separations was then assigned back to each source. 

Our second test statistic was computed between individual exposures of the \theli\ \rband\ images, which are used for
shape measurement. This metric ($\delta \mathrm{r}_{\mathrm{exp}}$) measures the systematic difference between the
centroids of targets extracted from individual \theli\ exposures, which are point-like and have high S/N.
This was achieved by matching sources extracted from \astrowise\ (used here purely
as a static reference) and the individual \theli\ exposures (within $2\arcsec$).  
For the \astrowise\ static reference, we applied selections on \textsc{Source Extractor} \citep{bertin/arnouts:1996}
output variables 
\begin{equation}\label{eqn: size}
  \mathtt{FLUX\_RADIUS} < 4.0
\end{equation}
and
\begin{equation}\label{eqn: snr}
  \mathtt{FLUX\_ISO}/\mathtt{FLUXERR\_ISO} > 10.0,\;
\end{equation}
to select point-like high-S/N sources. Settings used in the extraction of sources 
are provided in Table~4 of \citet{wright/etal:2024}.  
For the sources extracted from the individual \theli\ exposures, we performed the
same point source selection (Eq. \ref{eqn: size}) and a modified S/N criteria, to
account for the reduced depth of the individual exposures compared to the co-add:
\begin{equation}\label{eqn: snr exp}
  \mathtt{FLUX\_ISO}/\mathtt{FLUXERR\_ISO} > 10.0 / \sqrt(5)\;. 
\end{equation}
With these matches, we then computed the radial separation $r_{ij}$ between all exposures in $\{i,j \in [1,5]\,|\, i\neq
j\}$, computed the maximum of these separations in $2\times2$ arcminute on-sky bins, and smoothed these maps with a $1$
arcminute Gaussian kernel. We then combined these individual $15$ $i\leftrightarrow j$ maps into a final single map, by
selecting again the maximum separation per on-sky bin. This resulting map was then once-again smoothed using a $1$
arcminute Gaussian kernel. 

These maps provided us with two test statistics that could be used as an additional mask for the survey, given
specific thresholds in the statistic values. The first metric is relevant for possible failures in photometry: if the
astrometric solution between \astrowise\ and \theli\ differs, then the apertures extracted from \theli\ imaging will be
incorrectly placed on the \astrowise\ images.  The second metric is relevant for possible failures in 
shape measurement: the shape measurement algorithm \lensfit, which is used in \kids, assumes perfect astrometric
alignment of exposures within the \theli\ \rband\ imaging. As such, any residual astrometric misalignment will manifest as an
erroneously estimated shape, and coherent misalignments may manifest as an $E$- or $B$-mode cosmic-shear signal. 

For our first statistic, the degree of inconsistency between the imaging that can lead to systematic errors in
photometry is relatively benign: our photometric measurement code \gaap\ has a minimum aperture size of $0\farcs7$,
driven largely by the fact that our imaging has seeing that is consistently larger than $0\farcs6$ (see e.g. Figure 7
of \citealt{wright/etal:2024}). Systematic errors in aperture centroids at the level of the FWHM and larger will lead to
significant flux to be lost from the target aperture (assuming a point-like source), and thus may pose a problem for our
photometry. However, there are two additional relevant considerations. Firstly, in our cosmological analyses, all
sources that are used must be resolved and thus (in some direction) larger than the PSF FWHM. Secondly, even if
significant flux is missed, if the \astrowise\ imaging are all well aligned to each other, then the flux will be
consistently missed in all bands, thereby keeping the source SED largely consistent. As such, we expect the effect to be
rarely relevant, and (when it is relevant) only to have a minor impact on photometric redshifts. Nonetheless, we chose
to mask all areas of the survey where our first statistic exceeds the minimum PSF FWHM: that is, where
$\delta\mathrm{r}_{\mathrm{int}} > 0\farcs6$. This corresponds to approximately $0.7 \,{\rm deg}^2$ of removed data over the 
full \kidslegacy\ footprint. 

\begin{figure}
  \centering
  \includegraphics[clip=true,trim={0 0 0 0},width=\columnwidth]{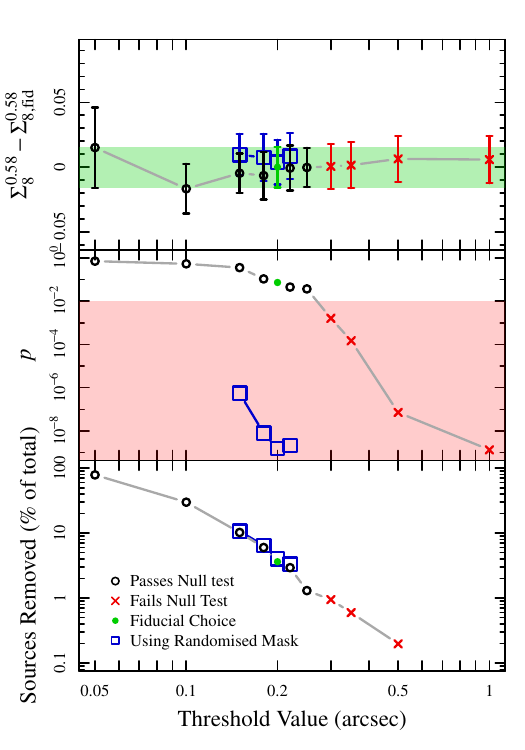}
  \caption{
  Summary of our blinded $B$-mode investigation where we measure the COSEBIs \cosebis\ cosmic shear statistics, masking
  survey area where our astrometric metric, $\delta\mathrm{r}_{\mathrm{exp}}$, exceeds a series of threshold values
  (shown on the $x$-axis).   The middle panel shows the $p$-value that quantifies the likelihood that the $B$-mode is
  consistent with zero.  As we reduce the $\delta\mathrm{r}_{\mathrm{exp}}$ threshold, the $p$-value increases above
  $p\leq0.01$ (red shaded region), and the masked survey passes the $B$-mode null test.  As the mask reduces the total
  number of sources (lower panel), the blue squares show the $p$-values from a random mask analysis, demonstrating that
  the improved $B$-mode is not driven by a reduction in area.  In the upper panel we find that the astrometric masking
  has a benign impact on our $\Sigma_8$ cosmological constraints.  The green data point and shaded regions shows the
  results from our fiducial analysis, demonstrating that failing data-vectors are consistent with our fiducial, all
  having less than $0.4\sigma$ changes in $\Sigma_8$.
  }\label{fig: bmodetest}
\end{figure}

For our astrometric statistic, there is much less intuition available regarding the level of astrometric error that may be
problematic for our analysis. As such, we chose to investigate (while remaining fully blinded; outcomes of the various 
tests were very similar for all blinds) the impact of various
masks (defined by thresholds in $\delta\mathrm{r}_{\mathrm{exp}}$) on our null test results, and on our cosmological
inference. Figure \ref{fig: bmodetest} shows the results of this investigation. On the $x$-axis of each panel in the
figure, we show the threshold in our astrometric metric $\delta\mathrm{r}_{\mathrm{exp}}$ that was used to define a mask.
This mask was then applied to the full \kidslegacy\ cosmic shear source sample, and the end-to-end cosmological analysis
pipeline was run (including \nz\ estimates, shape recalibration, two-point statistics, covariances, and cosmological
chains). For each threshold, we were therefore able to report the change in sample size (bottom panel of Fig. \ref{fig:
bmodetest}; similar to the change in area), the change in the $B$-mode null test $p$-value (middle panel), and the change in the inferred marginal value of
$\Sigma_8$ (top panel). 

Firstly, we see that the null test $p$-value passes our pre-defined threshold of $p=0.01$ with
$\delta\mathrm{r}_{\mathrm{exp}} \lesssim 0\farcs25$, corresponding to a masked fraction of sources of at least $1\%$. 
Secondly, we see that the value of $\Sigma_8$ is robust to the application of increasingly conservative
thresholds, demonstrating a $\Delta \Sigma_8 \lesssim 0.5\sigma$ variation when discarding up to 100 deg$^2$ of possibly
contaminated data and causing the $B$-mode significance to decrease to $p>0.3$. In the region around
where we first pass the null test, and do not throw away an overly excessive fraction of the source sample (i.e.
$\delta\mathrm{r}_{\mathrm{exp}} \in [0\farcs20,0\farcs25]$) our recovered marginal value of $\Sigma_8$ is at most 
$0.006$ (or $0.36\sigma$) smaller than in the case where we perform no masking at all (equivalent to the threshold of $1\arcsec$).
Thirdly, we can see that the null test $p$-value is correlated with the fraction of removed area, which begs the
question of whether the improvement in our null test might simply be the result of  reduced constraining power related
to the smaller sample. To test this possibility, we must construct a new mask that removes the same amount of area from
the survey at a given threshold, but which does not trace the underlying astrometric errors that we think are likely
responsible for our $B$-mode signal. To do this, we constructed a randomised version of our astrometric metric, by
randomising the individual masks (defined per square degree tile) between the 1347 tiles of the survey. Furthermore, we
also flip the shuffled masks in both the RA and Dec directions, to remove any possible correlations between the
constructed masks and the focal plane. We then repeated the analysis using the randomised mask, and found that the $B$-mode
signal remained significant (see the blue points in Fig.~\ref{fig: bmodetest}).  As such, we conclude that our
astrometric mask is removing a bone fide $B$-mode signal from the data vector, and not just removing sensitivity to the
$B$ mode by lowering the constraining power. 

As such, from these tests we can see that our astrometric statistic targets 
our $B$-mode signal, and that our recovered cosmology is fairly insensitive to our choice
of the mask. We therefore opt to
define a fiducial threshold in our test statistic that passes our $B$-mode test and makes some intuitive sense. To that
end, we define the fiducial mask to be where $\delta\mathrm{r}_{\mathrm{exp}} > 0\farcs2$, where $0\farcs2$ is chosen as
a round number that is smaller than the \theli\ pixel size ($0\farcs212$). This corresponds to approximately $4\%$ of the 
full \kidslegacy\ source sample, drawn from a previously unmasked area of $46.6$ square degrees.  

 \section{Cosmological parameters from \xipm}\label{sec:xipmcos}
\begin{figure}
  \centering
  \includegraphics[clip=true,trim={0 0 0 0},width=0.95\columnwidth]{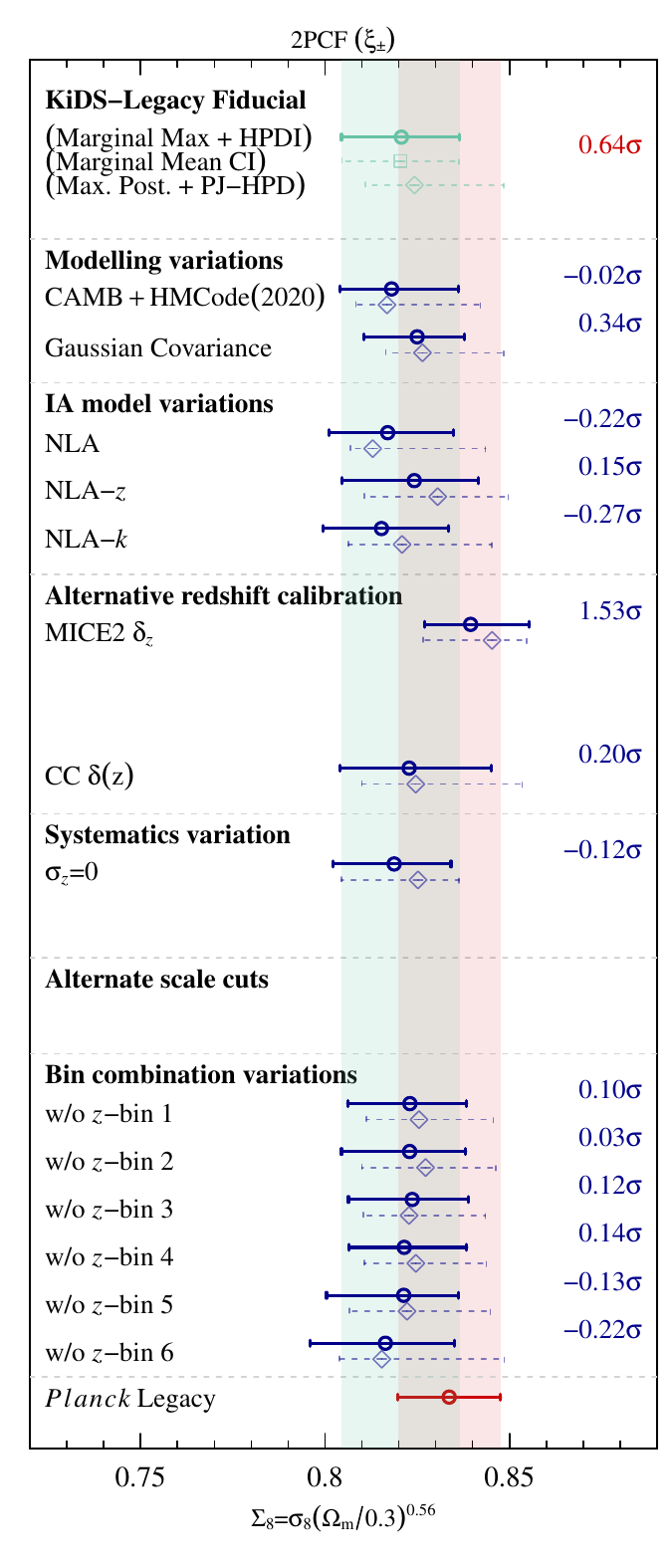}
  \caption{Whisker diagram of $\Sigma_8$ cosmological results from \xipm. 
  The figure is annotated as in \protect Fig.~\ref{fig:whisker_sig8}.}\label{fig:whisker_sig8_xipm}
\end{figure}

\begin{figure*}
  \centering
  \includegraphics[clip=true,trim={0 0 0 0},width=0.95\textwidth]{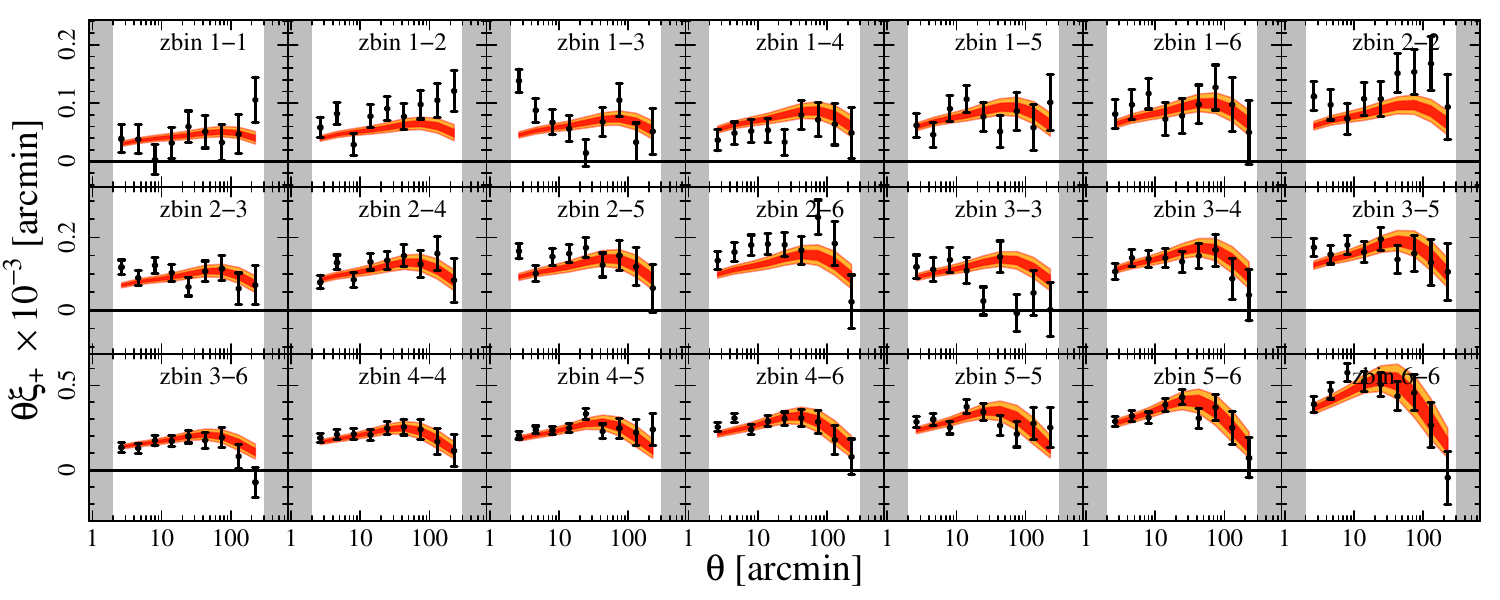}
  \includegraphics[clip=true,trim={0 0 0 0},width=0.95\textwidth]{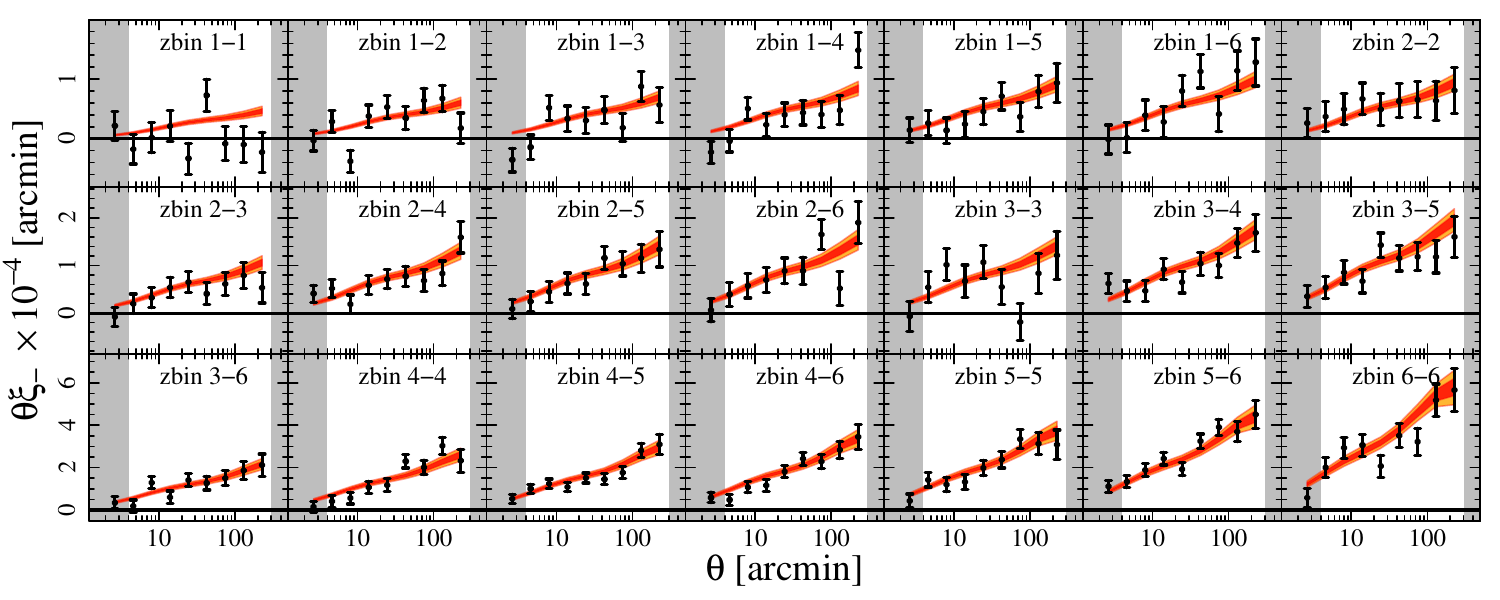}
  \caption{Data vector and posterior predictive distributions for the fiducial run of our \xipm. Scales in the 
  grey shaded region are excluded from our fiducial analysis. TPDs are shown for  
  $68^{\rm th}$ (red) and  $95^{\rm th}$ (orange) percentiles of the posterior models.}\label{fig:tpd_xipm}
\end{figure*}

\begin{table*}
  \caption{Constraints on $S_8$ ($\alpha=0.50$) and $\Sigma_8$ ($\alpha$ free) for \xipm\ under our fiducial setup.}\label{tab:S8constraints_xipm}
  \centering
  \begin{tabular}{cc|cccccccccc}
    \toprule
            & Setup    &  statistic & $\alpha$ & $\chi^2$  &   dof   & PTE   & $N_{\rm samp}^{\rm PJ-HPD}$ & Marginal                  & Max. Apost.               & Marginal                  \\
            &          &            &          &           &         &       &                             &          Mode+HPD         &               +PJ-HPD     &          Mean + CI        \\
    \midrule
               $S_8$ &             Fiducial &  $\xi_\pm$ & 0.50 & 413.1 & 351.5 & 0.013 &      48      & $0.825^{+0.017}_{-0.018}$ & $0.827^{+0.022}_{-0.015}$ & $0.825^{+0.018}_{-0.017}$ \\ [+0.1cm]
          $\Sigma_8$ &             Fiducial &  $\xi_\pm$ & 0.55 & 413.1 & 351.5 & 0.013 &      11      & $0.821^{+0.016}_{-0.016}$ & $0.825^{+0.025}_{-0.013}$ & $0.821^{+0.016}_{-0.016}$ \\ [+0.1cm]
    \bottomrule
  \end{tabular}
\end{table*}

Here, we present cosmological parameter constraints from \xipm, for comparison with previous work and external datasets. 
We followed the \kids\ tradition of using nine logarithmic bins in $\theta$, but constructed these over the same scale cuts 
as used by our fiducial \cosebis: $\theta \in [2,300]\arcmin$. We also removed the first $\theta$-bin from $\xi_-$,
in effect limiting it to $\theta\in[4,300]\arcmin$, following previous KiDS analyses. For our theoretical predictions we
must also apply this binning scheme, however we note that it is not sufficient to simply integrate $\xi_\pm(\theta)$ in
Eq.~(\ref{eq:xipm_th}) over the bin width. Instead, we must weight $\xi_\pm(\theta)$ by the weighted number of galaxy
pairs, $N_{\rm pair}(\theta)$ \citep[see Appendix A of][for details]{asgari/etal:2019}. 

We used the finely binned $\xi_\pm(\theta)$ and its corresponding number of galaxy pairs to make our measurements and our
predictions,
\begin{equation}
\label{eq:xipm_binning}
\xi_\pm^{\rm binned}(\theta_{\rm bin}) = \frac{\sum N_{\rm pair}(\theta) \;\xi_\pm(\theta)\; \Delta(\theta_{\rm bin}-\theta)}{\sum N_{\rm pair}(\theta) \;\Delta(\theta_{\rm bin}-\theta)}\;,
\end{equation}
where $\Delta(\theta_{\rm bin}-\theta)$ is the binning function, which is equal to $1$ if $\theta$ is within the
boundaries of the bin labelled $\theta_{\rm bin}$, and is $0$ otherwise. When predicting $\xi_\pm^{\rm
binned}(\theta_{\rm bin})$, we assumed that the distribution of the number of pairs and the value of $\xi_\pm(\theta)$
were both essentially constant within each of the 1000 bins, and used the predicted values of $\xi_\pm$ at the centres of
the bins from Eq.~\eqref{eq:xipm_th} in Eq.~\eqref{eq:xipm_binning}.

Figure~\ref{fig:whisker_sig8_xipm} presents the whisker diagram for our \xipm\ constraints under all analyses, and
Fig.~\ref{fig:tpd_xipm} presents our posterior predictive distributions for \xipm. Despite our 
concerns regarding the possible systematic contamination of our \xipm\ signals, our cosmological parameter constraints
from \xipm\ are in good agreement with the results from our other statistics. We note, however, that our posterior
predictive distributions (shown in Fig.~\ref{fig:tpd_xipm}) are much closer to the significance limit (PTE$<0.01$) than
our other statistics (see results for the fiducial setup in Table~\ref{tab:S8constraints_xipm}, and for the 
analysis variations in Table~\ref{tab: longtable}). Finally, we note that our companion paper
\citet{stoelzner/etal:2025} finds strong evidence for a scale-dependent tension between cosmological parameters modelled
with \xipm. Furthermore, \citet{oehl/2024} found that the likelihood of $\xi_\pm$ becomes non-Gaussian on large scales.
This is evidence that our choice to present constraints with \xipm\ as less reliable was warranted. 

\FloatBarrier
 
\section{Additional posterior constraints}\label{sec:posteriors}

In this appendix we provide additional posterior constraints, for various combinations of parameters and model setups. 

\subsection{Whisker diagram in $S_8$}

\begin{figure*}
  \centering
  \includegraphics[clip=true,trim={0 0 0 0},width=0.85\textwidth]{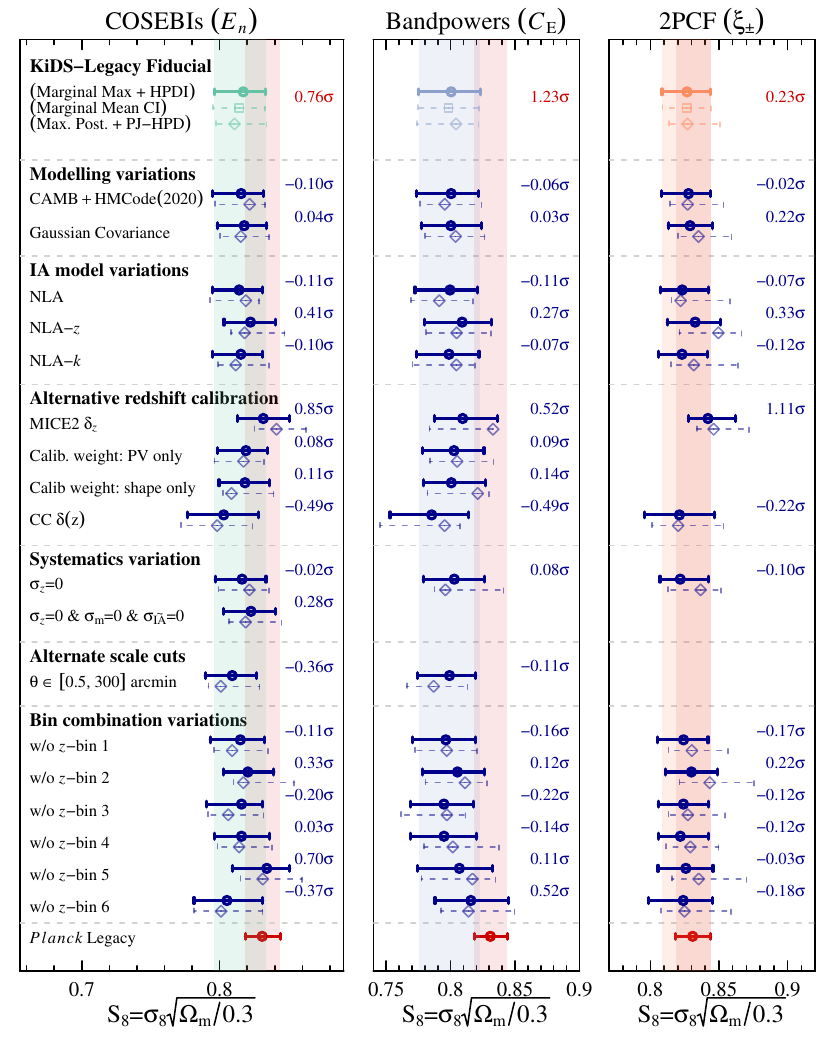}
  \caption{Whisker diagram of $S_8$ cosmological results presented in this work. Each panel shows the compilation 
  of results for one of our two-point statistics, as annotated. In each panel, we show the fiducial constraints in the 
  relevant statistic with coloured points (colour-coded by the statistic), and highlight the fiducial marginal HPDI 
  with a band of the equivalent colour. We also highlight the marginal HPDI in $S_8$ from {\it Planck}-Legacy (shaded red).
  All variations to our analysis are annotated, and values plotted in dark blue.
  We annotate each analysis variation with the difference (quantified using the Gaussian approximation to the Hellinger tension) 
  that the resulting marginal mode/HPDI has with respect to the fiducial (dark blue) or {\it Planck} (red). }\label{fig:whisker}
\end{figure*}

In this section we show the $S_8$ equivalent of our main whisker-diagrams, for comparison with previous work.
Figure~\ref{fig:whisker} shows the constraints for our \cosebisE, \bandpowersE, and \xipm\ statistics, with all analysis
variations included. The variability between the constraints is larger in this projection than in the $\Sigma_8$
projections, due to shifting of the posterior mode along the $\sigma_8$, $\Omega_{\rm m}$ degeneracy. Otherwise, we note
that all conclusions of this work are unchanged when analysing the $S_8$ space, rather than the $\Sigma_8$ space.

\subsection{Redshift distributions} 

\begin{figure*}
  \sidecaption
  \includegraphics[width=0.65\textwidth]{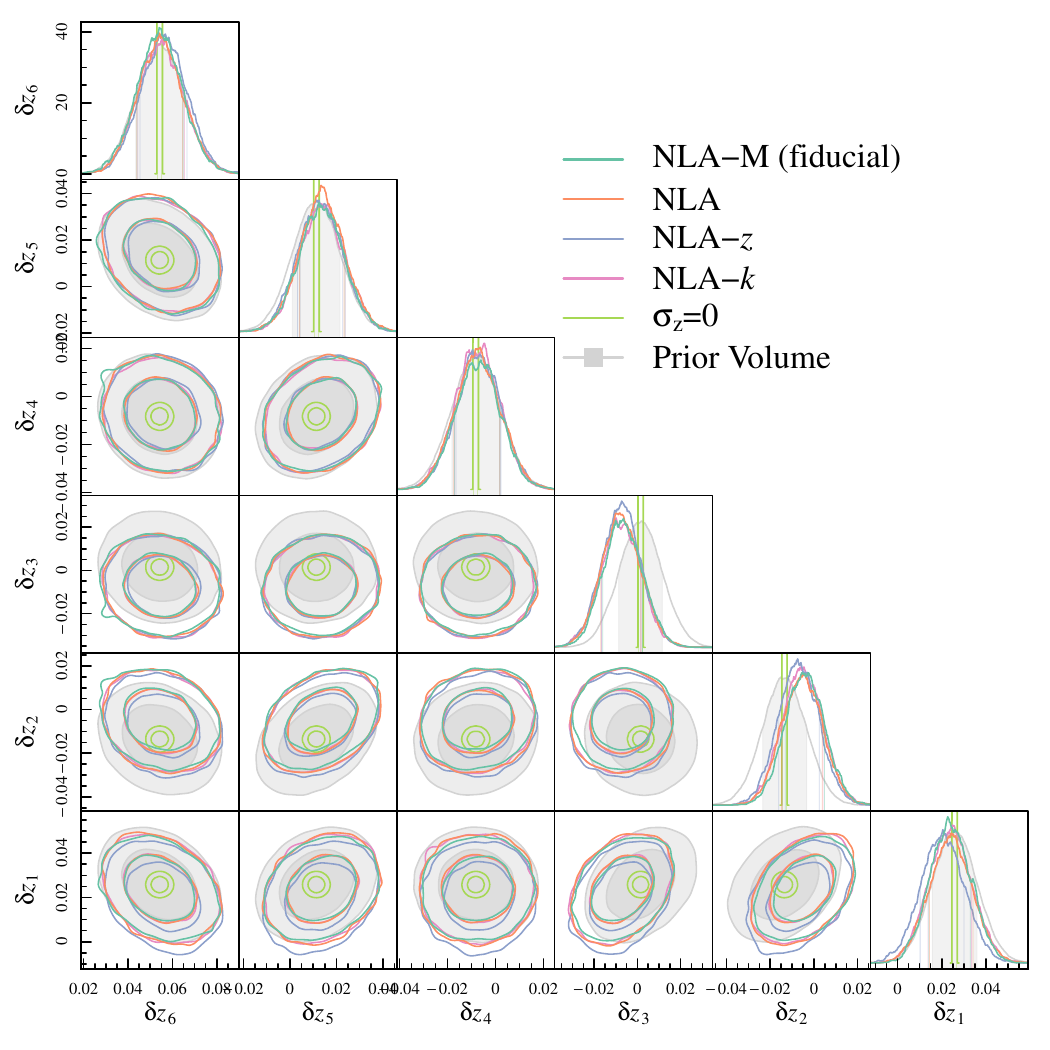}
  \caption{Posterior constraints on our redshift distribution bias parameters for chains utilising our various intrinsic
  alignment models. The filled contour shows the prior derived from our \skills\ simulations. Smoothing kernels are
  matched between all contours (and marginal distributions), and are shown by the non-zero width of the $\sigma_z=0$
  contours/marginals.}\label{fig:nzpost}
\end{figure*}

Figure \ref{fig:nzpost} shows the posterior constraints on our redshift distribution bias parameters, for chains
analysed using various intrinsic alignment models, compared to the prior. The chains are all consistent with the prior, 
with only the low-$z$ tomographic bins showing any tendency to shift away from the prior. Moreover, the posteriors are
all fully consistent between the various analyses, demonstrating that the choice of intrinsic alignment model does not
have a significant impact over the inferred redshift distribution biases. 

\subsection{Cosmological parameters}

\begin{figure*}
  \sidecaption
  \includegraphics[width=0.65\textwidth]{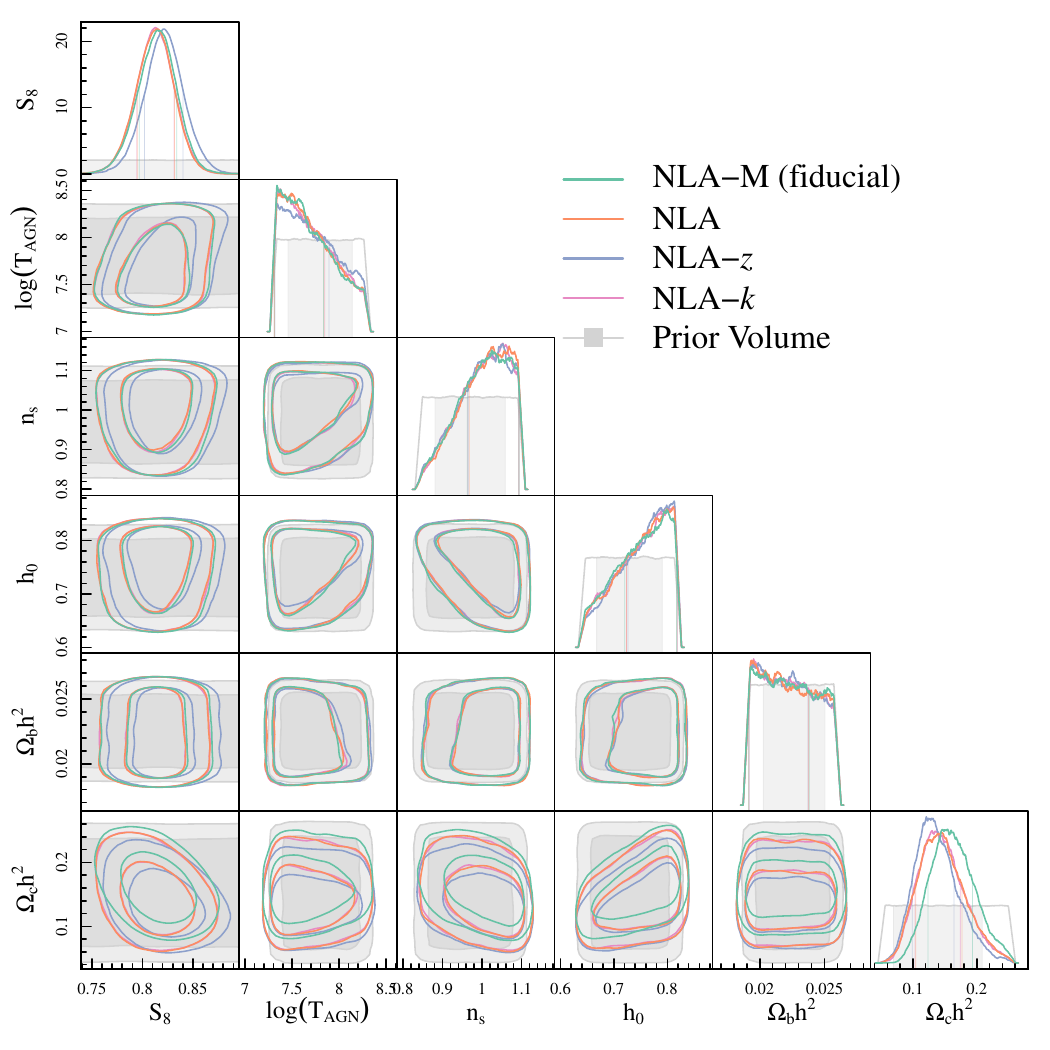}
  \caption{Posterior constraints on our cosmological parameters for chains utilising our various intrinsic
  alignment models. The filled contour shows the prior on each cosmological parameter.}\label{fig:cosmopost}
\end{figure*}

Figure \ref{fig:cosmopost} shows the posterior constraints on our cosmological parameters, again for chains
analysed using various intrinsic alignment models, and again compared to the prior. Only $S_8$ and $\Omega_{\rm m}$ 
can be reliably considered to be constrained, although there are interesting preferences in certain parameters. 

\FloatBarrier
\onecolumn

\section{Tabulated constraints}\label{sec:tables}

\begin{longtable}{cc|ccccccccc}
  \caption{Constraints on $\Sigma_8$ under our analysis variations.}\label{tab: longtable}\\
  \hline\hline
       &   Setup    &  Statistic & $\alpha$ & $\chi^2$  &   dof$^*$   & PTE   & PJ-HPD & Marginal                  & Max. Apost.               & Marginal                  \\
       &            &            &          &           &         &       &    $N_{\rm samp}$  &          Mode+HPD         &               +PJ-HPD     &          Mean + CI        \\
  \hline
  \endfirsthead

  \multicolumn{11}{c}{\tablename\ \thetable{} continued.}\\
  \hline\hline
       &   Setup    &  Statistic & $\alpha$ & $\chi^2$  &   dof$^*$   & PTE   & $N_{\rm samp}^{\rm PJ-HPD}$ & Marginal                  & Max. Apost.               & Marginal                  \\
       &            &            &          &           &         &       &                             &          Mode+HPD         &               +PJ-HPD     &          Mean + CI        \\
  \hline
  \endhead

  \hline
  \multicolumn{11}{r}{{Continued on next page}}\\
  \endfoot
  
  \hline
  \endlastfoot
  
          $\Sigma_8$ &          Iteration 1 &      $E_n$ & 0.58 & 127.3 & 120.5 & 0.318 &      23      & $0.821^{+0.014}_{-0.016}$ & $0.825^{+0.021}_{-0.014}$ & $0.820^{+0.015}_{-0.015}$ \\ [+0.1cm]
          $\Sigma_8$ &          Iteration 1 & $C_{\rm E}$ & 0.60 & 151.0 & 162.5 & 0.731 &      18      & $0.811^{+0.020}_{-0.014}$ & $0.824^{+0.006}_{-0.028}$ & $0.813^{+0.017}_{-0.017}$ \\ [+0.1cm]
          $\Sigma_8$ &          Iteration 1 &  $\xi_\pm$ & 0.55 & 413.1 & 351.5 & 0.013 &      23      & $0.823^{+0.014}_{-0.018}$ & $0.829^{+0.013}_{-0.020}$ & $0.821^{+0.016}_{-0.016}$ \\ [+0.1cm]
  \midrule 
          $\Sigma_8$ &             No Bin 1 &      $E_n$ & 0.58 & 89.8 & 84.5 & 0.326 &      31      & $0.823^{+0.014}_{-0.017}$ & $0.819^{+0.017}_{-0.013}$ & $0.821^{+0.015}_{-0.015}$ \\ [+0.1cm]
          $\Sigma_8$ &             No Bin 1 & $C_{\rm E}$ & 0.60 & 102.0 & 114.5 & 0.792 &      74      & $0.814^{+0.019}_{-0.015}$ & $0.818^{+0.020}_{-0.015}$ & $0.815^{+0.017}_{-0.017}$ \\ [+0.1cm]
          $\Sigma_8$ &             No Bin 1 &  $\xi_\pm$ & 0.56 & 282.5 & 249.5 & 0.074 &      33      & $0.823^{+0.016}_{-0.017}$ & $0.825^{+0.021}_{-0.014}$ & $0.822^{+0.016}_{-0.016}$ \\ [+0.1cm]
          $\Sigma_8$ &             No Bin 2 &      $E_n$ & 0.56 & 78.5 & 84.5 & 0.663 &      32      & $0.820^{+0.015}_{-0.017}$ & $0.814^{+0.030}_{-0.004}$ & $0.820^{+0.015}_{-0.015}$ \\ [+0.1cm]
          $\Sigma_8$ &             No Bin 2 & $C_{\rm E}$ & 0.60 & 116.4 & 114.5 & 0.433 &      49      & $0.811^{+0.019}_{-0.016}$ & $0.819^{+0.023}_{-0.016}$ & $0.813^{+0.017}_{-0.017}$ \\ [+0.1cm]
          $\Sigma_8$ &             No Bin 2 &  $\xi_\pm$ & 0.55 & 285.5 & 249.5 & 0.058 &      54      & $0.819^{+0.020}_{-0.013}$ & $0.829^{+0.019}_{-0.018}$ & $0.822^{+0.017}_{-0.017}$ \\ [+0.1cm]
          $\Sigma_8$ &             No Bin 3 &      $E_n$ & 0.59 & 93.6 & 84.5 & 0.234 &      10      & $0.824^{+0.015}_{-0.017}$ & $0.819^{+0.025}_{-0.008}$ & $0.823^{+0.015}_{-0.016}$ \\ [+0.1cm]
          $\Sigma_8$ &             No Bin 3 & $C_{\rm E}$ & 0.61 & 113.8 & 114.5 & 0.501 &      25      & $0.818^{+0.018}_{-0.017}$ & $0.828^{+0.011}_{-0.024}$ & $0.818^{+0.017}_{-0.017}$ \\ [+0.1cm]
          $\Sigma_8$ &             No Bin 3 &  $\xi_\pm$ & 0.56 & 288.8 & 249.5 & 0.044 &      18      & $0.824^{+0.015}_{-0.017}$ & $0.823^{+0.021}_{-0.012}$ & $0.822^{+0.016}_{-0.016}$ \\ [+0.1cm]
          $\Sigma_8$ &             No Bin 4 &      $E_n$ & 0.58 & 98.6 & 84.5 & 0.140 &      51      & $0.824^{+0.015}_{-0.016}$ & $0.820^{+0.019}_{-0.012}$ & $0.823^{+0.016}_{-0.015}$ \\ [+0.1cm]
          $\Sigma_8$ &             No Bin 4 & $C_{\rm E}$ & 0.60 & 109.1 & 114.5 & 0.625 &       5      & $0.814^{+0.016}_{-0.019}$ & $0.818^{+0.016}_{-0.020}$ & $0.812^{+0.017}_{-0.017}$ \\ [+0.1cm]
          $\Sigma_8$ &             No Bin 4 &  $\xi_\pm$ & 0.56 & 301.5 & 249.5 & 0.013 &      43      & $0.822^{+0.017}_{-0.015}$ & $0.825^{+0.019}_{-0.014}$ & $0.823^{+0.016}_{-0.016}$ \\ [+0.1cm]
          $\Sigma_8$ &             No Bin 5 &      $E_n$ & 0.56 & 99.6 & 84.5 & 0.125 &      31      & $0.823^{+0.019}_{-0.016}$ & $0.824^{+0.022}_{-0.014}$ & $0.825^{+0.017}_{-0.017}$ \\ [+0.1cm]
          $\Sigma_8$ &             No Bin 5 & $C_{\rm E}$ & 0.59 & 102.5 & 114.5 & 0.782 &       7      & $0.809^{+0.019}_{-0.019}$ & $0.821^{+0.016}_{-0.026}$ & $0.808^{+0.019}_{-0.019}$ \\ [+0.1cm]
          $\Sigma_8$ &             No Bin 5 &  $\xi_\pm$ & 0.56 & 296.1 & 249.5 & 0.023 &      25      & $0.822^{+0.015}_{-0.021}$ & $0.823^{+0.023}_{-0.016}$ & $0.818^{+0.018}_{-0.018}$ \\ [+0.1cm]
          $\Sigma_8$ &             No Bin 6 &      $E_n$ & 0.61 & 91.5 & 84.5 & 0.283 &      32      & $0.817^{+0.019}_{-0.017}$ & $0.814^{+0.020}_{-0.016}$ & $0.817^{+0.018}_{-0.018}$ \\ [+0.1cm]
          $\Sigma_8$ &             No Bin 6 & $C_{\rm E}$ & 0.59 & 105.5 & 114.5 & 0.715 &       4      & $0.817^{+0.019}_{-0.023}$ & $0.813^{+0.039}_{-0.010}$ & $0.815^{+0.021}_{-0.020}$ \\ [+0.1cm]
          $\Sigma_8$ &             No Bin 6 &  $\xi_\pm$ & 0.57 & 301.4 & 249.5 & 0.014 &      18      & $0.816^{+0.018}_{-0.021}$ & $0.814^{+0.033}_{-0.011}$ & $0.815^{+0.019}_{-0.019}$ \\ [+0.1cm]
  \midrule 
          $\Sigma_8$ &                 CAMB &      $E_n$ & 0.58 & 128.2 & 120.5 & 0.299 &      38      & $0.817^{+0.015}_{-0.015}$ & $0.830^{+0.002}_{-0.027}$ & $0.818^{+0.015}_{-0.015}$ \\ [+0.1cm]
          $\Sigma_8$ &                 CAMB & $C_{\rm E}$ & 0.60 & 151.7 & 162.5 & 0.718 &      35      & $0.812^{+0.017}_{-0.017}$ & $0.825^{+0.013}_{-0.024}$ & $0.812^{+0.017}_{-0.017}$ \\ [+0.1cm]
          $\Sigma_8$ &                 CAMB &  $\xi_\pm$ & 0.56 & 411.1 & 351.5 & 0.016 &      23      & $0.818^{+0.019}_{-0.014}$ & $0.818^{+0.025}_{-0.009}$ & $0.821^{+0.016}_{-0.016}$ \\ [+0.1cm]
  \midrule 
          $\Sigma_8$ &  Gaussian Cov &      $E_n$ & 0.60 & 127.6 & 120.5 & 0.312 &      19      & $0.820^{+0.014}_{-0.011}$ & $0.823^{+0.011}_{-0.014}$ & $0.821^{+0.013}_{-0.012}$ \\ [+0.1cm]
          $\Sigma_8$ &  Gaussian Cov & $C_{\rm E}$ & 0.61 & 151.9 & 162.5 & 0.714 &      24      & $0.817^{+0.014}_{-0.016}$ & $0.819^{+0.020}_{-0.013}$ & $0.816^{+0.015}_{-0.015}$ \\ [+0.1cm]
          $\Sigma_8$ &  Gaussian Cov &  $\xi_\pm$ & 0.56 & 411.3 & 351.5 & 0.015 &      18      & $0.825^{+0.013}_{-0.014}$ & $0.826^{+0.022}_{-0.010}$ & $0.825^{+0.013}_{-0.013}$ \\ [+0.1cm]
  \midrule 
          $\Sigma_8$ &        MICE2 $\delta_z$ &      $E_n$ & 0.60 & 123.2 & 120.5 & 0.415 &      18      & $0.836^{+0.013}_{-0.012}$ & $0.842^{+0.012}_{-0.015}$ & $0.837^{+0.012}_{-0.012}$ \\ [+0.1cm]
          $\Sigma_8$ &        MICE2 $\delta_z$ & $C_{\rm E}$ & 0.62 & 150.2 & 162.5 & 0.746 &       6      & $0.832^{+0.015}_{-0.016}$ & $0.840^{+0.009}_{-0.022}$ & $0.832^{+0.015}_{-0.015}$ \\ [+0.1cm]
          $\Sigma_8$ &        MICE2 $\delta_z$ &  $\xi_\pm$ & 0.57 & 400.3 & 351.5 & 0.037 &      12      & $0.839^{+0.015}_{-0.013}$ & $0.845^{+0.008}_{-0.019}$ & $0.841^{+0.014}_{-0.014}$ \\ [+0.1cm]
  \midrule 
          $\Sigma_8$ & Calib PV wgt &      $E_n$ & 0.58 & 137.8 & 120.5 & 0.134 &      26      & $0.820^{+0.015}_{-0.014}$ & $0.823^{+0.017}_{-0.013}$ & $0.821^{+0.014}_{-0.014}$ \\ [+0.1cm]
          $\Sigma_8$ & Calib PV wgt & $C_{\rm E}$ & 0.60 & 149.7 & 162.5 & 0.756 &      42      & $0.815^{+0.017}_{-0.017}$ & $0.818^{+0.025}_{-0.014}$ & $0.815^{+0.017}_{-0.017}$ \\ [+0.1cm]
  \midrule 
          $\Sigma_8$ & Calib shape wgt &      $E_n$ & 0.58 & 142.1 & 120.5 & 0.087 &      40      & $0.822^{+0.015}_{-0.013}$ & $0.829^{+0.008}_{-0.020}$ & $0.822^{+0.014}_{-0.014}$ \\ [+0.1cm]
          $\Sigma_8$ & Calib shape wgt & $C_{\rm E}$ & 0.60 & 150.2 & 162.5 & 0.746 &      39      & $0.816^{+0.019}_{-0.016}$ & $0.831^{+0.008}_{-0.028}$ & $0.816^{+0.017}_{-0.017}$ \\ [+0.1cm]
  \midrule 
          $\Sigma_8$ &           CC Delta z &      $E_n$ & 0.63 & 129.6 & 120.5 & 0.269 &      25      & $0.831^{+0.020}_{-0.017}$ & $0.831^{+0.018}_{-0.018}$ & $0.832^{+0.018}_{-0.018}$ \\ [+0.1cm]
          $\Sigma_8$ &           CC Delta z & $C_{\rm E}$ & 0.65 & 152.4 & 162.5 & 0.704 &      11      & $0.827^{+0.022}_{-0.020}$ & $0.829^{+0.012}_{-0.032}$ & $0.827^{+0.021}_{-0.021}$ \\ [+0.1cm]
          $\Sigma_8$ &           CC Delta z &  $\xi_\pm$ & 0.61 & 415.4 & 351.5 & 0.011 &      27      & $0.826^{+0.020}_{-0.018}$ & $0.828^{+0.027}_{-0.014}$ & $0.827^{+0.019}_{-0.019}$ \\ [+0.1cm]
  \midrule 
          $\Sigma_8$ &     [0.5,300] arcmin &      $E_n$ & 0.58 & 138.9 & 120.5 & 0.121 &      16      & $0.818^{+0.014}_{-0.016}$ & $0.816^{+0.013}_{-0.017}$ & $0.817^{+0.015}_{-0.015}$ \\ [+0.1cm]
  \midrule 
          $\Sigma_8$ &                  NLA &      $E_n$ & 0.54 & 128.2 & 120.5 & 0.299 &      63      & $0.810^{+0.017}_{-0.015}$ & $0.816^{+0.013}_{-0.020}$ & $0.811^{+0.016}_{-0.016}$ \\ [+0.1cm]
          $\Sigma_8$ &                  NLA & $C_{\rm E}$ & 0.58 & 153.4 & 162.5 & 0.683 &      66      & $0.809^{+0.018}_{-0.018}$ & $0.810^{+0.018}_{-0.018}$ & $0.808^{+0.018}_{-0.018}$ \\ [+0.1cm]
          $\Sigma_8$ &                  NLA &  $\xi_\pm$ & 0.53 & 414.4 & 351.5 & 0.012 &      32      & $0.821^{+0.017}_{-0.016}$ & $0.817^{+0.033}_{-0.006}$ & $0.820^{+0.016}_{-0.017}$ \\ [+0.1cm]
  \midrule 
          $\Sigma_8$ &              NLA-$z$ &      $E_n$ & 0.55 & 126.6 & 120.5 & 0.334 &      69      & $0.819^{+0.016}_{-0.018}$ & $0.817^{+0.021}_{-0.013}$ & $0.819^{+0.017}_{-0.017}$ \\ [+0.1cm]
          $\Sigma_8$ &              NLA-$z$ & $C_{\rm E}$ & 0.58 & 153.4 & 162.5 & 0.683 &      68      & $0.817^{+0.016}_{-0.023}$ & $0.813^{+0.018}_{-0.020}$ & $0.814^{+0.019}_{-0.019}$ \\ [+0.1cm]
          $\Sigma_8$ &              NLA-$z$ &  $\xi_\pm$ & 0.54 & 411.5 & 351.5 & 0.015 &      46      & $0.828^{+0.016}_{-0.021}$ & $0.837^{+0.018}_{-0.022}$ & $0.827^{+0.018}_{-0.018}$ \\ [+0.1cm]
  \midrule 
          $\Sigma_8$ &              NLA-$k$ &      $E_n$ & 0.54 & 127.5 & 120.5 & 0.314 &      69      & $0.813^{+0.014}_{-0.018}$ & $0.808^{+0.021}_{-0.011}$ & $0.811^{+0.016}_{-0.016}$ \\ [+0.1cm]
          $\Sigma_8$ &              NLA-$k$ & $C_{\rm E}$ & 0.59 & 152.5 & 162.5 & 0.702 &      99      & $0.810^{+0.016}_{-0.020}$ & $0.823^{+0.019}_{-0.025}$ & $0.808^{+0.018}_{-0.018}$ \\ [+0.1cm]
          $\Sigma_8$ &              NLA-$k$ &  $\xi_\pm$ & 0.53 & 413.5 & 351.5 & 0.013 &      90      & $0.819^{+0.018}_{-0.016}$ & $0.826^{+0.027}_{-0.015}$ & $0.820^{+0.017}_{-0.017}$ \\ [+0.1cm]
  \midrule 
          $\Sigma_8$ &            No Sigmaz &      $E_n$ & 0.58 & 131.9 & 120.5 & 0.225 &      40      & $0.818^{+0.015}_{-0.014}$ & $0.826^{+0.011}_{-0.019}$ & $0.819^{+0.015}_{-0.014}$ \\ [+0.1cm]
          $\Sigma_8$ &            No Sigmaz & $C_{\rm E}$ & 0.60 & 154.1 & 162.5 & 0.669 &      62      & $0.811^{+0.019}_{-0.015}$ & $0.815^{+0.020}_{-0.014}$ & $0.813^{+0.017}_{-0.016}$ \\ [+0.1cm]
          $\Sigma_8$ &            No Sigmaz &  $\xi_\pm$ & 0.56 & 418.6 & 351.5 & 0.008 &      41      & $0.819^{+0.015}_{-0.017}$ & $0.826^{+0.011}_{-0.021}$ & $0.819^{+0.016}_{-0.016}$ \\ [+0.1cm]
  \midrule 
          $\Sigma_8$ &       No Systematics &      $E_n$ & 0.60 & 133.7 & 122.5 & 0.231 &     135      & $0.829^{+0.012}_{-0.015}$ & $0.823^{+0.016}_{-0.010}$ & $0.827^{+0.013}_{-0.013}$ \\ [+0.1cm]
  \bottomrule
\end{longtable}

%\twocolumn

\section{Post-unblinding analyses and changes}\label{sec:updates}

In this appendix we detail all analyses that were taken, and any changes that were made to the analysis and/or
manuscript, after unblinding. 

\subsection{Comparison between Legacy and KiDS-1000}

As discussed in Sect.~\ref{sec:discussion}, \kidslegacy\ presents a decrease in the marginal Hellinger tension with
\planck\ of $2.2\sigma$ for $\Sigma_8$ ($1.65\sigma$ for $S_8$).  
After the unblinding of \kidslegacy\ we investigated the differences between the Legacy and KiDS-1000 samples and
analyses, to determine whether the change in cosmological parameters could be attributed to systematic or random
variations. To do this, we reanalysed the KiDS-1000 and KiDS-Legacy shape catalogues with a series of analysis variations,
which changed the sample used for cosmological inference and/or the analysis choices that were used in the inference. 
This required us, in each case, to rerun our \nz\ estimation, our correlation function measurement, covariance
computation, and posterior inference. We did not, however, remeasure the redshift distribution bias parameters or the
multiplicative shape measurement biases for every setup. 

\begin{figure*}[h!]
  \sidecaption
  \includegraphics[width=0.7\textwidth]{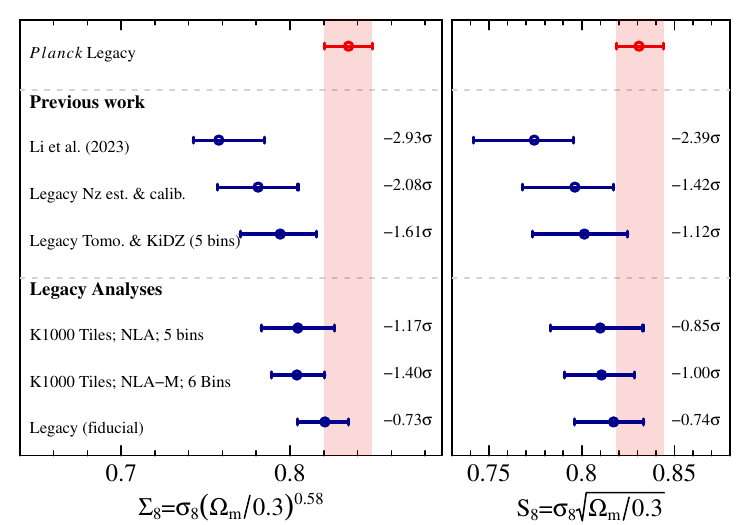}
  \caption{Variation in recovered \cosebisE\ cosmological constraints for $\Sigma_8$ ({\em left}) and $S_8$ ({\em
  right}) when moving between KiDS-1000 and \kidslegacy\ analyses, relative to the constraints from \planck\ (red).
  Each point is annotated with the Hellinger tension between the marginal constraint and that of \planck.    
  }\label{fig:changes_whisker}
\end{figure*}

Figure \ref{fig:changes_whisker} presents a graphical summary of the changes we found in our recovered
cosmological constraints when analysing KiDS-1000 using Legacy analysis choices 
and calibration methods, and when analysing Legacy using subsets defined by the KiDS-1000 footprint and photo-z
baseline. 
Our investigation of the consistency between the data from \citet{li/etal:2023b} and Legacy was focussed on two areas:
the impact of Legacy analysis choices and calibration methods, and the impact of the new on-sky area available to Legacy. 
We quantified the effect of
each analysis variation by recomputing the Hellinger tensions with respect to \planck, as this allows us to circumvent
the complex covariance between various analyses of KiDS-1000 and \kidslegacy. 

As a result of this investigation, we found no evidence of statistically significant systematic biases in the sample and
analysis of \kidslegacy. We determined that the primary differences between the cosmological results of
\citet{li/etal:2023b} and Legacy are attributable to changes to our redshift calibration sample and methodology, and to
statistical noise driven by the new area available to \kidslegacy\ outside the KiDS-1000 footprint.  We have made no
changes to the analysis methodology of Legacy as presented in the main text after this investigation, and thus conclude
that this suite of tests do not represent a possible source of bias in our blinded analysis. 

\subsubsection{Analysing KiDS-1000 with Legacy calibration methods}
We started by investigating the impact of the updated redshift estimation and calibration methods of
\citet{wright/etal:2025} on the cosmological constraints of \citet{li/etal:2023b}. To do this we reran the entirety of 
the cosmic shear measurement from KiDS-1000, from raw shape catalogues to cosmological inference, incorporating our new redshift
estimation and calibration methodologies (leveraging e.g. gold-weight and simulation construction with sample
matching). In this analysis we computed revised redshift bias estimates using the original KiDS-1000 redshift calibration
sample, but updating to SKILLS image simulations and gold-weight redshift distribution estimation. 
The result of this end-to-end reanalysis are provided in Fig.~\ref{fig:changes_whisker} as `Legacy Nz est. \& calib.'. 
We found that the updated redshift distribution calibration method brings the cosmological estimates from
the \citet{li/etal:2023b} sample into better agreement with \planck, at $2.08\sigma$ consistency in $\Sigma_8$ and
$1.42\sigma$ consistency in $S_8$. Importantly, both of these measurements satisfy the typical KiDS threshold for
consistency of $2.3\sigma$ (equivalent to a one-sided probability to exceed of $0.01$). 

\subsubsection{Updating KiDS-1000 to the Legacy calibration sample} \label{sec:k1000_as_legacy}
We also measured the contribution to the difference between KiDS-1000 and KiDS-Legacy caused by the differences to
our analysis choices and calibration sample, including our new tomographic bin limits and our greatly expanded redshift 
calibration dataset. To do this we again performed a full end-to-end reanalysis of KiDS-1000, this time with our new
tomographic bin limits (excluding the sixth bin), and new redshift calibration sample from KiDZ. In this analysis we
assume the redshift distribution biases and multiplicative shear biases are the same as the fiducial KiDS-Legacy
analysis.  We find that the use of the larger KiDZ calibration sample also slightly reduces the offset with \planck, to
$1.61\sigma$ consistency in $\Sigma_8$ and $1.12\sigma$ consistency in $S_8$ (`Legacy Tomo. \& KiDZ $($5 bins$)$').
This demonstrates that the KiDS-1000 sample, when analysed with our larger spectroscopic calibration sample and new
redshift calibration methods (but without adding higher redshift sources, or using updated imaging, or new photo-$z$, or
additional area) is consistent with the results from \planck\ and with the results from Legacy. 

\subsubsection{Legacy analysis in the KiDS-1000 volume}
We explored the consistency between the constraints from KiDS-1000 and \kidslegacy\ by analysing the two samples in 
the same volume, and with the same analysis choices. We were able to do this fairly easily thanks to our analysis from 
Sect.~\ref{sec:k1000_as_legacy}, which homogenised the tomography and \nz\ estimation and calibration between the two
samples. By restricting \kidslegacy\ to only the footprint of KiDS-1000, excluding the sixth tomographic bin, and 
using the NLA IA model, we were able to perform a comparison that is mostly free of statistical fluctuations. In
fact there are still differences in the source samples due to changes in source extraction, image masking, source selection,
and more, which we assumed add only a small amount of statistical noise to the comparison. 
We find that the Legacy sample in the same volume as KiDS-1000 is slightly closer to \planck\ than KiDS-1000, with a 
$1.17\sigma$ consistency in $\Sigma_8$ and $0.85\sigma$ consistency in $S_8$. This indicates that the sum total of
changes (such as updated imaging calibrations, photo-$z$, masking, etc.) lead to a $0.44\sigma$ ($0.27\sigma$) decrease in
the Hellinger distance between \planck\ and KiDS; significantly less than the change introduced by our new calibration
methods.  

\subsubsection{Sixth tomographic bin and NLA-$M$}
We further investigated the influence of adding the sixth tomographic bin and our new NLA-$M$ IA model to
the analysis of Legacy in the footprint of KiDS-1000 (`K1000 Tiles; NLA-M; 6 bins'), finding that these changes caused a
significant reduction in our marginal uncertainties (as expected) but did not lead to a significant change in the
estimated cosmological parameters. As such, our Hellinger distance with \planck\ increases with these additions, to 
$1.40\sigma$ consistency in $\Sigma_8$ and $1.00\sigma$ consistency in $S_8$. 

\subsubsection{Additional Legacy area}
Finally, we added the data back into Legacy that resides outside the footprint of KiDS-1000, and recover our fiducial
result, which exhibits a $0.73\sigma$ consistency in $\Sigma_8$ and $0.74\sigma$ consistency in $S_8$. 
This demonstrates that there is a non-negligible statistical noise component in the difference between the sources probed
within the footprint of KiDS-1000 and over the final \kidslegacy\ area. 

\subsection{Expanded scale cuts}
Post unblinding, we discovered a typographical error in the pipeline which performed the measurement of correlation
functions with expanded scale cuts (Sect.~\ref{sec:scalecuts}). This error meant that the pre-unblinding measurements
were made with uncalibrated shapes directly from our shape measurement code \lensfit, rather than with the recalibrated 
shapes (Sect.~\ref{sec: rowestats}). We reran the expanded scale cut analysis with the correct recalibrated shapes
post-unblinding, finding that the consistency between the analyses with fiducial and expanded scale cuts improved from
$0.3\sigma$ to $0.15\sigma$ with the use of the correct shapes. No conclusions were changed as a result of this error. 

\subsection{Iterative covariances} 
All fiducial chains were recomputed using our iterative covariance framework post-unblinding. This was a conscious
pre-unblinding choice, as the iterative covariances require non-negligible CPU time for the computation of MAP and new
covariances. We found that this process had a negligible impact on our constraints (see Sect.~\ref{sec:iterative}), and
thus does not represent a possible source of bias in our blinded analysis.

\end{appendix}

\end{document}